%% file: ms.tex
\documentclass[apj]{emulateapj}
\usepackage{apjfonts}
\usepackage{graphicx}
\usepackage{longtable}

\newcommand{\teff}{T_{\rm eff}}
\renewcommand{\bv}{B\, -\, V}

\newcommand{\vk}{V\, -\, K_s}

\newcommand{\ebv}{E(B\, -\, V)}
\newcommand{\dmn}{(m\, -\, M)_0}

\slugcomment{Accepted for Publication in the Astrophysical Journal Supplement Series}
\shorttitle{Spectroscopic Survey of G and K Dwarfs in the {\it Hipparcos} Catalog. I.} \shortauthors{Kim et~al.}

\begin{document}

\title{Spectroscopic Survey of G and K Dwarfs in the {\it Hipparcos} Catalog. I.\\ Comparison between the {\it Hipparcos} and Photometric Parallaxes}

\author{Bokyoung Kim\altaffilmark{1},
Deokkeun An\altaffilmark{1,2},
John R.\ Stauffer\altaffilmark{2,3},
Young Sun Lee\altaffilmark{4},\\
Donald M.\ Terndrup\altaffilmark{2,5},
Jennifer A.\ Johnson\altaffilmark{5}
}

\altaffiltext{1}{Department of Science Education, Ewha Womans University, 52 Ewhayeodae-gil, Seodaemun-gu, Seoul 03760, Republic of Korea; deokkeun@ewha.ac.kr}
\altaffiltext{2}{Visiting astronomer, Kitt Peak National Observatory, National Optical Astronomy Observatory, which is operated by the Association of Universities for Research in Astronomy (AURA) under a cooperative agreement with the National Science Foundation.}
\altaffiltext{3}{Spitzer Science Center, California Institute of Technology, Pasadena, CA 91125}
\altaffiltext{4}{Department of Astronomy and Space Science, Chungnam National University, 99 Daehak-ro, Daejeon 34134, Republic of Korea}
\altaffiltext{5}{Department of Astronomy, The Ohio State University, 140 West 18th Avenue, Columbus, OH 43210}

\begin{abstract}

The tension between the {\it Hipparcos} parallax of the Pleiades and other independent distance estimates continues even after the new reduction of the {\it Hipparcos} astrometric data and the development of a new geometric distance measurement for the cluster. A short Pleiades distance from the {\it Hipparcos} parallax predicts a number of stars in the solar neighborhood that are sub-luminous at a given photospheric abundance. We test this hypothesis using spectroscopic abundances for a subset of stars in the {\it Hipparcos} catalog, which occupy the same region as the Pleiades in the color-magnitude diagram. We derive stellar parameters for $170$ nearby G and K type field dwarfs in the {\it Hipparcos} catalog based on high-resolution spectra obtained using KPNO 4-m echelle spectrograph. Our analysis shows that, when the {\it Hipparcos} parallaxes are adopted, most of our sample stars follow empirical color-magnitude relations. A small fraction of stars are too faint compared to main-sequence fitting relations by $\Delta M_V \ga 0.3$~mag, but the differences are marginal at a $2\sigma$ level partly due to relatively large parallax errors. On the other hand, we find that photometric distances of stars showing signatures of youth as determined from lithium absorption line strengths and $R'_{\rm HK}$ chromospheric activity indices are consistent with the {\it Hipparcos} parallaxes. Our result is contradictory to a suggestion that the Pleiades distance from main-sequence fitting is significantly altered by stellar activity and/or the young age of its stars, and provides an additional supporting evidence for the long distance scale of the Pleiades.

\end{abstract}


\section{Introduction}

The determination of accurate distances to stars is the key to understanding how stars and the Galaxy have formed and evolved. The {\it Hipparcos} mission \citep{perryman:97} was especially valuable, providing trigonometric parallaxes for $\sim10^5$ stars to a precision of $1$--$2$~mas. Therefore, it was a big surprise when the {\it Hipparcos} distance to the Pleiades open cluster was in disagreement with distances from the main-sequence (MS) fitting at more than a $3\sigma$ level \citep{pinsonneault:98}. To reconcile a short Pleiades from {\it Hipparcos}, it has been suggested that the metal abundance of the Pleiades is significantly lower than the solar \citep{percival:03}, thereby decreasing a distance from MS fitting. However, there are a large number of spectroscopic studies on the cluster's metallicity in the literature, which essentially indicate a near-solar metallicity of the cluster \citep[see references in][]{an:07b}. In addition, the enhanced helium abundance of the cluster was suggested \citep{belikov:98}, but the expected amount of helium has to be enormous ($Y\approx0.34$) in this solar metallicity cluster. An argument was also made that distance estimates from theoretical stellar models have been overestimated for young clusters due to yet unknown, age-related physics \citep{vanleeuwen:99}.

The discrepant {\it Hipparcos} result for the Pleiades has subsequently led to many efforts to determine the cluster's distance from binaries and independent parallax measurements \citep[e.g.,][]{munari:04,pan:04,soderblom:05}. These results confirm the long distance scale from MS fitting, supporting the hypothesis that the {\it Hipparcos} result was in error. The most likely explanation is related to the {\it Hipparcos} parallaxes themselves. \citet{pinsonneault:98} showed that a dozen bright stars near the center of the Pleiades all had virtually the same parallax ($\sim9$~mas), which is more than $1$~mas larger than the mean parallax for other cluster stars. They attributed this to a local zero-point error of individual stellar parallaxes that are correlated over the {\it Hipparcos}' $0.9\deg$ field of view \citep[see also][]{narayanan:99}. By re-reducing part of the {\it Hipparcos} data, \citet{makarov:02,makarov:03} were able to demonstrate that such correlated errors could explain the discrepant Pleiades parallax estimate. Additional effects may result from the way the {\it Hipparcos} data were obtained and analyzed \citep{vanleeuwen:05a,vanleeuwen:05b}, and it was hoped that a new reduction of the {\it Hipparcos} raw data might resolve the issue.

However, the new reduction of the {\it Hipparcos} data \citep{vanleeuwen:07a,vanleeuwen:07b} still leads to a short distance to the Pleiades. In the most recent analysis, \citet{vanleeuwen:09} found $8.32\pm0.13$~mas for the average parallax of the Pleiades, or $\dmn=5.40\pm0.03$ ($120.2\pm1.9$~pc), which is significantly shorter than the weighted average distance $\dmn=5.63\pm0.02$ ($133.7\pm1.2$~pc) from independent astrometric and binary solutions \citep[see references in][]{an:07b}. Moreover, \citet{melis:14} recently used the Very Long Baseline Interferometry (VLBI) to directly measure a geometric distance to the Pleiades, and found $\dmn=5.67\pm0.02$ ($136.2\pm1.2$~pc), which agrees with the long distance scale.

An alternative, but indirect test of the Pleiades distance can be made by examining nearby field stars that occupy the same region on a color-magnitude diagram (CMD) as those in the Pleiades. The short {\it Hipparcos} distance to the Pleiades predicts a number of stars in the solar neighborhood that are sub-luminous at a given photospheric abundance. Since absolute magnitudes ($M_V$) of stars are sensitive to the photospheric abundance, it is possible to distinguish sub-luminous stars (with the hypothesized Pleiades-like phenomenon) from normal ones, when accurate metallicity measurements are available. Here, our core assumption in this paper is that parallaxes for the majority of stars in the {\it Hipparcos} catalog are correct, but only a small fraction of these stars (such as those in the Pleiades) have incorrect parallaxes due to large, hidden systematic errors.

Previously, \citet{soderblom:98} performed such test using a set of nearby field stars, but found no stars of similar characteristics with the Pleiades members. However, the interpretation of their result is somewhat complicated by the fact that stars in their sample mostly have spectral types later than K2. Late-type, young, chromospherically active stars can be heavily spotted and hence variable, and their optical colors can differ significantly from those of older field stars of the same spectral type \citep{stauffer:03}. On the other hand, rapid rotation does not cause significant photometric anomalies of stars with spectral types earlier than K2. 

The goal of this paper is to obtain accurate metal abundances for a subset of G and early K-type stars in the {\it Hipparcos} catalog, and look for sub-luminous field stars at a given metallicity. Such stars will have longer distances from MS fitting than those computed from the {\it Hipparcos} parallaxes. Furthermore, if the assertion by \citet{vanleeuwen:09} is correct, and the young age of the Pleiades is responsible for the long MS-fitting distance to the cluster, young field stars, such as those selected from strong lithium absorptions or \ion{Ca}{2} H and K emissions, will be fainter than older counterparts by $\Delta M_V\ga0.2$~mag at a common [Fe/H]. Assuming a constant star formation rate and an age of $\sim8$~Gyr for the thin disk stellar population \citep[e.g., see][and references therein]{casagrande:16}, about $2.5\%$ of stars in the solar neighborhood were formed in the last $\sim200$~Myr. A couple of young stars should be present in a sample of $\sim100$ field stars. The expected number of such stars will decrease if an exponentially decreasing star formation rate is assumed \citep[e.g.,][]{aumer:09}, but a relatively large number of young open clusters in the solar neighborhood implies the presence of many young stars near the Sun. Furthermore, a vertical age gradient in the disk \citep[e.g.,][]{casagrande:16} would yield more young stars near the Galactic plane. A well-defined set of field stars can be used to disprove a null hypothesis that the short distance to the Pleiades is caused by the young age of the cluster.

This paper is organized as follows. In \S~\ref{sec:observation} we describe spectroscopic observations and data reductions. Derivation of stellar parameter is given in \S~\ref{sec:param}. In \S~\ref{sec:result} we derive MS-fitting distances for individual stars using spectroscopic metallicities, and compare them to the {\it Hipparcos} parallaxes. A summary of our results is given in \S~\ref{sec:summary}.

\section{Spectroscopic Observations and Data Reductions}\label{sec:observation}

\subsection{Sample Selection}\label{sec:sample}

\begin{figure*} \centering
\includegraphics[scale=0.65]{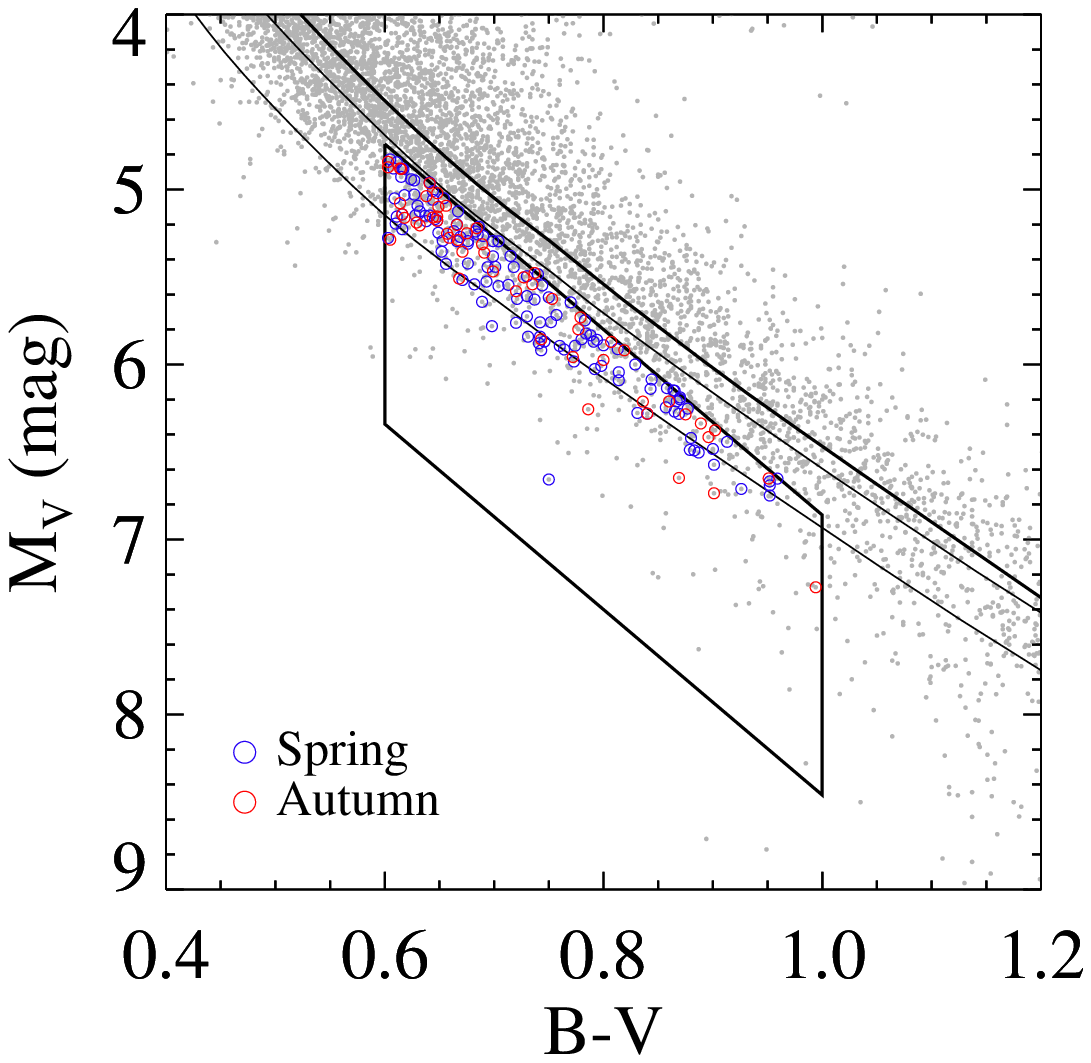}
\includegraphics[scale=0.65]{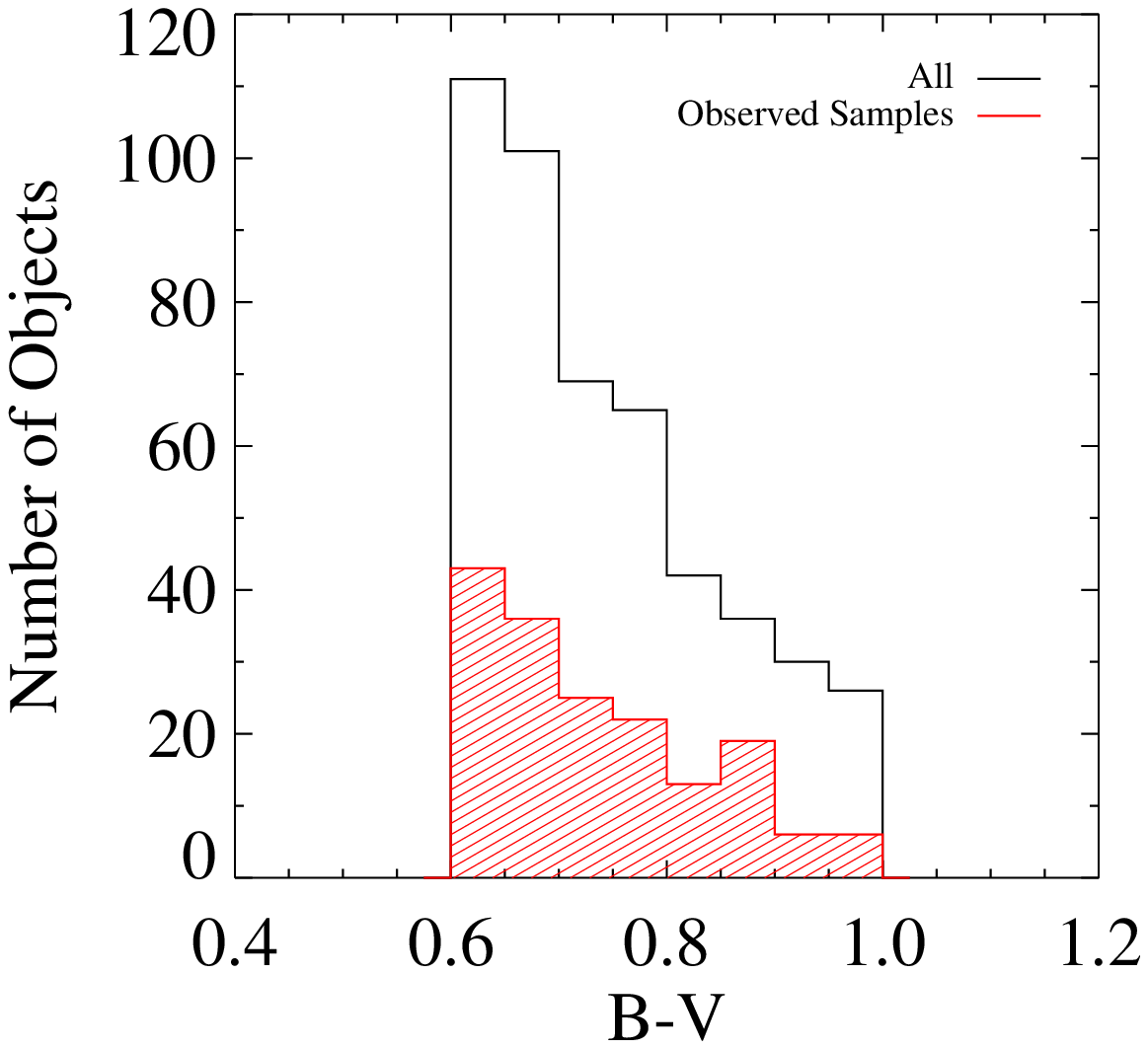}
\includegraphics[scale=0.65]{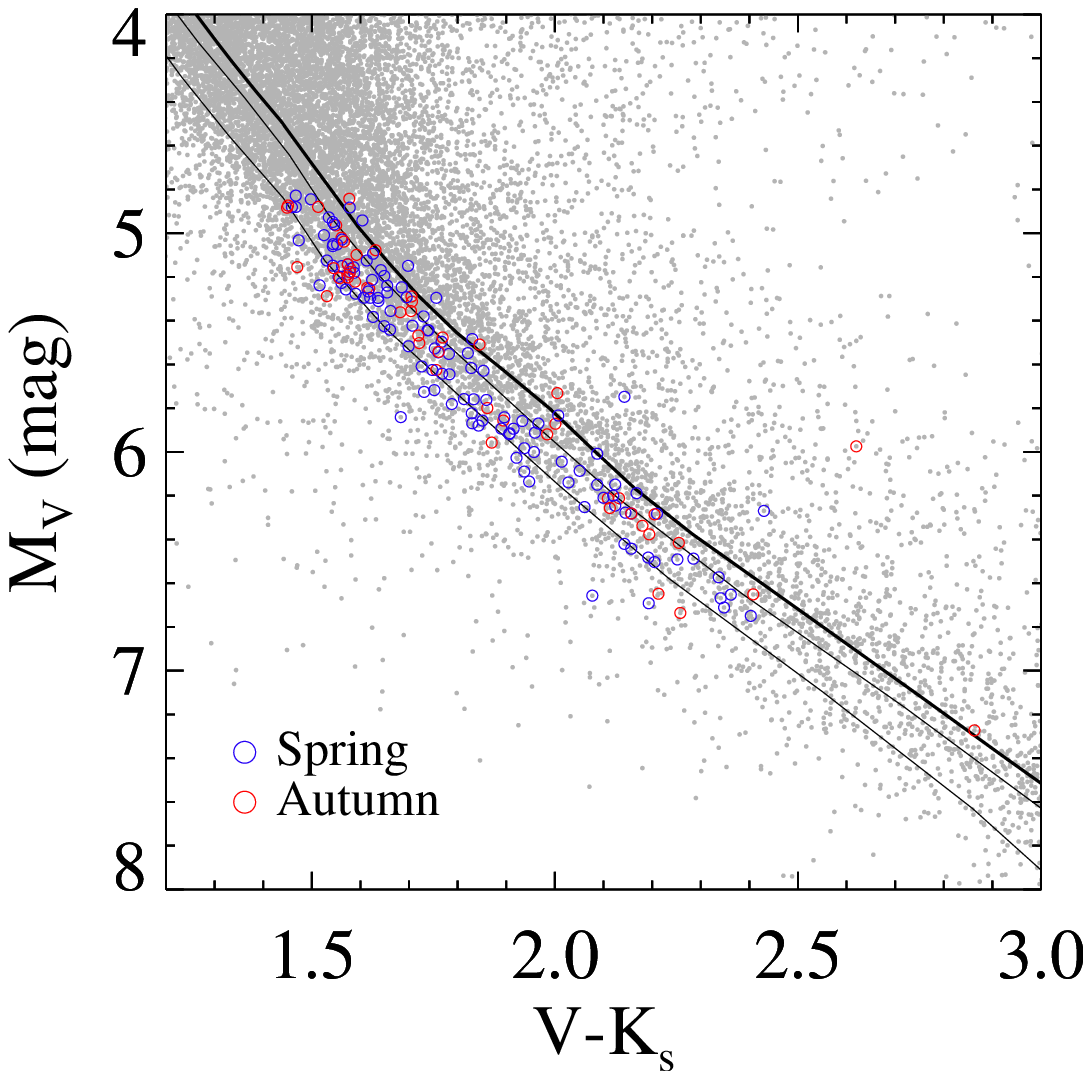}
\includegraphics[scale=0.65]{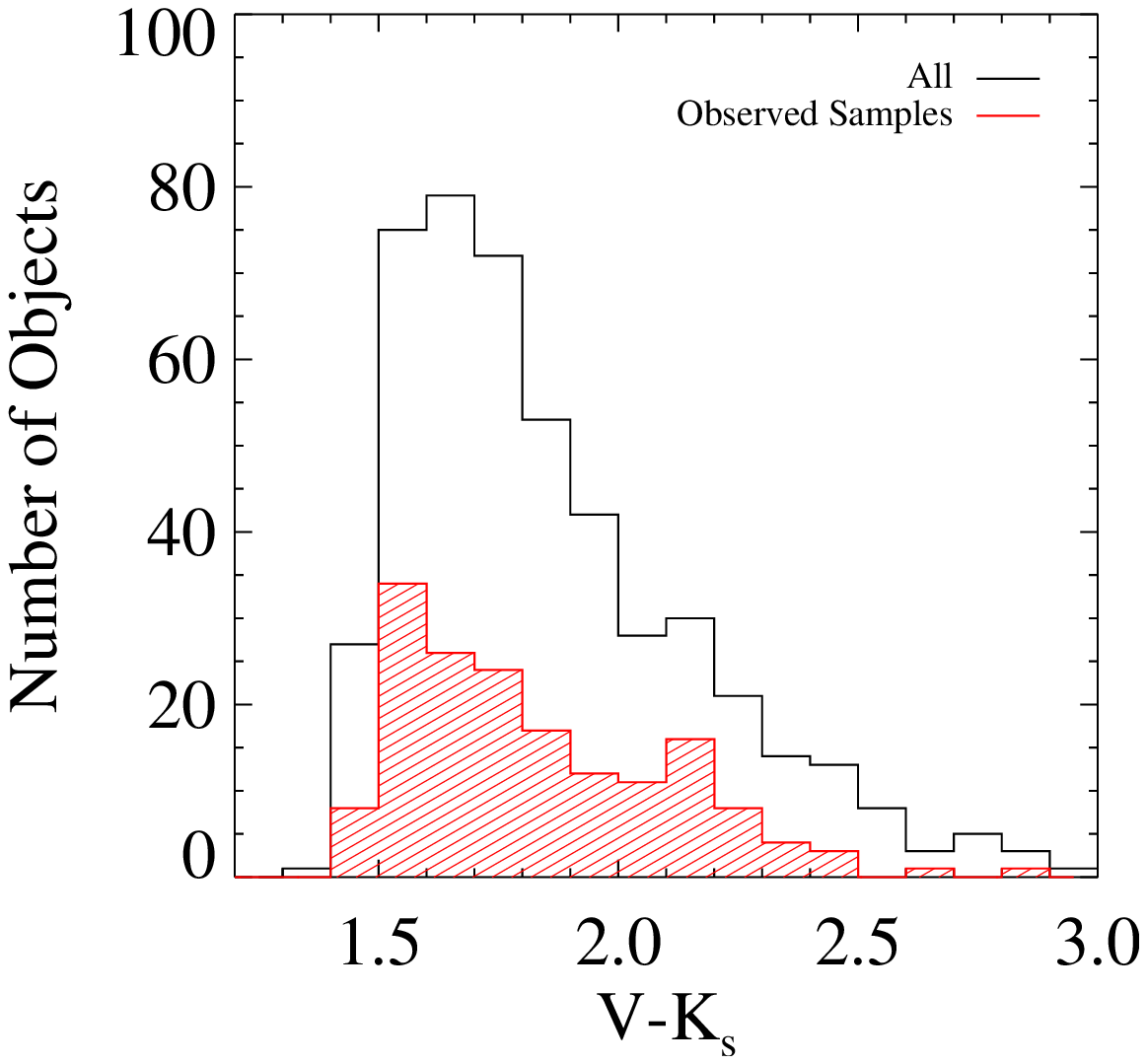}
\caption{{\it Top left:} Color-magnitude diagram of field stars in the solar neighborhood with {\it Hipparcos} parallaxes ($\sigma_\pi/\pi \leq 0.07$). The parallelogram represents a color-magnitude selection of spectroscopic targets in this study, where the KPNO samples observed in the spring ($N=120$) and autumn ($N=53$) runs are shown in blue and red circles, respectively. The thick solid line is the observed MS of the Hyades ([Fe/H]$=+0.13$) at the {\it Hipparcos} distance to the cluster \citep{perryman:98}, while thin lines are theoretical isochrones with empirical colors at [Fe/H]$=-0.3$ and $0.0$, respectively \citep{an:15}. {\it Bottom left:} Same as in the top left-hand panel, but with $\vk$ colors. {\it Right panels:} Color distributions of the KPNO sample stars (red histogram) in $\bv$ (top) and $\vk$ (bottom), respectively. The black histogram represents a distribution of all stars in the {\it Hipparcos} catalog ($\sigma_\pi/\pi \leq 0.07$) that are found within the parallelogram in the top left-hand panel.\label{fig:cmd}} \end{figure*}

The top left-hand panel in Figure~\ref{fig:cmd} illustrates our sample selection based on a $\bv$ CMD. Grey points are stars with good parallaxes ($\sigma_{\pi} / \pi \leq 0.07$) in the revised {\it Hipparcos} catalog \citep{vanleeuwen:07a,vanleeuwen:07b}. We took $BV$ photometry of these stars from the NASA Star and Exoplanet Database (NStED), most of which are those transformed from $B_TV_T$ in the Tycho-2 catalog \citep{hog:00} using transformation equations found in \citet{mamajek:02,mamajek:06}. We neglected foreground extinctions of these stars since they are mostly found within $\sim50$~pc from the Sun.

The parallelogram in the top left-hand panel of Figure~\ref{fig:cmd} indicates our color-magnitude selection, corresponding to $4.74 \leq M_{V} \leq 6.34$ at $\bv=0.60$~and $6.86 \leq M_{V}\leq 8.46$ at $\bv=1.0$. The bluer color limit was set to minimize the evolutionary effect on stellar luminosity and to perform a reliable line absorption analysis. We selected stars near or below the MS of the Pleiades on the absolute $V$ magnitude ($M_V$) versus $\bv$ CMD, assuming a distance of the cluster from the recent {\it Hipparcos} parallax, $\dmn=5.40$ \citep{vanleeuwen:09}. We retained stars if they are found within $\Delta M_V \sim1.4$~mag below and $\Delta M_V \sim 0.2$~mag above the MS of the Pleiades, as shown by the top and the bottom sides of the parallelogram. We applied a color limit $\bv=1.0$ to exclude chromospherically active low-mass stars with large color anomalies \citep{stauffer:03}.

In the top left-hand panel of Figure~\ref{fig:cmd}, the thick solid line represents the observed MS of the Hyades \citep{pinsonneault:04}. We adopted a distance to the cluster's center of mass $\dmn=3.33\pm0.01$ from the {\it Hipparcos} catalog \citep{perryman:98}. The Hyades covers a large area on the sky, which makes parallax measurements of its individual members less vulnerable to the suspected spatial correlation of the {\it Hipparcos} parallax \citep{pinsonneault:98,narayanan:99,debruijne:01}. The cluster is approximately $550$~Myr old \citep{perryman:98}, and has ${\rm [Fe/H]}=+0.13\pm0.01$ \citep{paulson:03} with negligible foreground reddening \citep[e.g.,][]{taylor:80}. The two thin solid lines are $550$~Myr old theoretical models at ${\rm [Fe/H]} = -0.3$ and ${\rm [Fe/H]}=0.0$ \citep{an:07b}, of which colors were calibrated using the observed MS of the Hyades. All together, these lines show a typical metallicity sensitivity of colors and magnitudes of MS stars. In this study, however, we avoided using theoretical isochrones and relied on the observed MS of the Hyades to empirically derive a MS-fitting distance to individual field stars.

Meanwhile, the top side of the parallelogram in Figure~\ref{fig:cmd} is not exactly parallel to the solar metallicity isochrone. This is because the Pleiades' MS, which was used to set our sample color-magnitude cut, is known to become progressively fainter than those for older stars or standard stellar models as one moves toward redder colors \citep{stauffer:03,an:07b}, although the observed magnitude offset in the $\teff$ range of our sample is not as severe as those seen for stars with $\bv \ga 1$ \citep[see Figure~20 in][]{an:07b}. Nevertheless, our search for anomalously faint stars in the solar neighborhood is almost insensitive to how we set the upper limit in stellar brightness since such stars would be significantly fainter than the above brightness limit in any case (see discussions in \S~\ref{sec:kpno}).

In total $480$ G- and K-type dwarfs satisfy the above selection criteria. We combined $V$-band magnitude with $K_{s}$ in the Two Micron All Sky Survey \citep[2MASS;][]{skrutskie:06}\footnote{This publication makes use of data products from the Two Micron All Sky Survey, which is a joint project of the University of Massachusetts and the Infrared Processing and Analysis Center/California Institute of Technology, funded by the National Aeronautics and Space Administration and the National Science Foundation.}. Except five stars (HIP~9172, HIP~56837, HIP~91438, HIP~96100, and HIP~101997), our targets have valid $K_s$-band measurements, which are not saturated, undetected, blended, or contaminated. The bottom left-hand panel in Figure~\ref{fig:cmd} shows the same set of stars in the $\vk$ CMD, although we did not employ $\vk$ colors in our sample selection.

\subsection{Observations and Data Reductions}\label{sec:obs}

We selected a random subset of {\it Hipparcos} field stars, which satisfy our color-magnitude cut, and obtained their high-resolution ($\lambda/\Delta\lambda\sim60,000$) spectra with the echelle spectrograph on the Mayall 4-m telescope at Kitt Peak National Observatory (KPNO). Most of our targets are bright ($V \sim 9$ mag), and are spread all around sky, except those with low declinations ($\delta \la -30\arcdeg$) due to observing restrictions at KPNO. Our observing campaign was composed of 5 nights in May 2010 and 4 nights in September 2010. In the left-hand panels of Figure~\ref{fig:cmd}, targets observed in spring and autumn are marked in blue and red circles, respectively. Unresolved binaries are cooler than single stars and therefore better represented in $\vk$ with its longer wavelength baseline. Our color-magnitude cut is likely smeared out in the $\vk$ CMD by these unresolved binaries along with photometric color errors. The $\bv$ and $\vk$ color distributions of our observed sample are shown by a red shaded histogram in the right-hand panels in Figure \ref{fig:cmd}. The open histogram represents a distribution of the initial sample selected above from the {\it Hipparcos} catalog (those within a parallelogram in the top left-hand panel). The overall shapes of the red shaded and open histograms in each panel are similar with each other, as expected from a random selection of stars in our spectroscopic observations.

The sky was clear in spring, but the dome was closed during two nights in autumn due to bad weather conditions. We used a red long camera and settled on the $58.5$-$63$ grating with a $\sim1\arcsec$ slit width. With the $2048\times2048$ T2KB CCD and 226-1 cross disperser, the wavelength coverages were set to $4340$~\AA--$7670$~\AA\ in May, and $4400$~\AA--$7870$~\AA\ in September. The spectra are of high quality, with high signal-to-noise (S/N) ratios ($>100$ per pixel).

In total $120$~stars were observed in spring, and $53$~stars were observed in autumn. Three stars were observed in both spring and autumn runs (HIP~113884, HIP~104733, HIP~98792). From our observing runs, spectra for $170$ field stars in the {\it Hipparcos} catalog were obtained. Table~\ref{tab:phot} lists our sample stars with $V$, $\bv$, and $\vk$ colors and their errors from the {\it Hipparcos} catalog. A minimum error in $V$ and $\bv$ was set to $0.02$~mag. The $9^{th}$ and $10^{th}$ columns show revised {\it Hipparcos} parallaxes ($\pi$) and their errors \citep{vanleeuwen:07a,vanleeuwen:07b}. The $M_V$ and its error in the next two columns were computed using $V$ and $\pi$ and their associated errors. Seven targets were identified as a spectroscopic binary system in the $10^{th}$ catalogue of spectroscopic binary orbits \citep{pourbaix:04}\footnote{\tt http://sb9.astro.ulb.ac.be}, and are marked in Table~\ref{tab:phot}. A number of repeat observations for each target is indicated in the following columns. About one third of the sample stars ($N=69$) were observed $2$--$3$ times, and $8$ stars were observed in two different nights to check for systematics in our abundance measurements.

We reduced raw data frames using standard data processing packages in IRAF\footnote{IRAF is distributed by National Optical Astronomy Observatories (NOAO), which is operated by the Association of Universities for Research in Astronomy, Inc., under arrangement with the National Science Foundation, United States.}. This included a bias correction, bad pixel correction, wavelength calibration using Th-Ar lamp, spectral extraction, and radial velocity correction. For data taken in autumn, we found a uniformly increasing bias pattern toward one side of the CCD, which consistently appeared over the entire observing run. We combined all of the zero-exposure frames, made an average bias frame, and subtracted it from our science data frames. We corrected for a wavelength shift from a line-of-sight velocity using the H$\alpha$ line profile. When the H$\alpha$ line was not available, the \ion{Na}{1} $5890.0$~\AA\ line was used for a radial velocity correction.

\section{Derivation of Stellar Parameters}\label{sec:param}

Our conclusions about stars with the {\it Hipparcos} parallaxes are sensitive to the adopted sizes of errors in metallicity, while stellar parameters -- effective temperatures ($\teff$), surface gravity ($\log{g}$), and metallicity ([Fe/H]) -- derived from spectroscopy are subject to various systematic errors. Since such errors could originate from different spectroscopic analysis techniques \citep[e.g.,][]{torres:12}, we derived stellar parameters in two parallel approaches. We employed a spectral synthesis code MOOG\footnote{~http://www.as.utexas.edu/$\sim$chris/moog.html} \citep{moog} based on equivalent width (EW) measurements of iron lines (\S~\ref{sec:moog}). Because this procedure was performed by hand, we restricted our MOOG analysis to a subset of stars in the sample ($N=74$ out of $170$) with highly precise parallax measurements ($\sigma_\pi/\pi \leq 0.03$), and those ($N=13$) having largest $M_V$ differences between {\it Hipparcos} and MS fitting despite their less accurate parallaxes ($\sigma_\pi/\pi > 0.03$). For the entire sample (including those analyzed using MOOG), we employed an automated spectral matching technique (SMT), which iteratively fits synthetic spectra to an observed spectrum to search for the best matching parameters with least human intervention. For an efficient estimation of stellar parameters with SMT, we degraded our spectra to a medium resolution ($R=10,000$), while keeping the same precision in the stellar parameter estimates. Below we describe each of these two approaches, along with external checks with previous work in the literature.

\subsection{Stellar Parameters from MOOG}\label{sec:moog}

We selected stars with accurate parallaxes ($\sigma_\pi/\pi \leq 0.03$) in our sample ($N=74$), and used MOOG to derive their precise atmospheric parameters, taking advantage of high-resolution $(R \sim60,000)$ spectra of high quality (S/N $>100$). We selected a list of Fe lines, which were commonly included in both \citet{bensby:03} and \citet{boesgaard:13}. Within the wavelength range of our echelle data, there were $37$ \ion{Fe}{1} lines and $5$ \ion{Fe}{2} lines. These lines are listed in Table~\ref{tab:line} along with their central wavelengths, excitation potentials (E.P.), and oscillator strengths ($\log{gf}$). We determined a local continuum for each absorption line using a polynomial with a degree of $3$, and measured its EW using SPECTRE\footnote{http://www.as.utexas.edu/$\sim$chris/spectre.html} (Sneden, version 2003).

We constructed Kurucz stellar model atmospheres based on newly computed opacity distribution functions (ODFs) with updated opacities and abundances \citep{castelli:04}\footnote{http://kurucz.harvard.edu/grids.html}. We interpolated ODFs at a given set of [Fe/H] and microturbulence $\xi$, and constructed desired models using the ATLAS9 code of stellar atmosphere. We assumed solar abundance ratios in \citet{grevesse:99} with an enhancement in the $\alpha$-elements ratios (O, Mg, Si, Ca, and Ti) relative to Fe: $+0.4$~for ${\rm [Fe/H]} \leq -1.0$, $+0.3$~for ${\rm [Fe/H]} = -0.75$, $+0.2$ for ${\rm [Fe/H]} = -0.5$, $+0.1$ for ${\rm [Fe/H]} = -0.25$, and $0.0$ for ${\rm [Fe/H]} \geq +0.0$.

We used the {\it abfind} driver in MOOG to constrain $\teff$, $\log{g}$, [Fe/H], and micro-turbulence ($\xi$), self-consistently. The effective temperature was derived by requiring that individual line abundances be independent of excitation potential and that $\xi$ be independent of line strength. Insisting on ionization equilibrium between Fe I and Fe II allowed for a simultaneous determination of $\log{g}$ with $\teff$ and $\xi$. Our sample stars are relatively cool, and therefore non-LTE corrections are likely negligible \citep[see][]{bensby:14}.

Table~\ref{tab:moog} lists $\teff$, $\log{g}$, [Fe/H], and $\xi$ of stars analyzed using MOOG. In addition to $74$ stars with accurate parallax measurements, we included $13$ stars in our MOOG analysis, which show large differences in distance between {\it Hipparcos} and MS fitting (see \S~\ref{sec:kpno}). We adopted a solar Fe abundance $A{\rm (Fe)}$\footnote{$A{\rm (Fe)} \equiv \log[N{\rm (Fe)}/N{\rm (H)}]+12$.}$=7.52$ in \citet{anders:89}. The [Fe/H]$_{\rm corr}$ is a corrected [Fe/H] value, which was put on the metallicity scale adopted in this paper, and is described below in detail. We computed an error in [Fe/H] by propagating errors in \ion{Fe}{1} and \ion{Fe}{2} abundances. If there exist repeat measurements ($N_{\rm obs}$), we took a standard deviation of these measurements divided by a square root of $N_{\rm obs}$ or the one propagated from individual abundance errors, whichever is larger. A typical root-mean-square (rms) dispersion of \ion{Fe}{1} abundance is approximately $0.15$~dex for each star. Given the large number of iron lines used in this study ($N=37$), this results in an unrealistically small error in [Fe/H] ($\sigma\approx0.02$~dex). To estimate more realistic errors in abundance, we ran additional models with different input stellar parameters ($\Delta T_{\rm eff} = \pm100$~K, $\Delta \log g = \pm0.3$~dex, $\Delta {\rm [Fe/H]} = \pm0.1$~dex, and $\Delta \xi$~by $\pm0.3$~km~s$^{-1}$). Table~\ref{tab:sys} shows mean abundance errors obtained from all stars analyzed using MOOG. A quadrature sum of these errors yields a systematic error in [Fe/H] of the order of $0.08$~dex.

For an external check on our derived stellar parameters, we compared our values with atmospheric parameters for a large number of field dwarfs in \citet[][hereafter VF05]{valenti:05}. They used a software package Spectroscopy Made Easy \citep[SME;][]{valenti:96} to derive $\teff$, $\log{g}$, and individual elemental abundances based on high-resolution spectra. They obtained both an overall metallicity ([M/H]) and an elemental abundance of iron ([Fe/H]) for each star, but we utilized their [Fe/H] as we derived metallicities using MOOG from EW measurements of Fe.

\begin{figure*}
\centering
\includegraphics[scale=0.45]{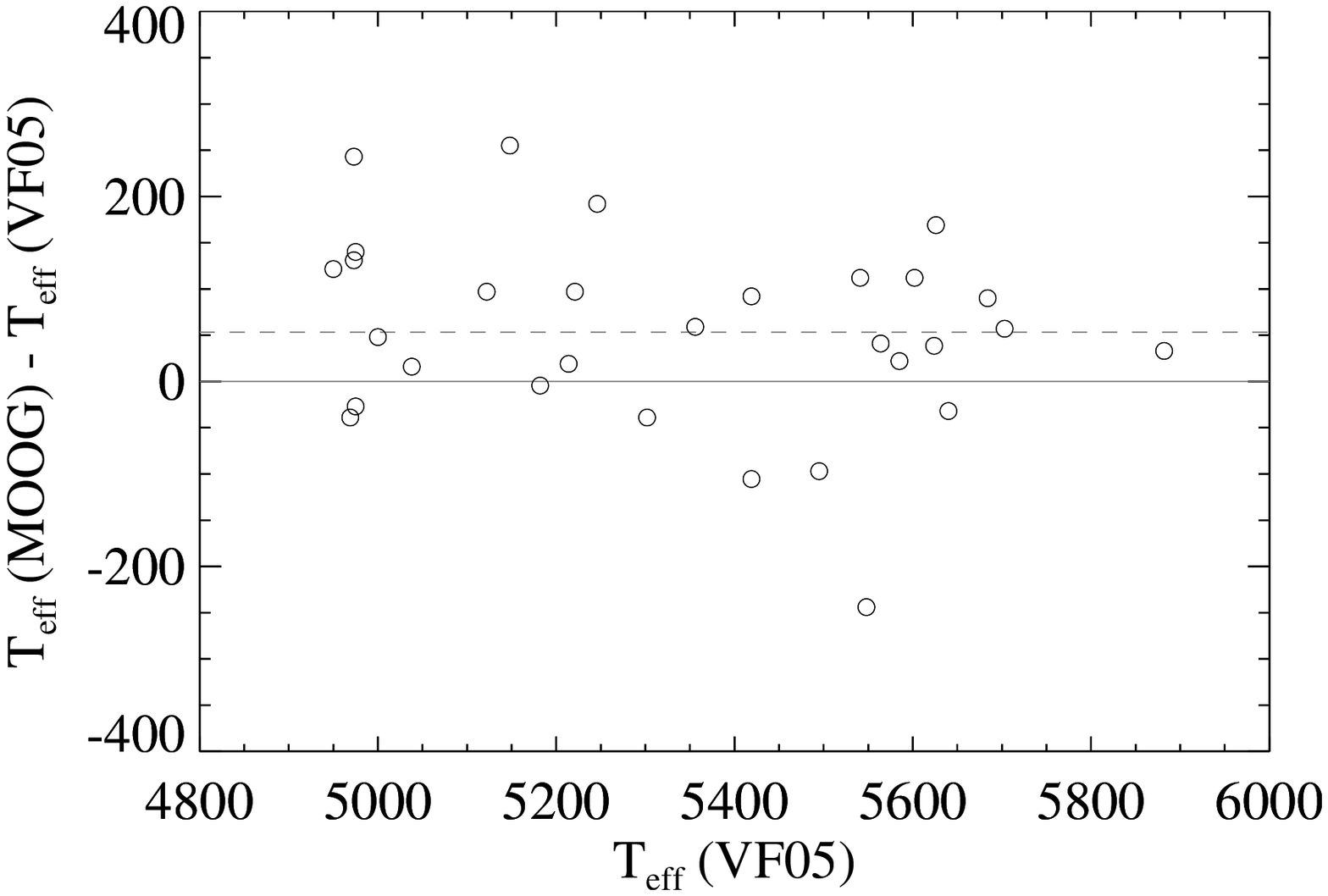}\includegraphics[scale=0.45]{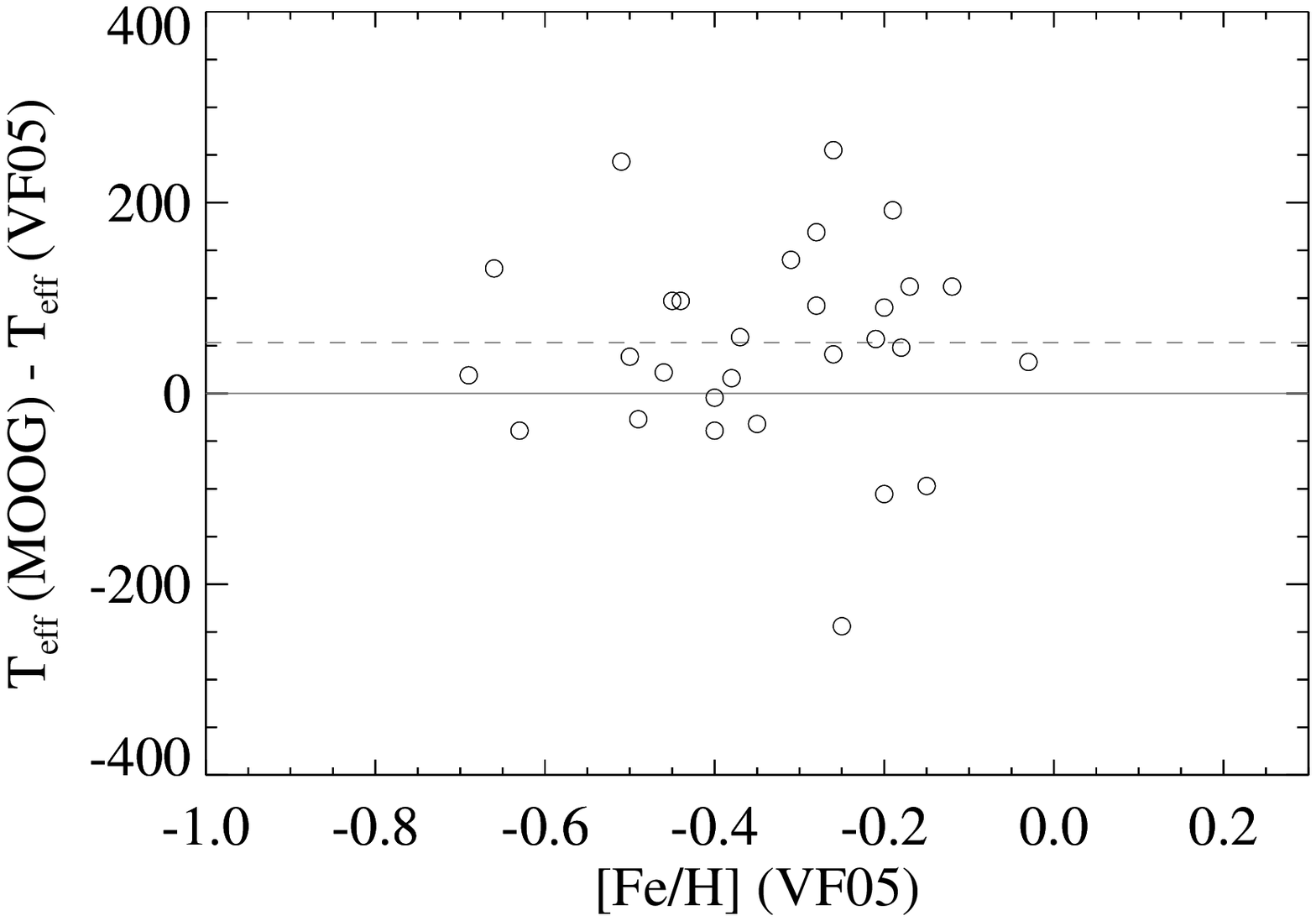}
\includegraphics[scale=0.45]{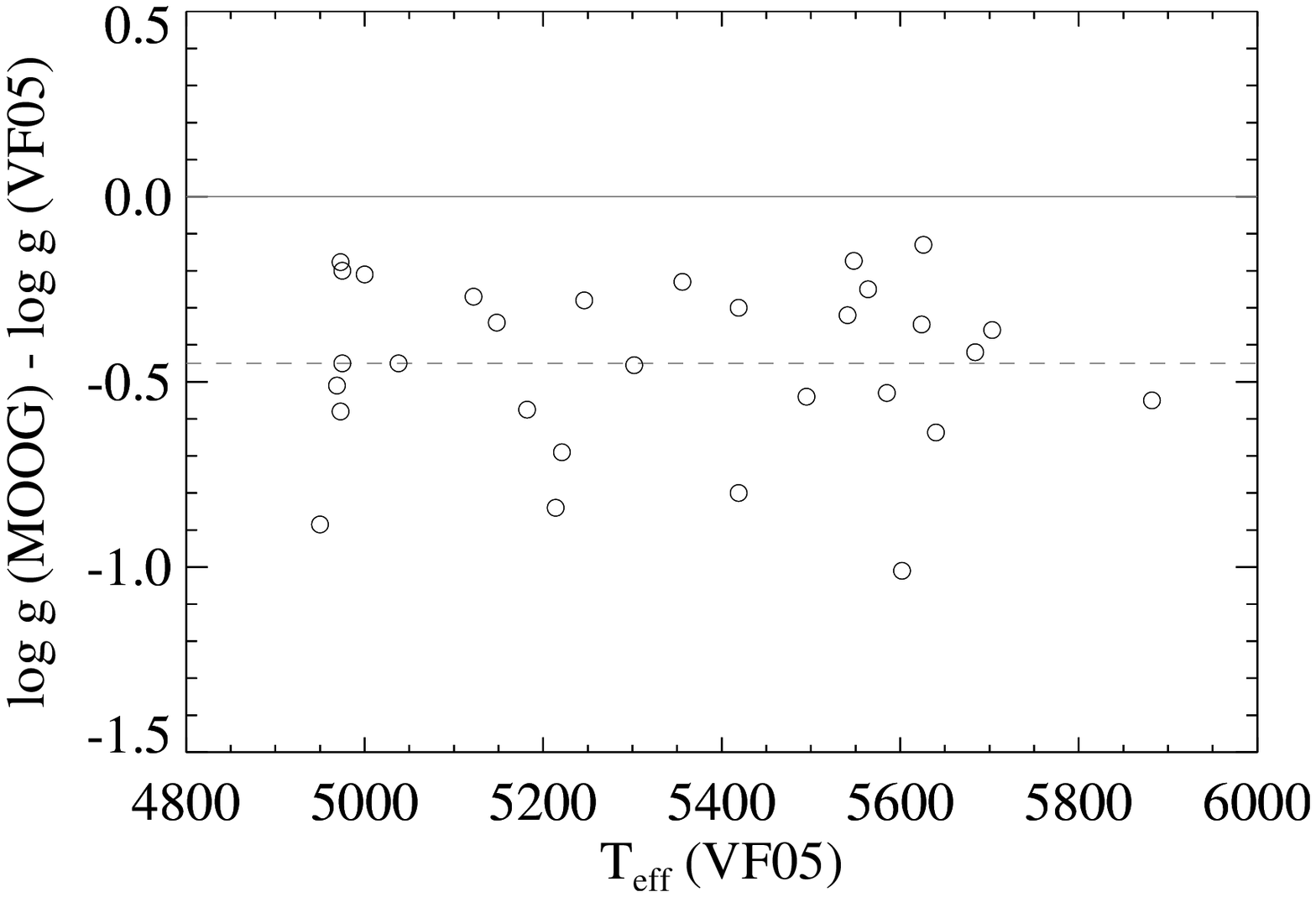}\includegraphics[scale=0.45]{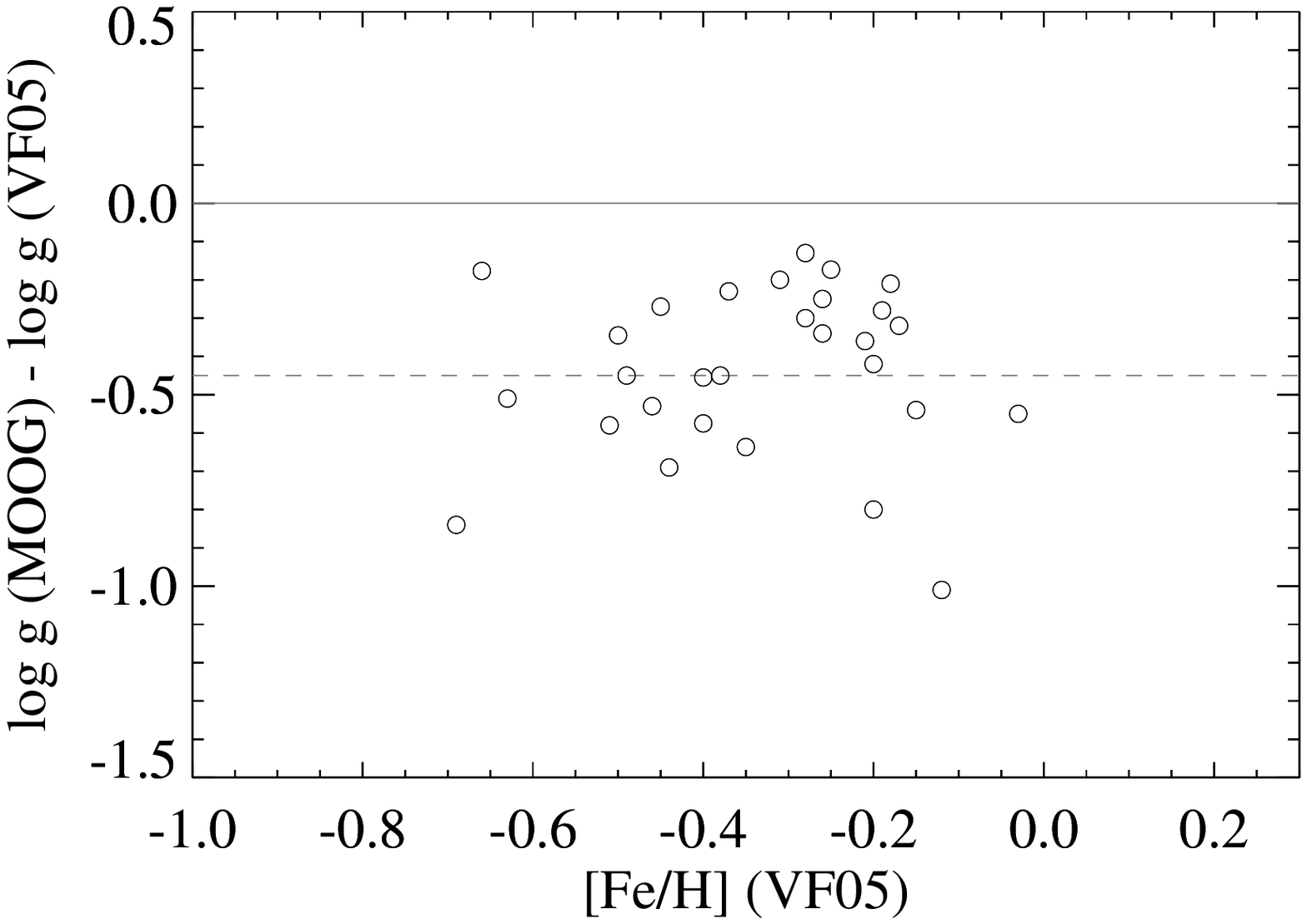}
\includegraphics[scale=0.45]{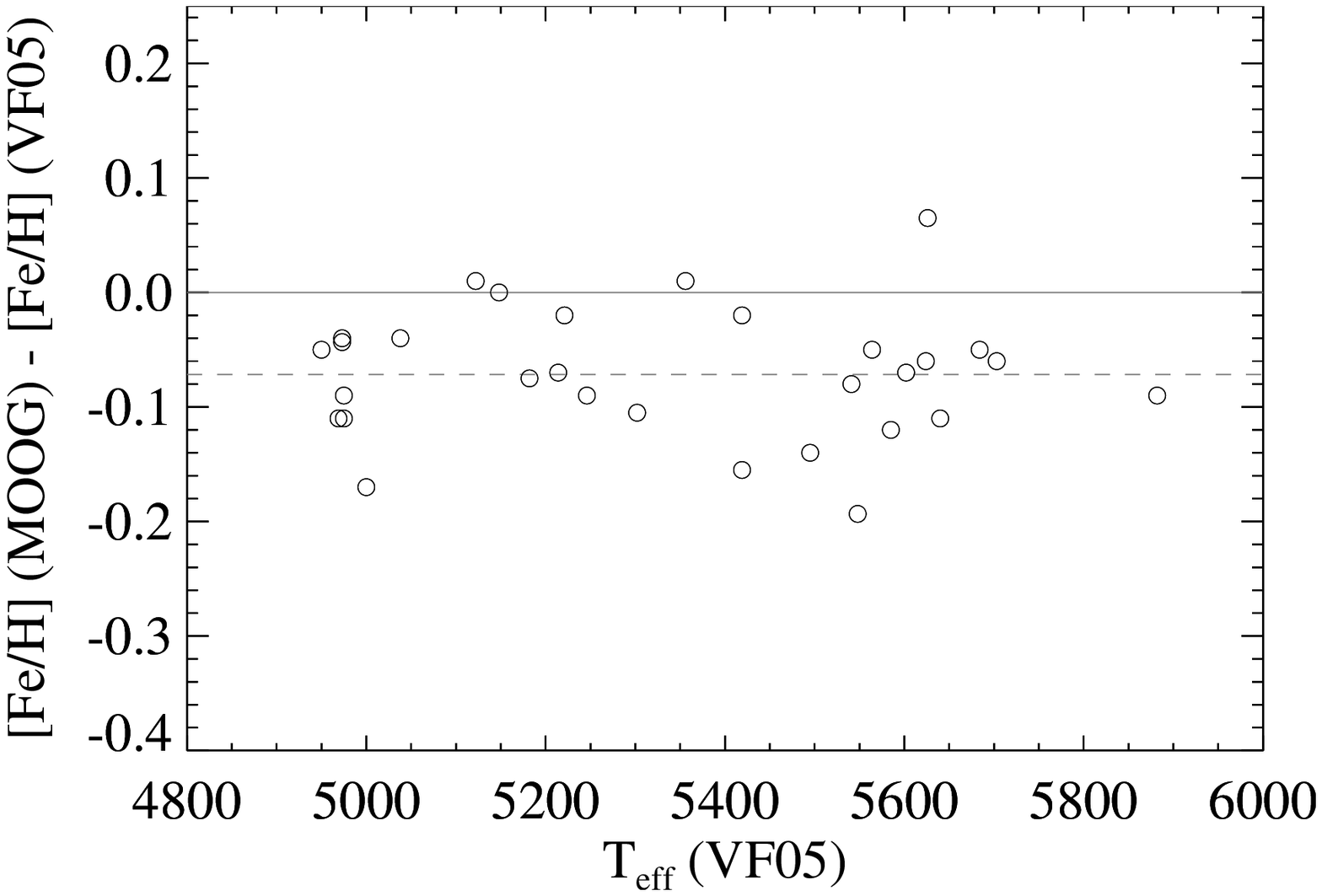}\includegraphics[scale=0.45]{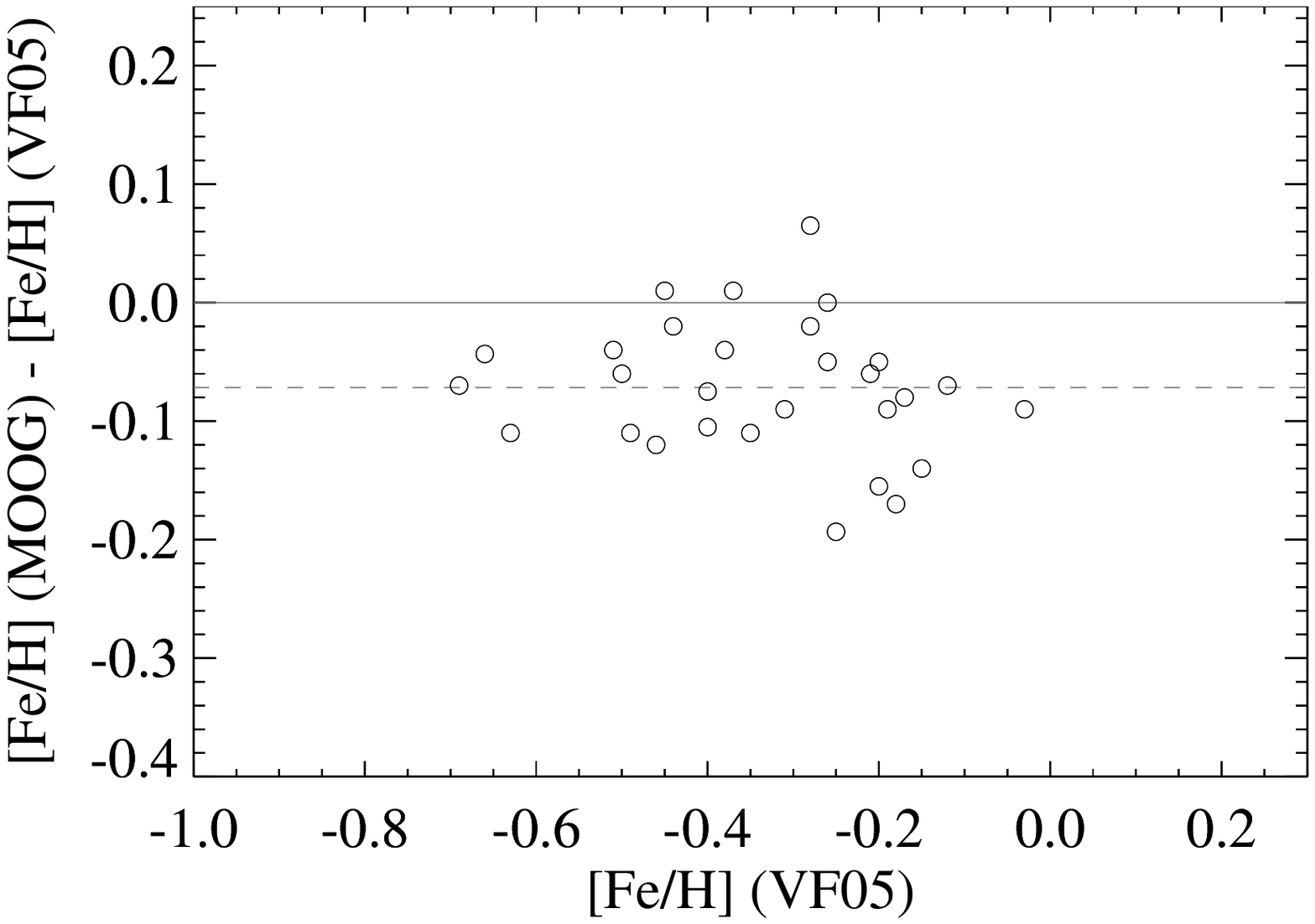}
\caption{Comparisons in spectroscopic parameters between VF05 and MOOG. The dashed line indicates a mean difference, while the solid line is a zero difference. In the bottom panels, metallicities from MOOG represent raw estimates, before applying a zero-point adjustment (see text).\label{fig:comp_moog_sme}} \end{figure*}

Figure~\ref{fig:comp_moog_sme} shows comparisons in $\teff$ (top), $\log{g}$ (middle), and [Fe/H] (bottom) between MOOG and VF05. Statistical properties of the parameter comparisons are summarized in the first row in Table~\ref{tab:stat}. The MOOG $\teff$ is on average $53$~K higher than VF05 $\teff$, which is not unexpected from the two independent analyses. However, our spectroscopic $\log{g}$ estimates are systematically smaller than those from VF05. The observed trend in $\log{g}$ is similar to those seen from a large number of disk stars in \citet{bensby:14}, who found that $\log{g}$ determinations from \ion{Fe}{1} and \ion{Fe}{2} ionization equilibrium are systematically smaller than the one based on the {\it Hipparcos} parallax. The offset was seen for dwarfs over a wide range of temperature in their study, but there was no convincing trend observed for giants. Theoretical models \citep{an:07b,an:15} also suggest that MS stars within a narrow range of color ($0.6 \leq \bv \leq 1.0$) have surface gravities $4.4 \la \log{g} \la 4.6$ over a range of metallicity covered by our sample stars, but our MOOG estimates are about $0.3$~dex smaller than these. Nevertheless, the $\log{g}$ dependence of our metallicity estimate from MOOG is weak (Table~\ref{tab:sys}), and an adjustment of $\log{g}$ would hardly affect our [Fe/H] estimates.

In Figure~\ref{fig:comp_moog_sme} observed $1\sigma$ dispersions in the parameter comparisons are $\sigma (\Delta \teff) = 104$~K and $\sigma (\Delta {\rm [Fe/H])} = 0.06$. VF05 computed formal $1\sigma$ uncertainties of their parameter estimates as $\sigma (\teff) = 44$~K and $\sigma {\rm ([Fe/H])} = 0.03$, suggesting that $\teff$ and [Fe/H] determined from our MOOG analysis have a precision of $\sigma (\teff) \approx 90$~K and $\sigma {\rm ([Fe/H])} \approx 0.05$~dex. The latter is close to our expectation ($\sigma\approx0.08$~dex) from errors computed by varying stellar input parameters in the models (Table~\ref{tab:sys}).

We also checked the accuracy of stellar parameters derived using MOOG against a calibration sample in \citet{casagrande:10}, which have been used in the derivation of their empirical color-$\teff$ relations based on the Infrared Flux Method (IRFM). They provided a large compilation of stellar parameters from a high-resolution spectroscopy in the literature. We cross-identified stars in their and our catalogs based on both coordinates and $V$ magnitudes. There were $9$ stars in common, and the comparison with their stellar parameters is included in Table~\ref{tab:stat}. The average difference in $\teff$ between \citet{casagrande:10} and our MOOG-based estimates is $32$~K.

Our [Fe/H] estimates from MOOG are systematically lower than those in VF05 and in \citet{casagrande:10} by $0.07\pm0.01$~dex and $0.02\pm0.01$~dex, respectively. However, the above metallicity estimates from MOOG were not adjusted for an instrumental correction to the solar Fe abundance, because we did not obtain a solar spectrum during the observing runs. For this reason, we adjusted our original [Fe/H] estimates from MOOG to match published values in VF05 by adding a constant offset ($\Delta {\rm [Fe/H]} = 0.07$~dex). The last column ([Fe/H]$_{\rm corr}$) in Table~\ref{tab:moog} lists [Fe/H] values after this correction.

\subsection{Spectral Matching Technique (SMT)}\label{sec:smt}

In addition to the EW analysis using MOOG, we employed a SMT to derive stellar parameters ($\teff$, $\log{g}$, and [Fe/H]) for all of our sample stars. The SMT routine is essentially a modified version of the NGS1 technique adopted in the SEGUE \citep[Sloan Extension for Galactic Understanding and Exploration;][]{yanny:09} Stellar Parameter Pipeline \citep[SSPP;][]{lee:08a,lee:08b}. It was originally designed to match a grid of synthetic models to low-resolution stellar spectra in the Sloan Digital Sky Survey \citep[SDSS;][]{york:00,stoughton:02}. We modified and optimized the original NGS1 code to analyze our echelle spectra. We found that the accuracy of stellar parameters derived from SMT remains almost unaffected with varying spectral resolutions. For this reason, we linearly re-binned observed spectra with Gaussian smoothing to $0.25$~\AA~per pixel ($R\approx10,000$) for a fast and efficient analysis of our extensive data set. The smoothing was also helpful for improving S/N ratios of the spectra.

We generated a grid of models using the Kurucz stellar model atmospheres based on the new ODFs as described above. We utilized a pre-computed set of models\footnote{Available at http://kurucz.harvard.edu/grids.html.} to construct a finer grid by linearly interpolating these models over a wide range of parameter space, covering $4000~{\rm K} < T_{\rm eff} < 10000~ {\rm K}$~in steps of $250$~K, $0.0 < \log {g} < 5.0$~in steps of $0.25$~dex, and $-5.0 < {\rm [Fe/H]} < +1.0$~in steps of $0.25$~dex. Synthetic model spectra were then generated using the {\it synthe} code at $3,000$~\AA\ $\leq \lambda \leq 10,000$~\AA\ at a resolution of $0.01$~\AA, where we adopted $\xi [{\rm km\ s}^{-1}] = -0.345 \times \log {g} + 2.22$, which was derived from the SSPP calibration with high-resolution spectra. After constructing a full set of synthetic spectra, we re-sampled model spectra to $0.25$~\AA~wide linear pixels, corresponding to $R = 10,000$ at $5000$~\AA.

We matched synthetic spectra with observed data at $4500$~\AA\ $< \lambda < 5500$~\AA\ and $5900$~\AA\ $< \lambda < 6900$~\AA, which contain a large number of isolated Fe lines along with other various metallic lines. In each of the above wavelength ranges, we normalized model spectra using a pseudo-continuum, which was constructed by iteratively rejecting data points that are more than $1\sigma$~below and $4\sigma$~above a fitted polynomial curve. The degree of a polynomial was set to $21$ to cover each echelle order. We took the same normalization step for the observed spectra, which produced the same level of line-strength suppression. We used the $\chi^{2}$ minimization routine MPFIT\footnote{~http://www.physics.wisc.edu/$\sim$craigm/idl/fitting.html} \citep{markwardt:09} to search the grid of synthetic spectra for the best-fitting model parameters. In this step, we generated a synthetic spectrum at intermediate values of $T_{\rm eff}$, $\log {g}$, and ${\rm [Fe/H]}$ from the model grid using a spline interpolation. Errors in the best-fitting model parameters were determined by the square root of the diagonal elements in the covariance matrix.

\begin{figure}
\epsscale{1.15}
\plotone{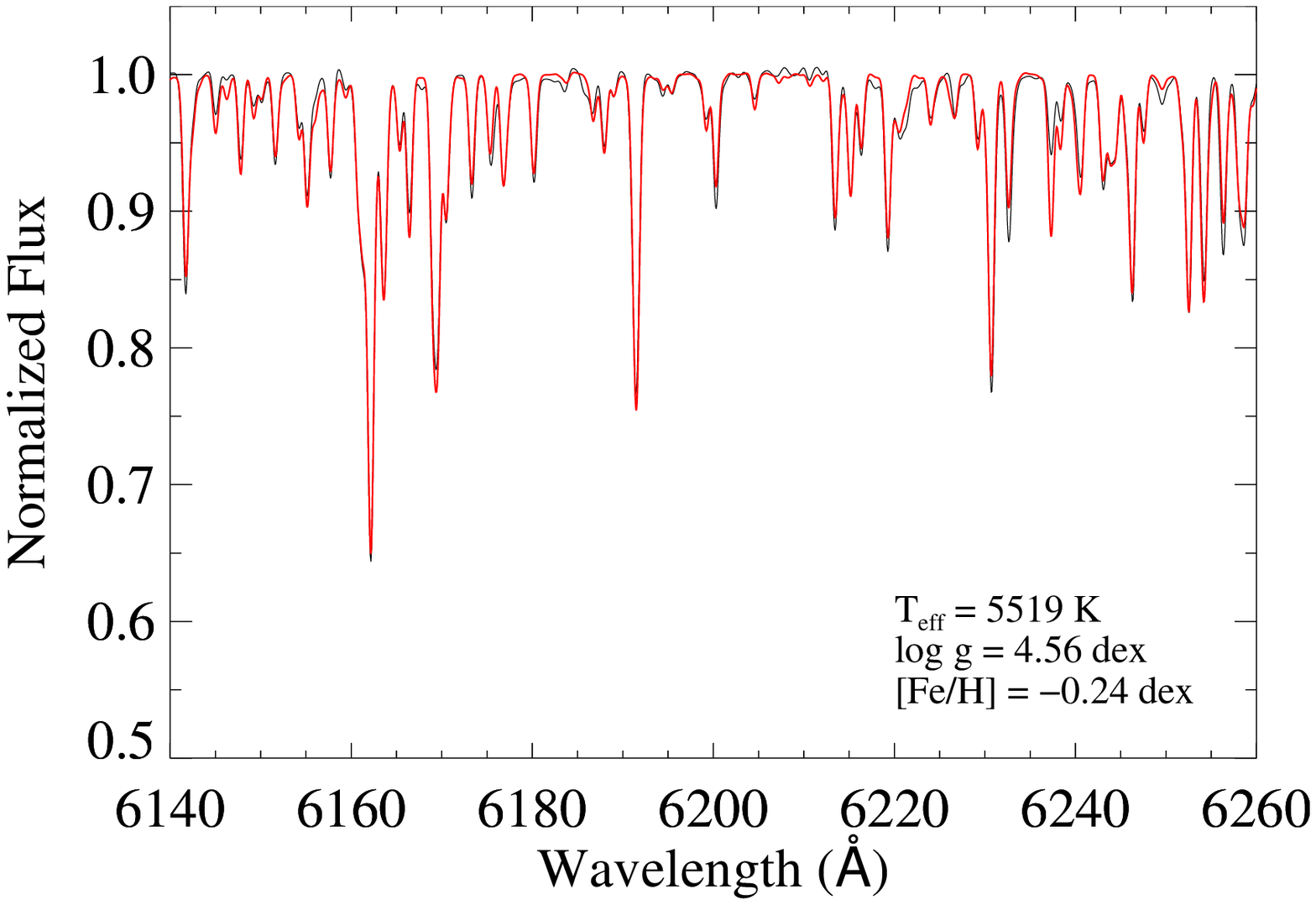}
\caption{A segment of an observed spectrum of HIP~98677 (grey line). The red line represents the best-fitting synthetic model spectrum in the SMT analysis. The spectral resolution was degraded to $R\sim10,000$ (see text).\label{fig:smtmatching}} \end{figure}

Figure~\ref{fig:smtmatching} shows an example of the results from SMT. The black solid line is an observed spectrum of HIP~98677 after downgrading its resolution to $R=10,000$. The red solid line represents our best-fitting synthetic spectrum with $T_{\rm eff} = 5519$~K, $\log {g} = 4.56$~dex, ${\rm [Fe/H]} = -0.24$ dex, and $\xi = 0.65~{\rm km~s}^{-1}$, which shows an excellent match to the observed data. The second through $7^{\rm th}$ columns in Table~\ref{tab:smt} summarize results from SMT with $1\sigma$ errors in $\teff$, $\log{g}$, and [Fe/H]. For stars observed multiple times, average values of individual parameter estimates are listed. Errors in these quantities are either random errors derived from SMT or a standard deviation of measurements from multiple observations divided by a square root of $N_{\rm obs}$, whichever is larger. The mean errors are $79$~K, $0.097$ dex, and $0.067$~dex in $\teff$, $\log{g}$, and [Fe/H], respectively.

We repeated the above search for stellar parameters ($\log{g}$ and [Fe/H]) while holding $\teff$ fixed using the IRFM relation \citep{casagrande:10}. We refer to this approach as SMT2 as opposed to SMT3 based on the full three-parameter fitting above. The polynomial equation in the IRFM relation includes metallicity terms, so we derived $\teff$ using $\bv$ photometry at an initial [Fe/H] derived above. The photometric temperature led to the new $\log{g}$ and [Fe/H] estimates. With these values, we took one more iteration to estimate photometric $\teff$ and found best-fitting metallicity and surface gravity for our target stars. The $9^{\rm th}$ through $14^{\rm th}$ columns in Table~\ref{tab:smt} show these values and their errors. The error in $\teff$ was computed assuming $0.02$~mag error in the $\bv$ photometry. The typical errors in $\teff$, $\log{g}$, and [Fe/H] from this approach are $60$~K, $0.020$~dex, and $0.016$~dex, respectively. As summarized in Table~\ref{tab:stat}, SMT2 results are generally consistent with those from SMT3: Differences in the derived parameters are $\Delta \teff=51$~K, $\Delta \log{g} = 0.09$, and $\Delta {\rm [Fe/H]} = 0.03$.

\begin{figure*}
\centering
\includegraphics[scale=0.45]{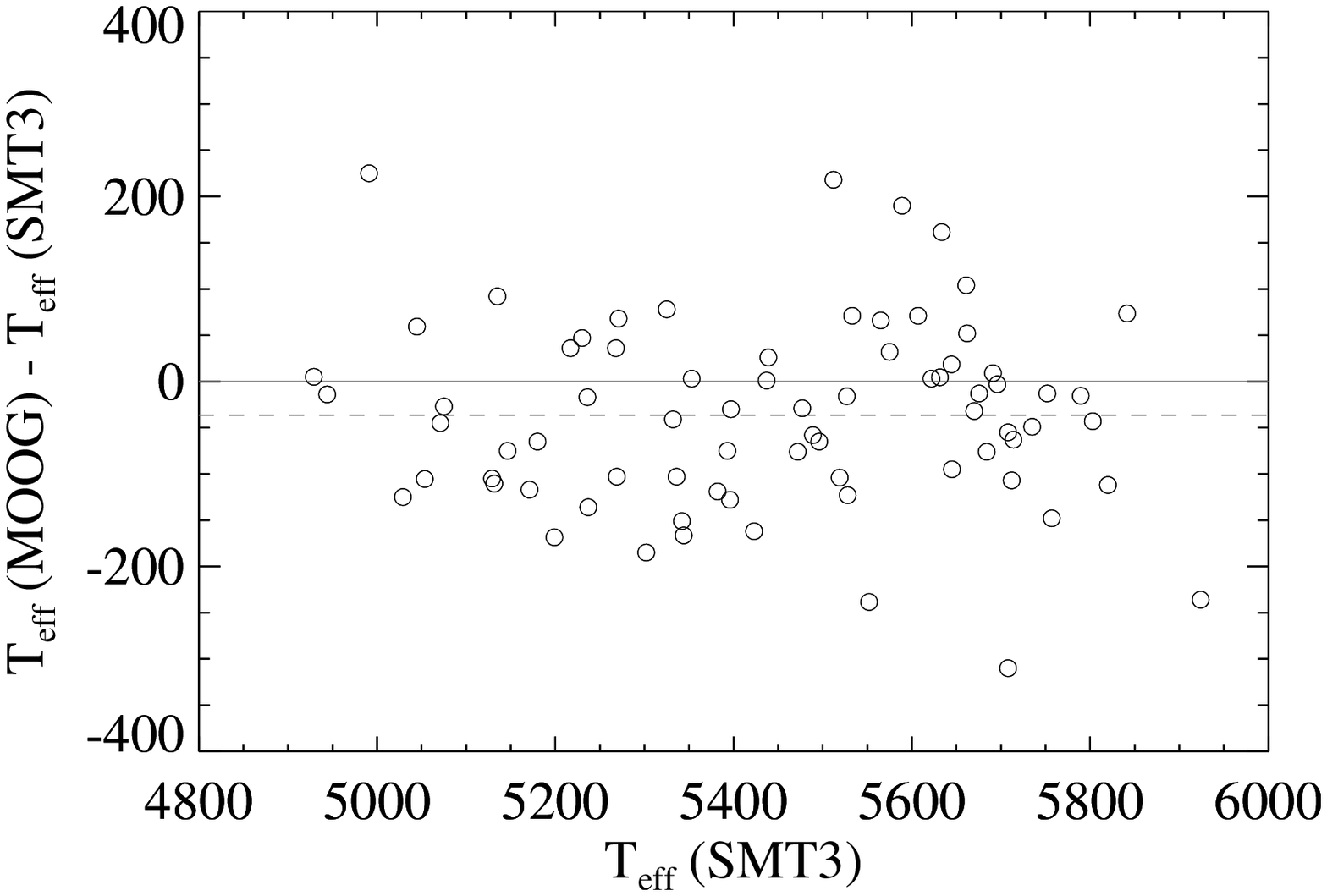}\includegraphics[scale=0.45]{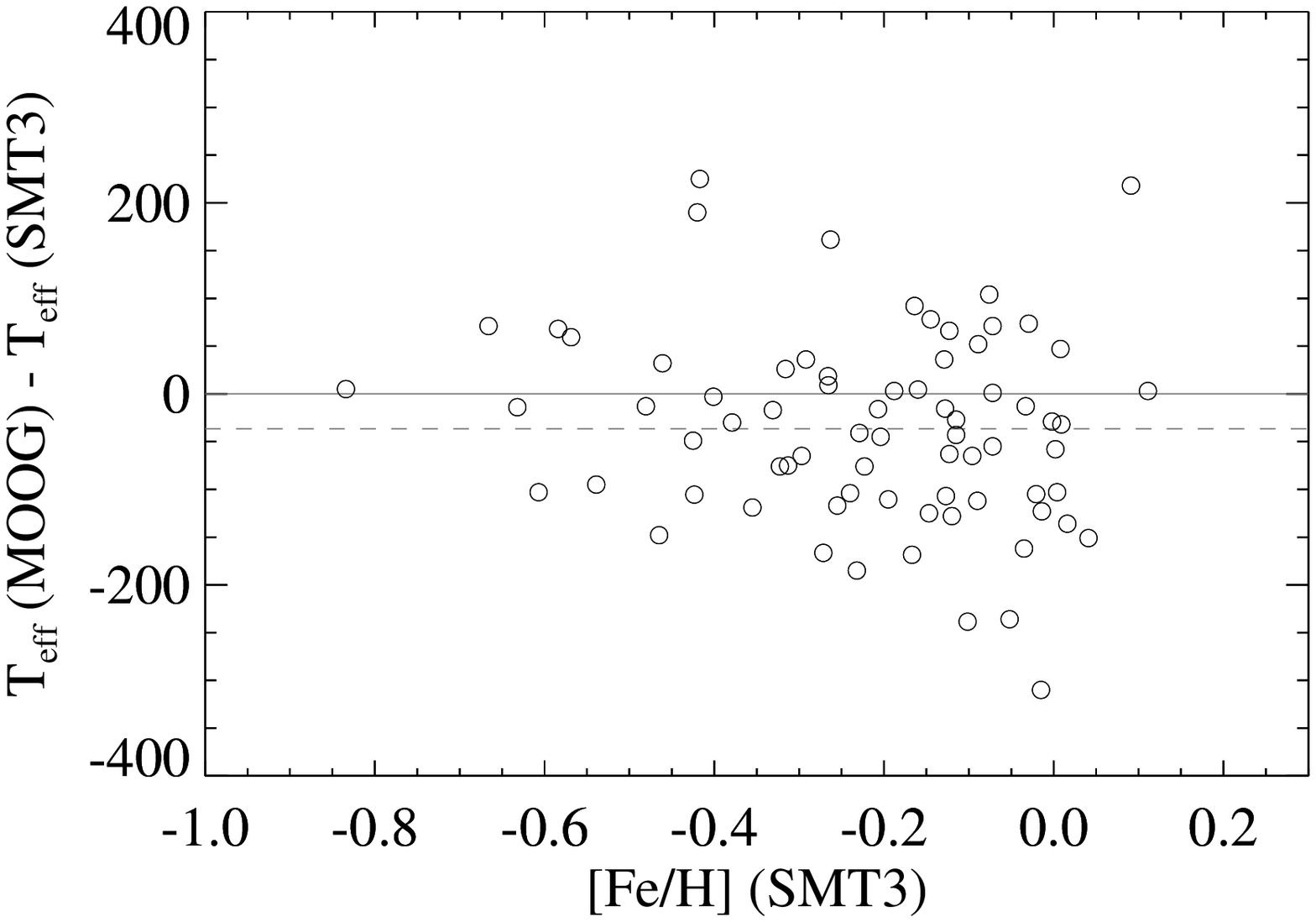}
\includegraphics[scale=0.45]{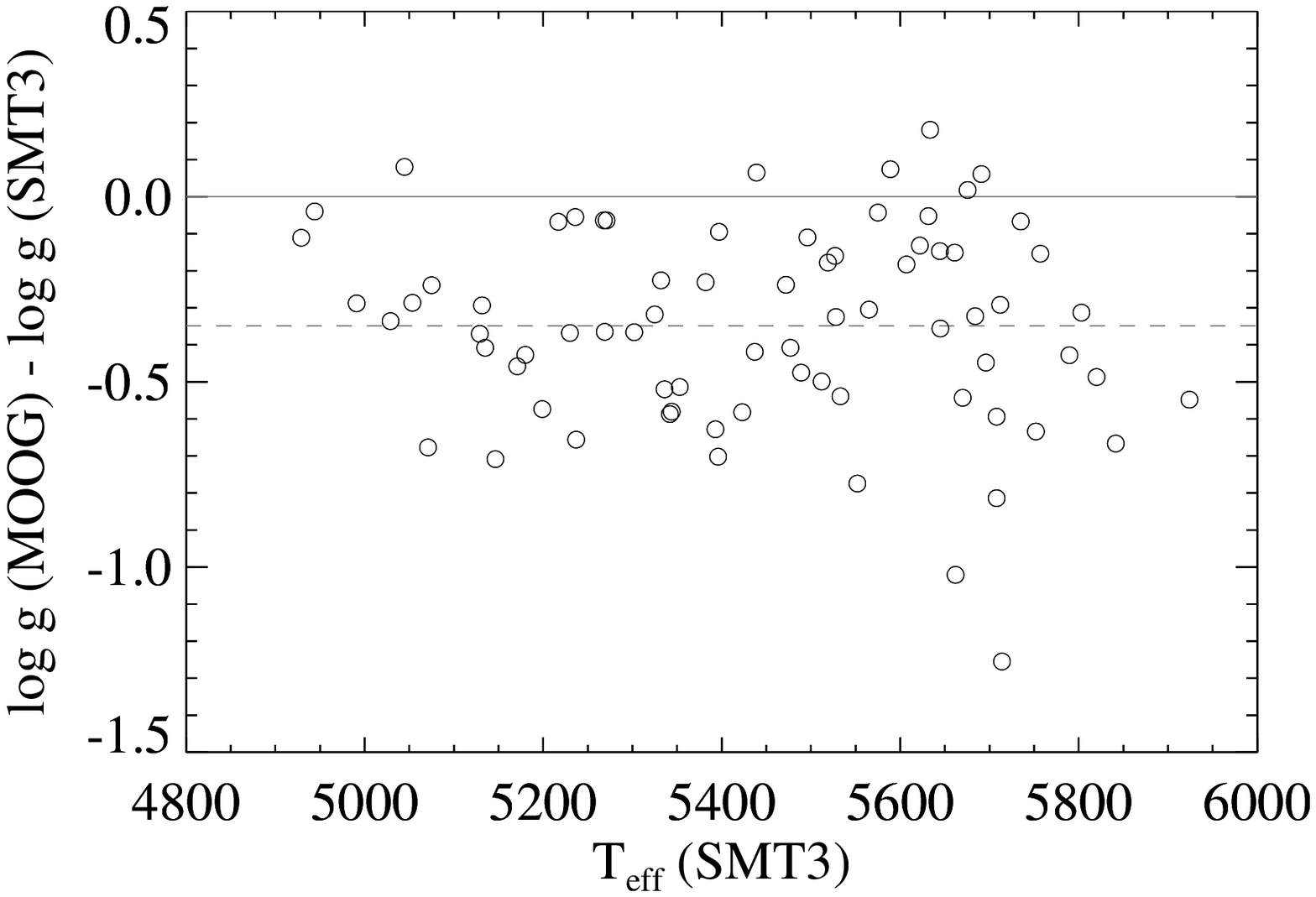}\includegraphics[scale=0.45]{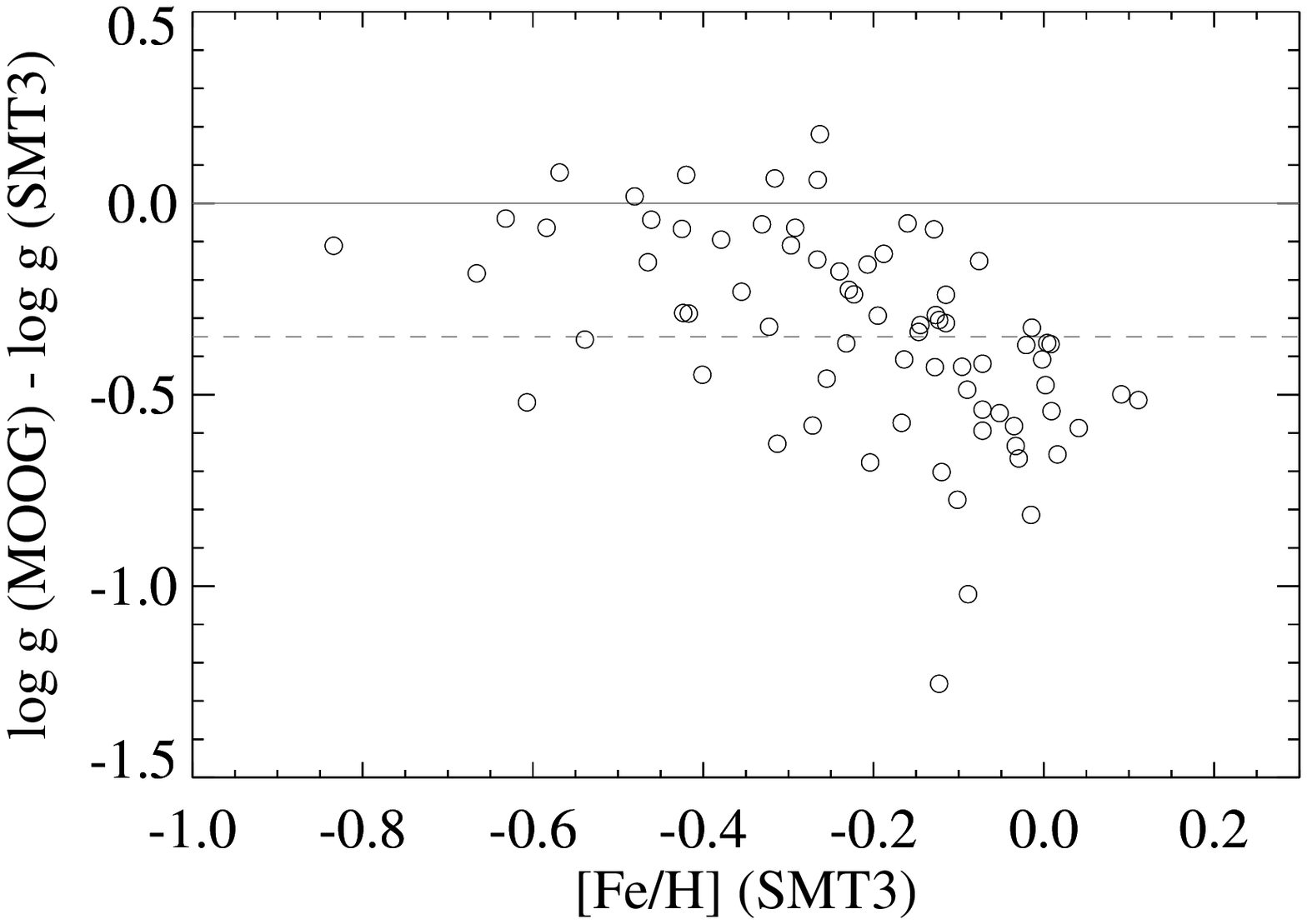}
\includegraphics[scale=0.45]{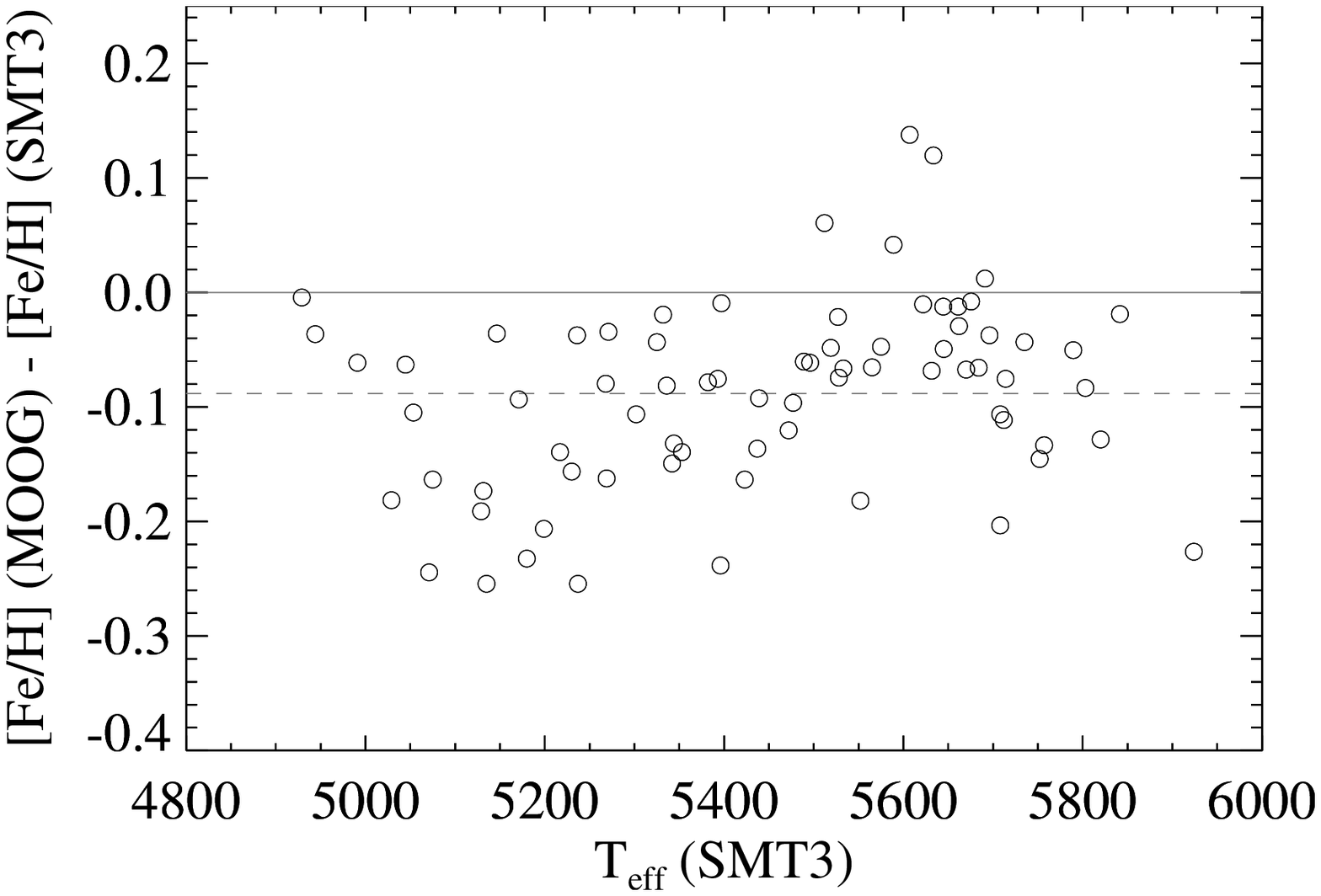}\includegraphics[scale=0.45]{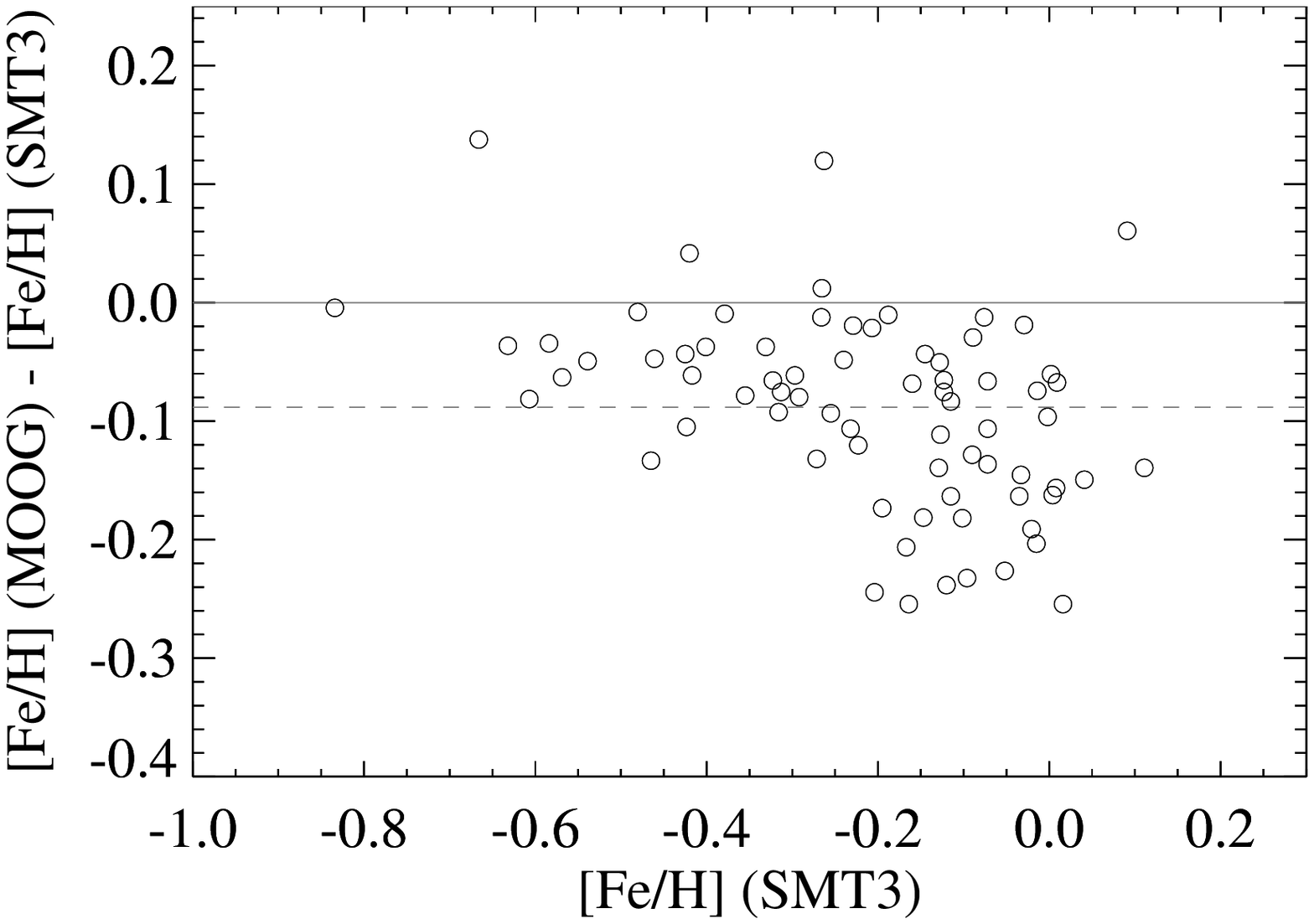}
\caption{Comparisons of spectroscopic parameters between MOOG and SMT with the full three-parameter fitting (SMT3). The dashed line indicates a mean difference, while the solid line is a zero difference. In the bottom panels, metallicities from SMT3 represent raw values, while MOOG estimates are those with a zero-point adjustment (see text).\label{fig:comp_moog_smt3}} \end{figure*}

\begin{figure*}
\centering
\includegraphics[scale=0.45]{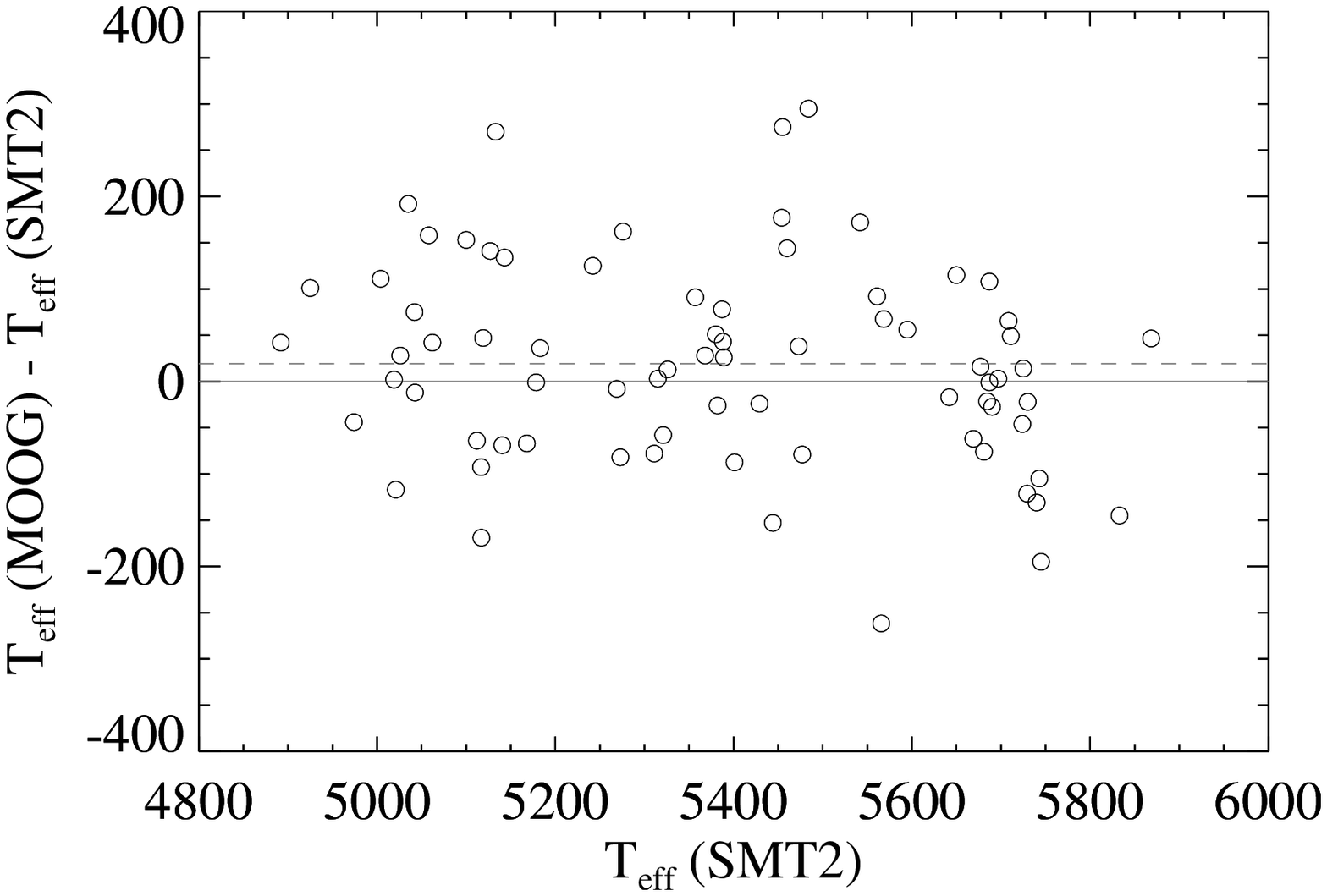}\includegraphics[scale=0.45]{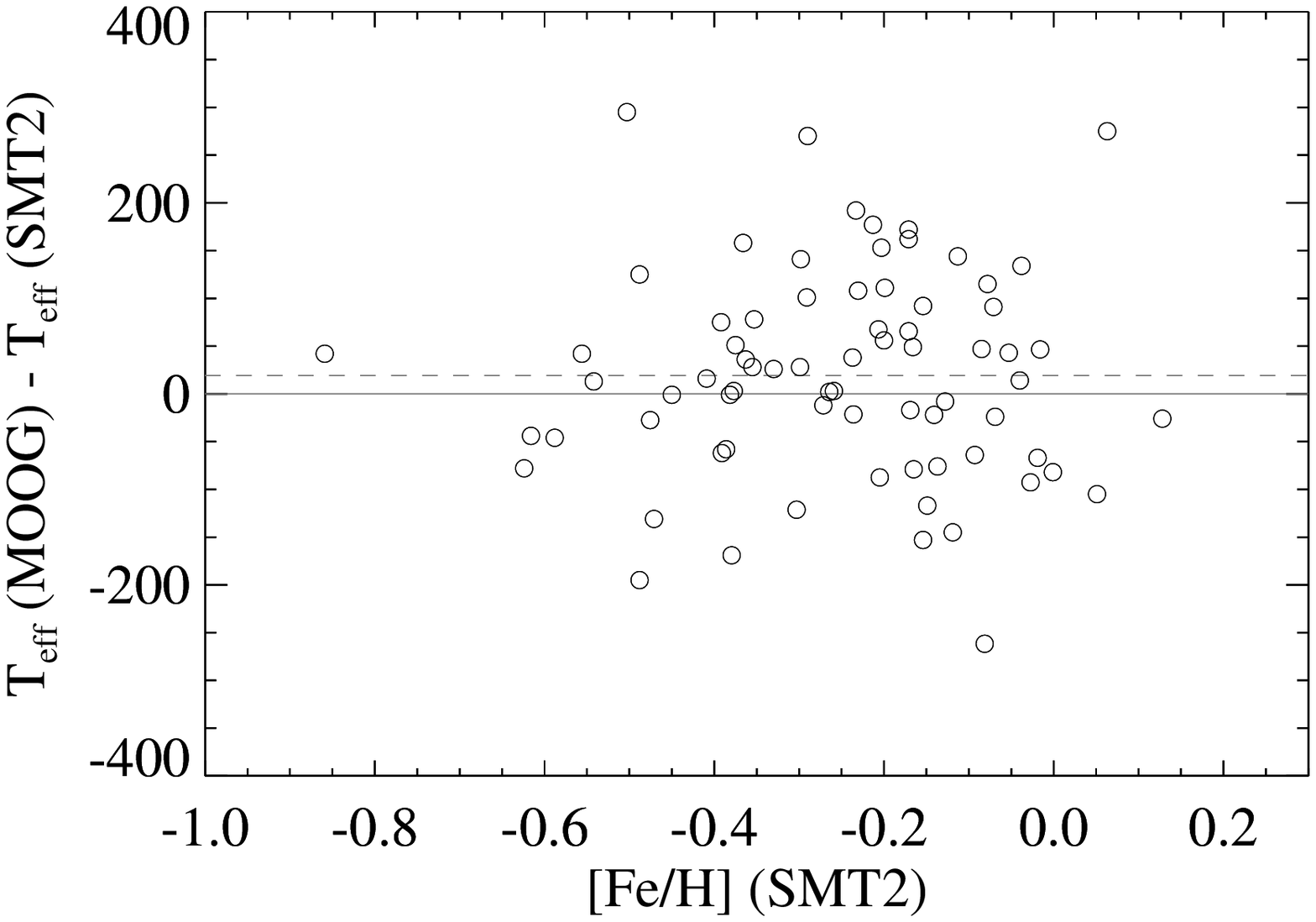}
\includegraphics[scale=0.45]{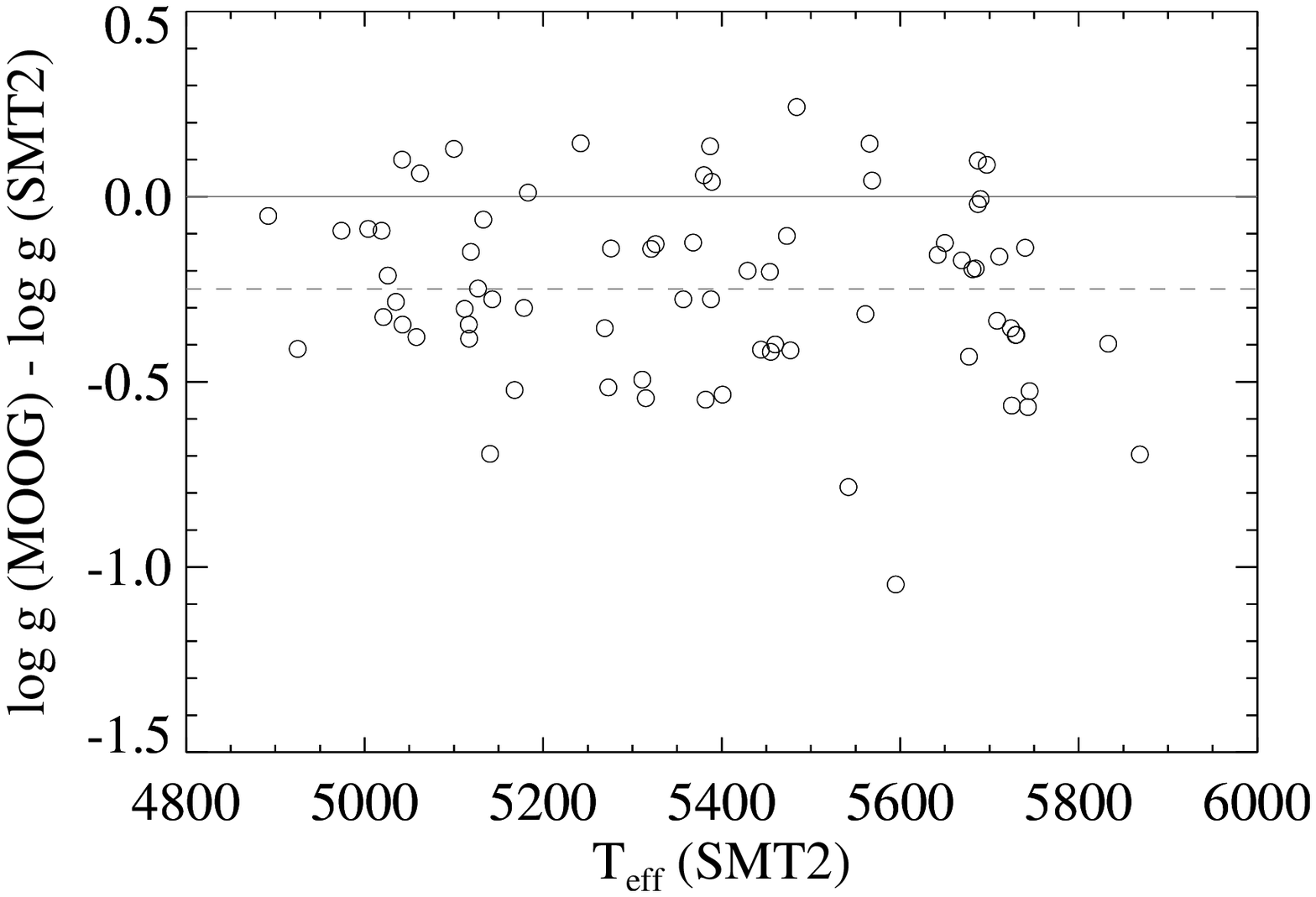}\includegraphics[scale=0.45]{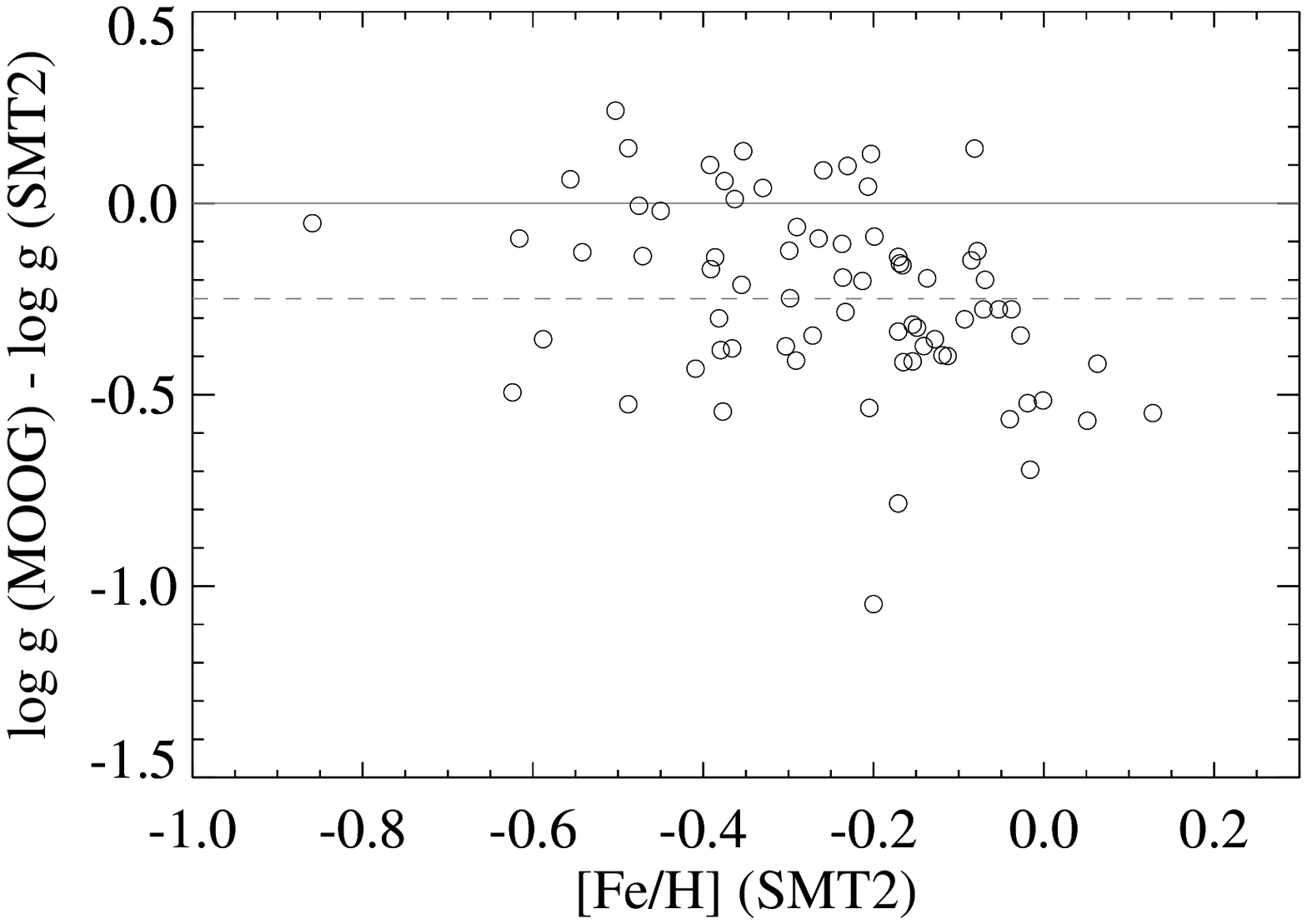}
\includegraphics[scale=0.45]{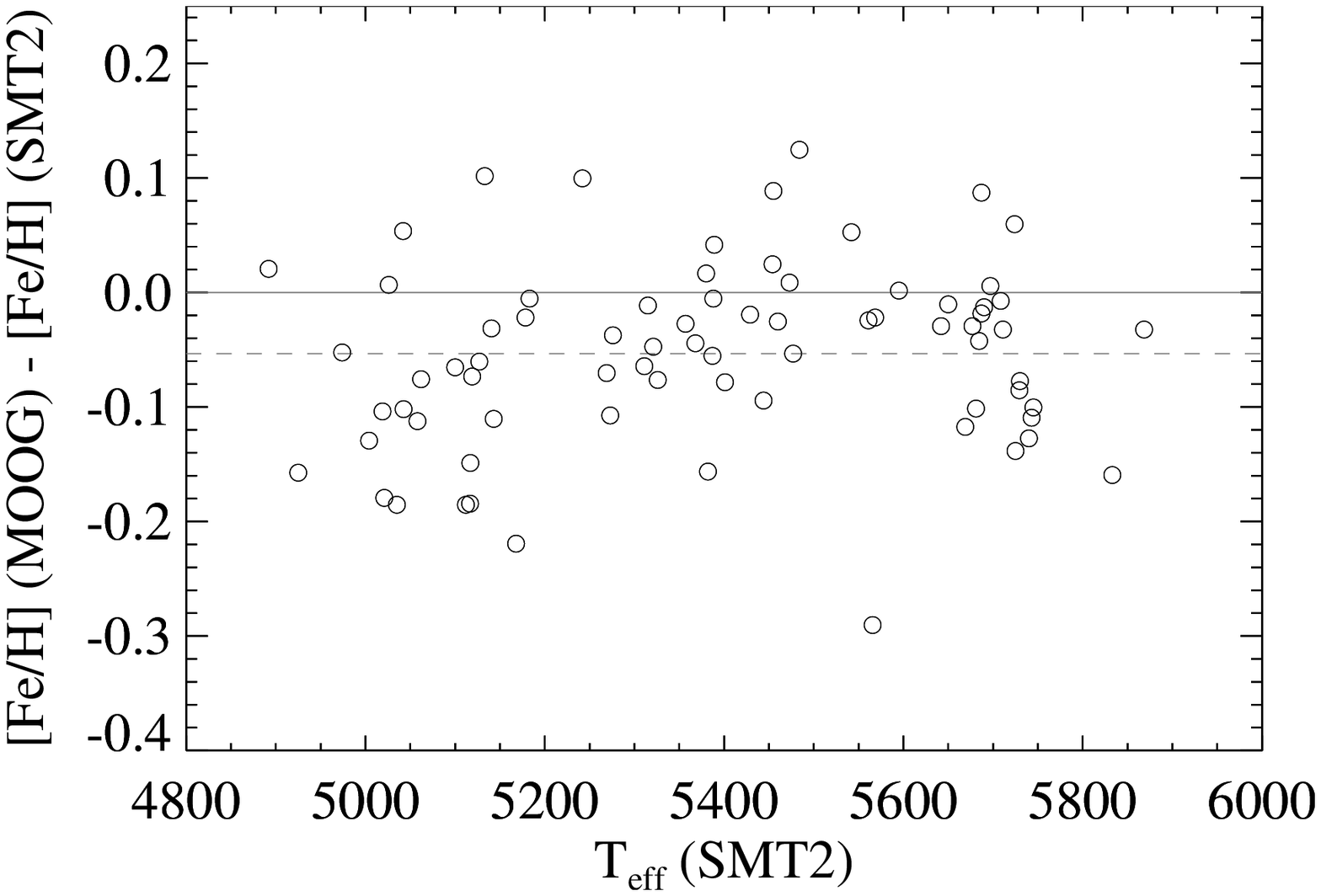}\includegraphics[scale=0.45]{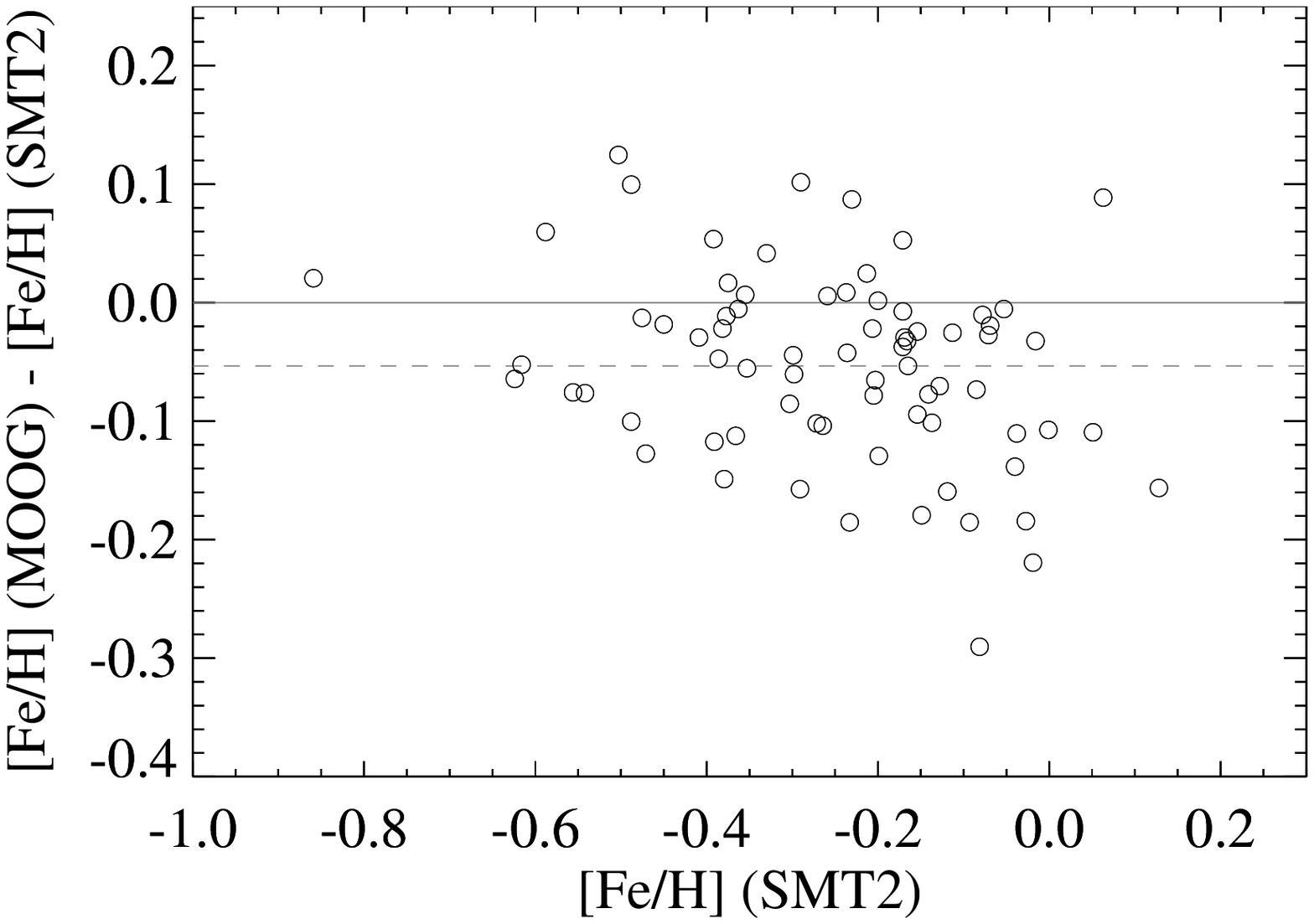}
\caption{Same as in Figure~\ref{fig:comp_moog_smt3}, but comparisons with SMT analysis based on IRFM temperatures (SMT2).\label{fig:comp_moog_smt2}} \end{figure*}

Comparisons of stellar parameters with MOOG are shown in Figures~\ref{fig:comp_moog_smt3} and \ref{fig:comp_moog_smt2} for SMT3 and SMT2, respectively. In each panel,  a solid line represents a zero difference, and a dashed line is an average difference between MOOG and SMT3/SMT2. Corrected values of [Fe/H] from MOOG were used in all comparisons. The SMT3 $\teff$ estimates are on average higher than those from MOOG, but the difference is small ($\Delta T_{\rm eff} = 37$~K). Similarly, the MOOG $\teff$ estimates are higher than those from SMT2, but the average difference is negligible ($19$~K). As discussed above, $\log{g}$ values from MOOG were probably underestimated, and the comparison with SMT suggests that the difference in $\log{g}$ is correlated with its metallicity. The metallicities from both SMT3 and SMT2 are systematically higher than those from MOOG by $\Delta{\rm [Fe/H]} = 0.09$~dex and $0.05$~dex, respectively.

\begin{figure*}
\centering
\includegraphics[scale=0.45]{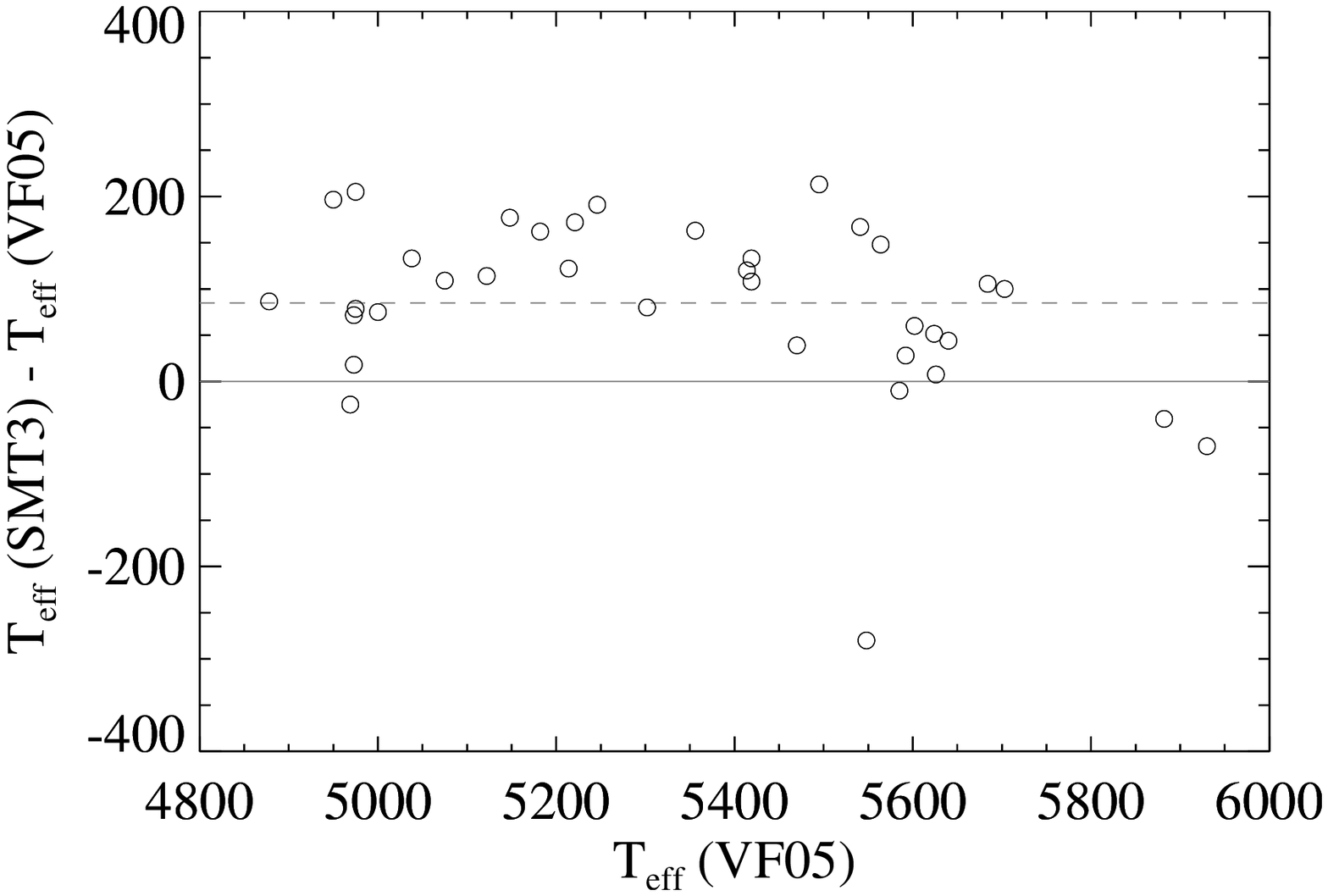}\includegraphics[scale=0.45]{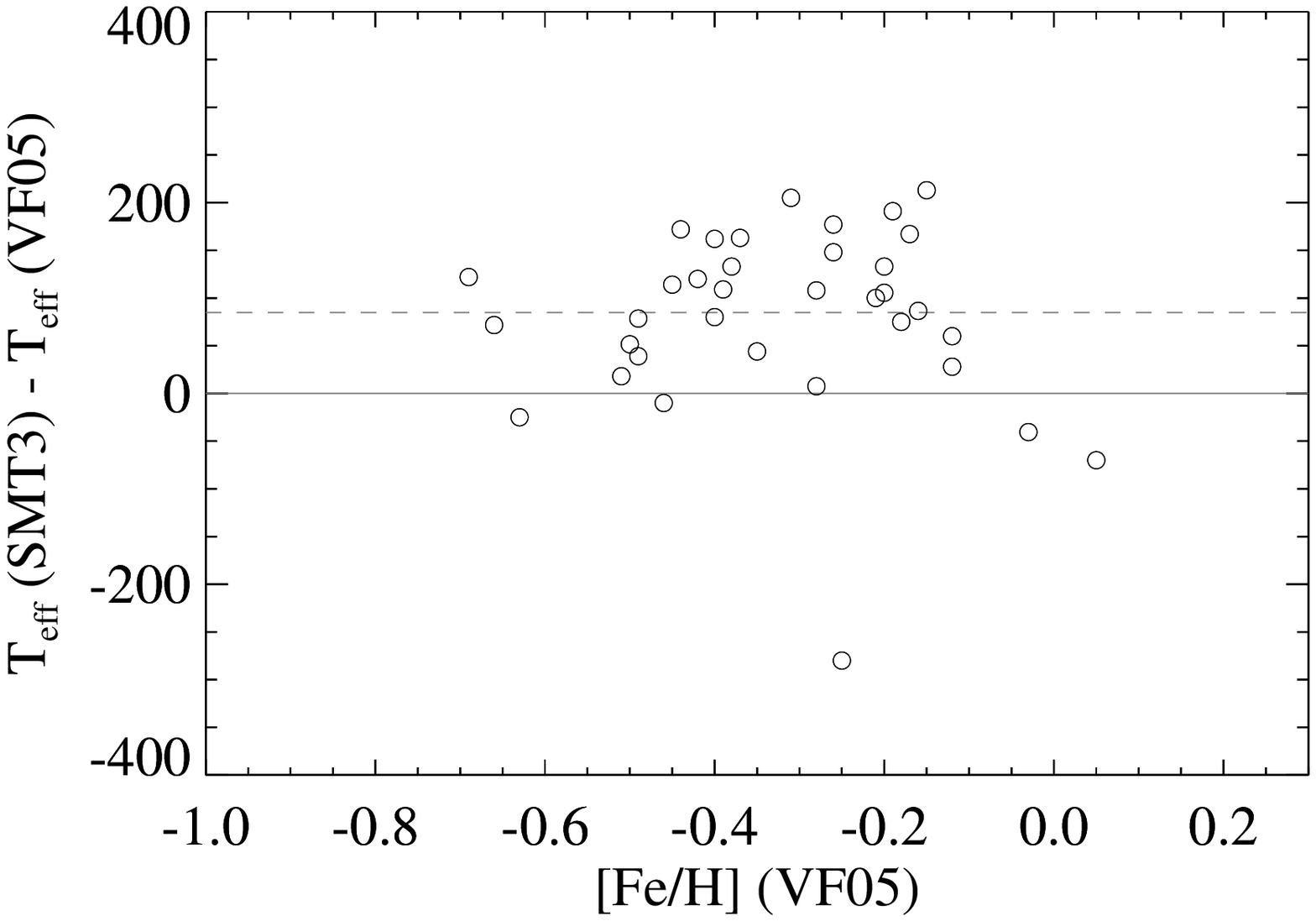}
\includegraphics[scale=0.45]{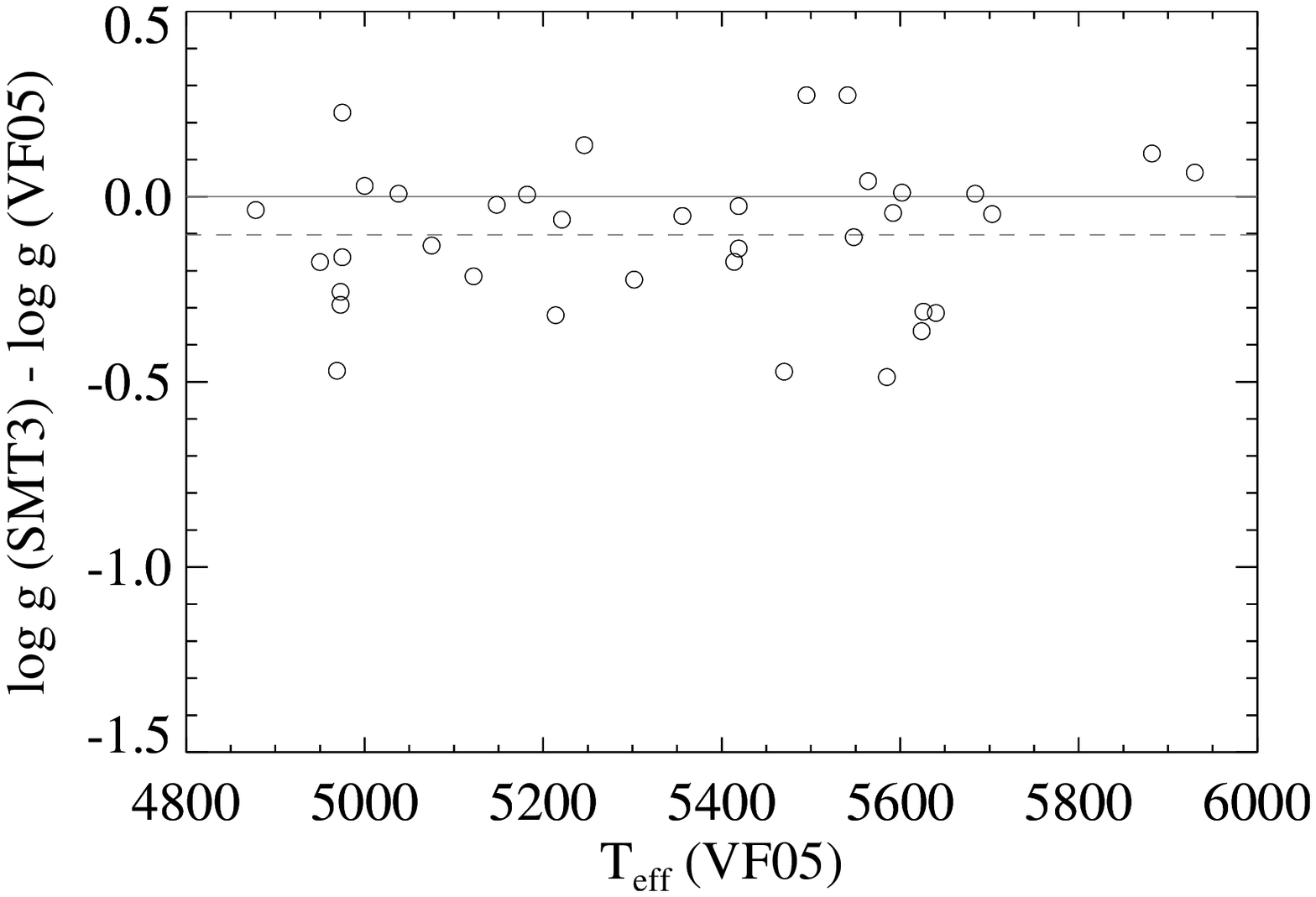}\includegraphics[scale=0.45]{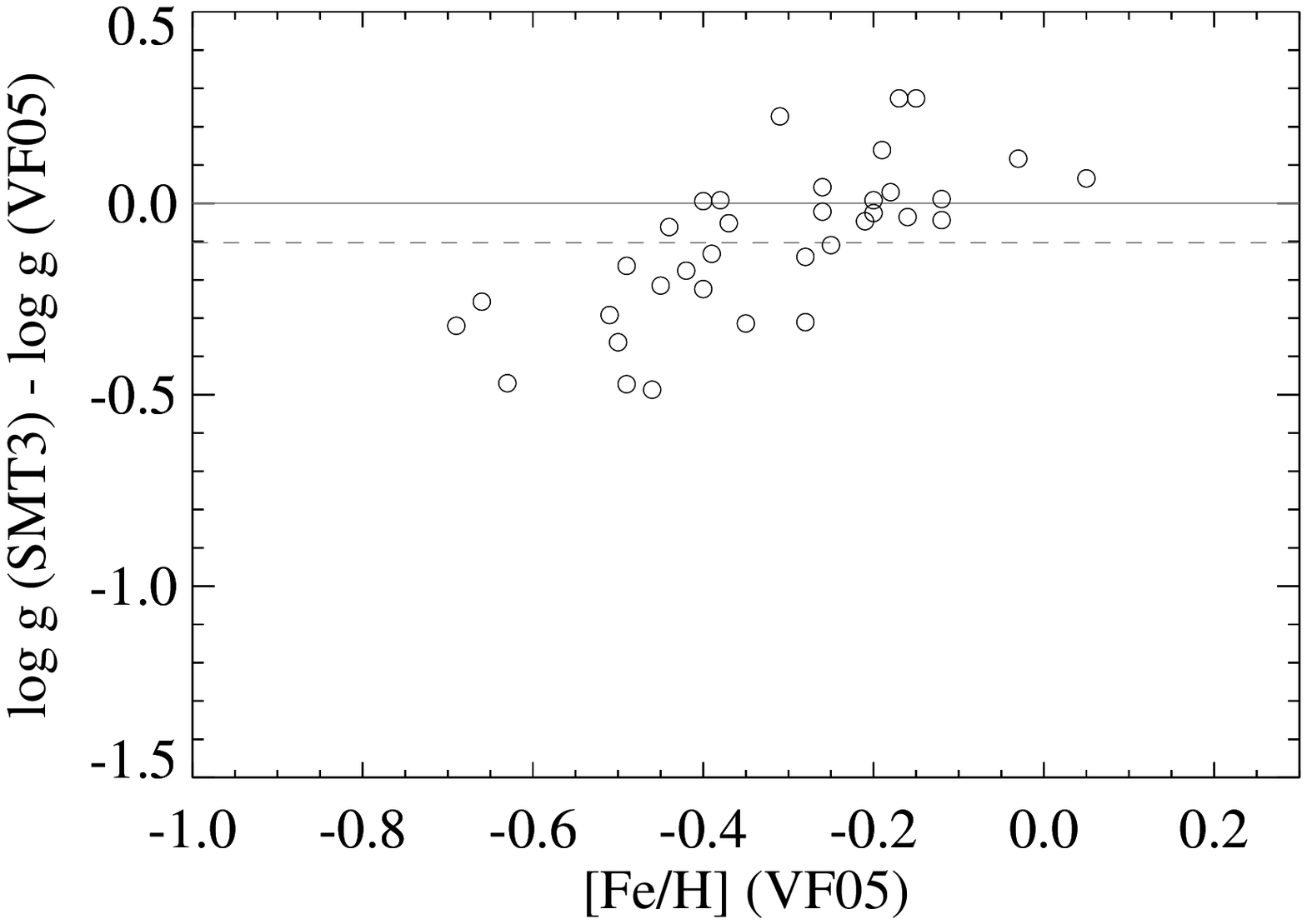}
\includegraphics[scale=0.45]{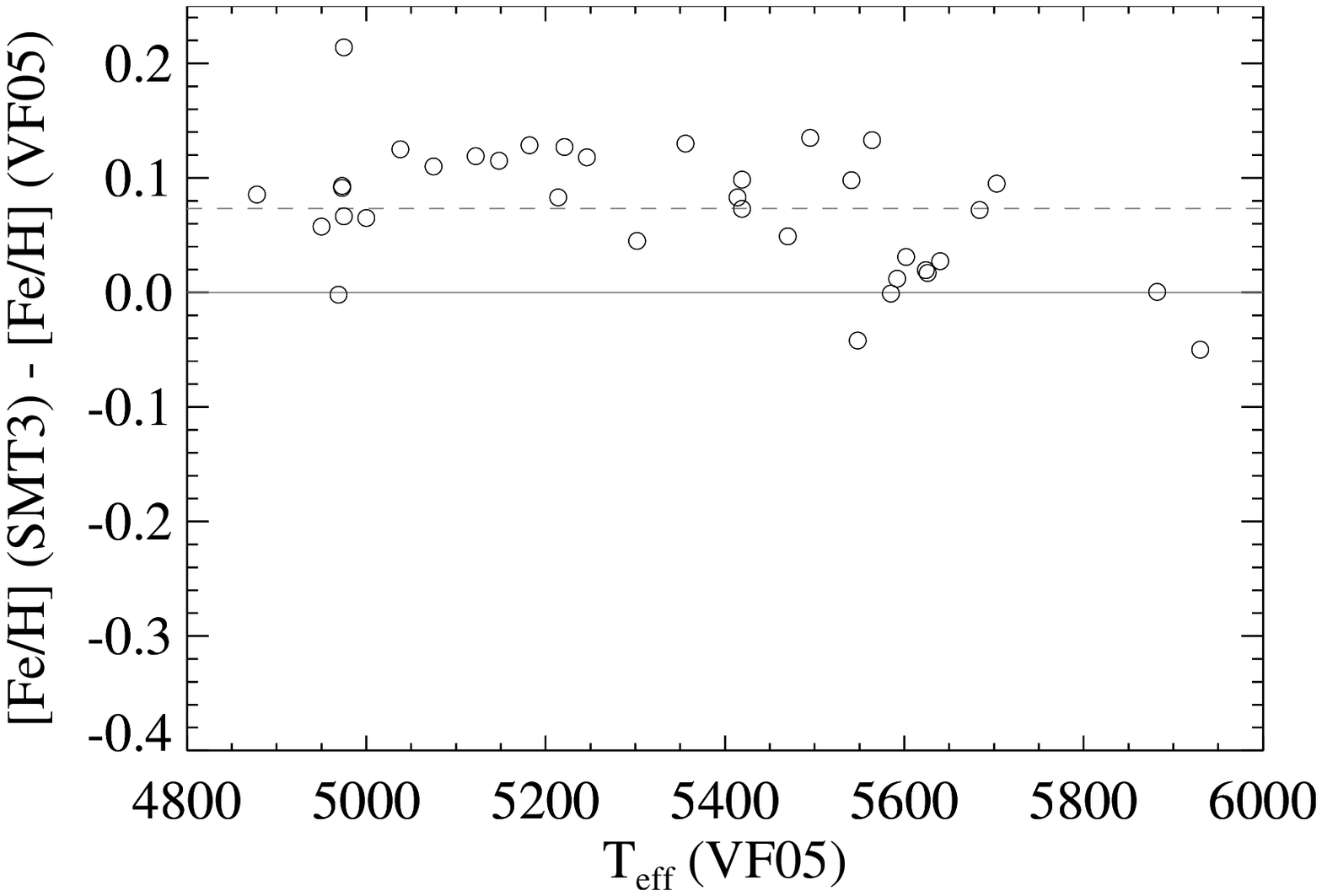}\includegraphics[scale=0.45]{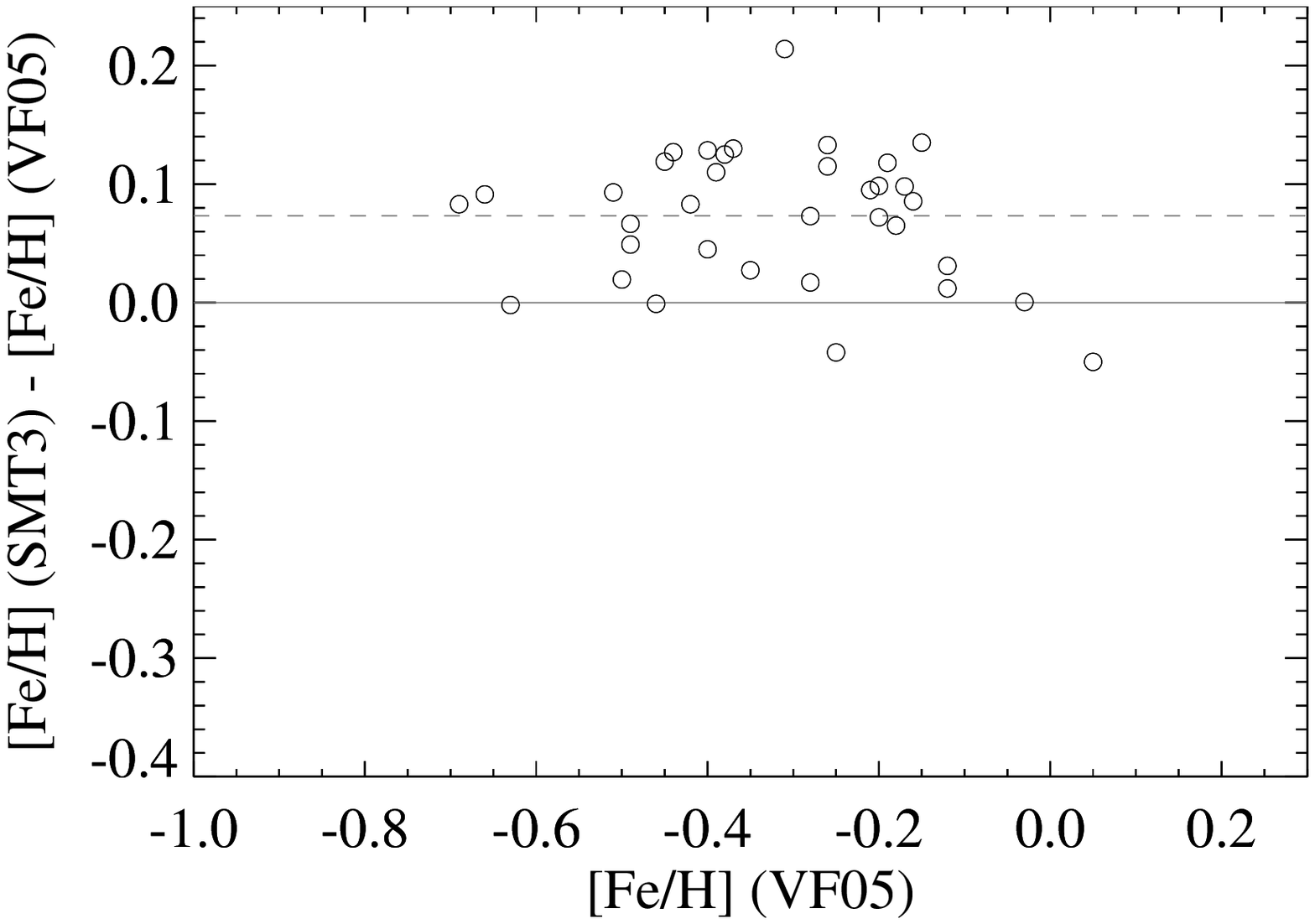}
\caption{Same as in Figure~\ref{fig:comp_moog_smt3}, but showing comparisons between SMT3 and VF05. In the bottom panels, metallicities from SMT3 represent raw estimates, before applying a zero-point adjustment (see text).\label{fig:comp_smt3_sme}} \end{figure*}

Figure~\ref{fig:comp_smt3_sme} displays comparisons in $\teff$, $\log{g}$, and [Fe/H] between SMT3 and VF05. The temperatures and metallicities from SMT3 are systematically higher than those from VF05 by $\Delta T_{\rm eff} = 85$~K and $\Delta{\rm [Fe/H]} = 0.07$~dex (see Table~\ref{tab:stat}). Given the correlation between the two quantities, it seems likely that a higher $\teff$ has led to a higher [Fe/H]. There also appears a correlation of the $\log{g}$ difference with metallicity, although the trend is much weaker than those seen from the comparisons of VF05 with MOOG.

The $1\sigma$ dispersion in $\teff$ comparison between SMT3 and VF05 is $95$~K. This is consistent with a reported internal precision in each of the methods ($44$~K and $79$~K for VF05 and SMT3, respectively). Similarly, a $1\sigma$ dispersion in $\teff$ comparison between SMT3 and MOOG is $101$~K. This suggests that MOOG temperatures have internal errors of $\sim90$~K, and is consistent with our earlier estimate based on a comparison with VF05. Similarly, a $1\sigma$ dispersion in [Fe/H] comparison between SMT3 and VF05 is $0.06$~dex, which is not far from a quadrature sum of individual internal precision measurements ($0.030$~dex for VF05 and $0.067$~dex for SMT3). A $1\sigma$ dispersion in [Fe/H] comparison between SMT3 and MOOG is $0.08$~dex, and is broadly consistent with $\sigma {\rm ([Fe/H])} = 0.05$--$0.08$ for our MOOG results (see \S~\ref{sec:moog}).

In terms of a zero point in metallicity, all of the above comparisons suggest that metallicity estimates from SMT3 are about $0.1$~dex higher than those from VF05. The level of the systematic offset is not alarmingly large, and an $\sim0.1$~dex systematic offset in [Fe/H] is not uncommon among different spectroscopic analyses \citep[e.g.,][]{torres:12}. A similar offset was also found from a comparison with spectroscopic metallicities in \citet[][see Table~\ref{tab:stat}]{casagrande:10}: Although there are only $11$ stars available in the comparison, the mean difference is $\Delta{\rm [Fe/H]} = 0.14$~dex in the sense that SMT3 predicts higher metallicities. Given the systematic nature of the difference, we decided to adjust SMT3 metallicities for our sample stars by a constant offset ($\Delta {\rm [Fe/H]} = 0.07$~dex), to be consistent with the metallicity scale of VF05 (and MOOG).

\begin{figure}
\centering
\includegraphics[scale=0.65]{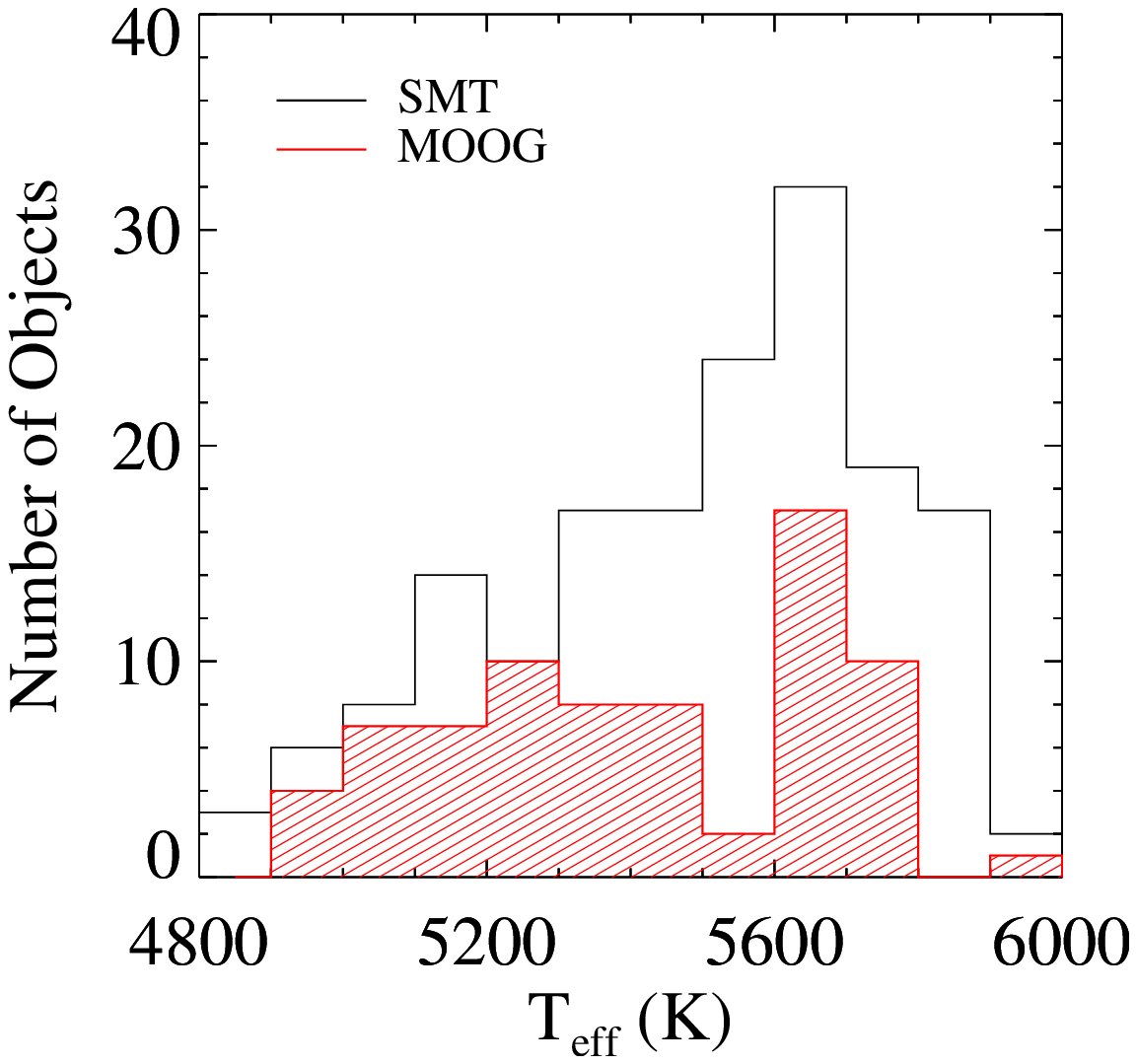} \includegraphics[scale=0.65]{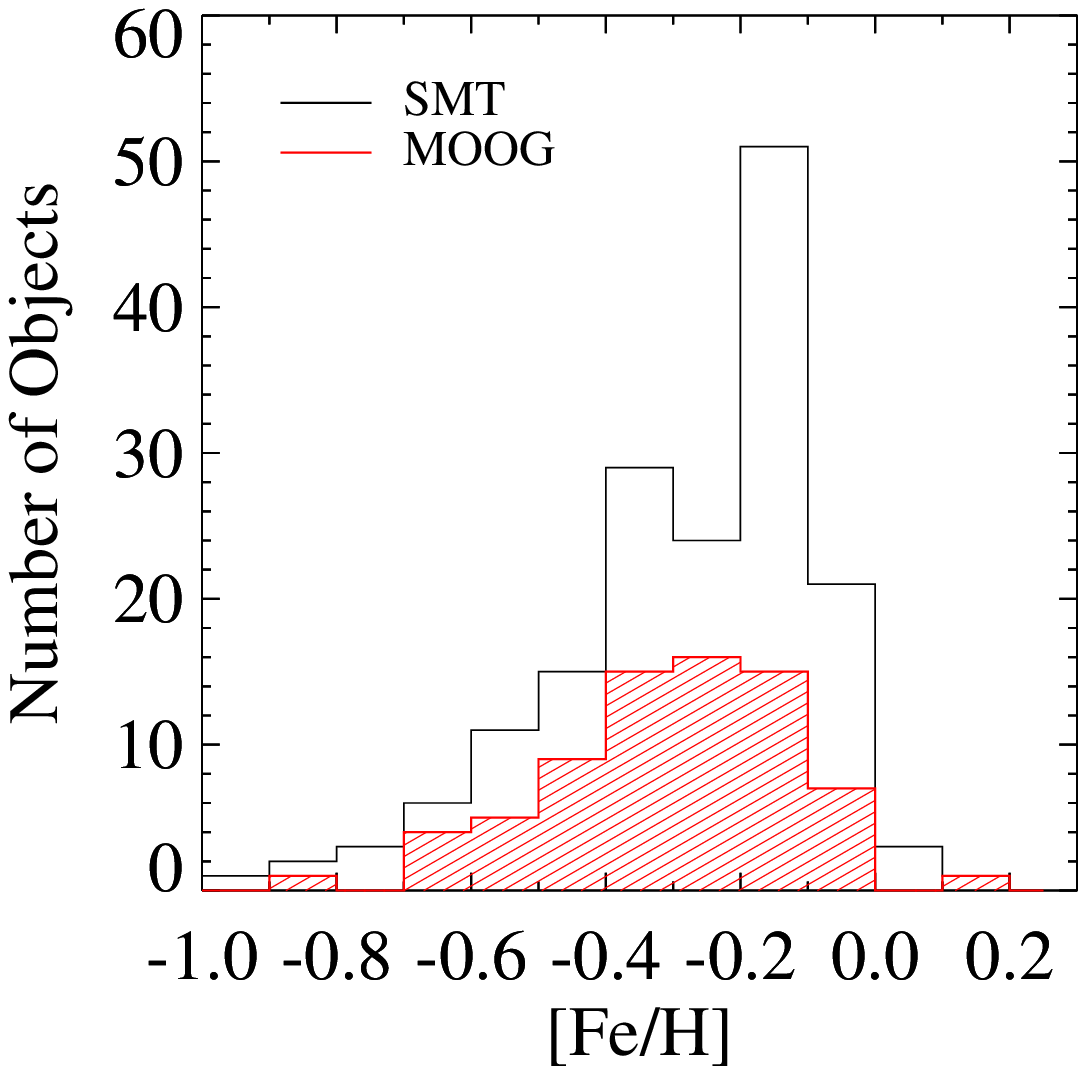}
\caption{Distributions of effective temperature (left) and metallicity (right) of the KPNO samples as obtained using SMT3 (black histogram) and those from MOOG (red shaded histogram). Metallicities are those after applying a zero-point adjustment. \label{fig:pardist}} \end{figure}

The $8^{\rm th}$ column in Table~\ref{tab:smt} (``[Fe/H]$_{\rm corr}$'') lists metallicities from SMT3 after the above zero-point correction. In Figure~\ref{fig:pardist} a black histogram shows a distribution of effective temperature (left panel) and that of metallicity (right panel) for all of our sample stars as obtained from SMT3. The red shaded histogram represents a subset of these stars, which was analyzed using MOOG. Our sample covers $4800 \la \teff {\rm (K)} \la 5900$ and $-0.8 \la {\rm [Fe/H]} \la 0.2$, and there is no correlation found between $\teff$ and [Fe/H] in each of these two parallel approaches.

\section{Results}\label{sec:result}

The goal of this work is to identify hypothesized sub-luminous field stars in the solar neighborhood, which have similar photometric properties with those in the Pleiades. Such stars are sub-luminous in the sense that they would reveal themselves as having fainter absolute magnitudes from {\it Hipparcos} than those inferred from MS fitting. If the {\it Hipparcos} parallaxes are correct, luminosities of these stars were overestimated in MS fitting for still unknown reasons. Since the luminosity of MS stars is strongly dependent on metallicity, below we first establish a metallicity sensitivity of stellar colors and magnitudes using previous star samples in the literature with accurate parallaxes and metallicities (\S~\ref{sec:sens}). The comparison in $M_V$ between {\it Hipparcos} and MS fitting is presented in \S~\ref{sec:kpno} for our KPNO sample. We utilize our metallicity measurements from both MOOG and SMT analyses to check our results against potential systematic errors in the adopted metallicity values. Although our sample selection was not designed to find young stars in the solar neighborhood, there are a small number of stars found in our sample with strong lithium absorptions and/or chromospheric activity levels that are characteristic of stars in the Pleiades. In \S~\ref{sec:activity} we use these stars to test a hypothesis that the Pleiades' distance is over-estimated in MS fitting due to young ages of its members. Finally, we present our best estimate of a MS-fitting distance to the Pleiades based on the observed MS of the Hyades and empirical metallicity sensitivities of stellar colors and magnitudes (\S~\ref{sec:plfit}).

\subsection{Metallicity Sensitivity Function}\label{sec:sens}

An absolute magnitude of an $i^{th}$ star in our sample can be computed either based on the {\it Hipparcos} parallax ($M^{\rm HIP}_{V,i}$) or MS fitting ($M^{\rm MS}_{V,i}$). The difference between the two $M_V$ estimates is equal to a difference in distance modulus:
\begin{eqnarray}
\Delta M_{V,i}
&\equiv& M^{\rm HIP}_{V,i} - M^{\rm MS}_{V,i} \label{eq:mv1} \\
&=& (V\, -\, M_V)^{\rm MS}_{0,i} - (V\, -\, M_V)^{\rm HIP}_{0,i}.\label{eq:mv2}
\end{eqnarray}
While $M^{\rm HIP}_{V,i}$ can be directly estimated from a $V$-band magnitude and a trigonometric parallax for individual stars, MS-fitting approach requires a well-defined set of color-magnitude relations over a wide range of metallicity. More specifically, $M^{\rm MS}_{V,i}$ in the above equations is a function of both temperature (or color) and metallicity, and should be known a priori to derive a distance from MS fitting. Within the temperature range of our sample, metal-poor MS stars are fainter than metal-rich stars at a given $\bv$ color, and the amount of offset in $M_V$ as a function of metallicity can be constrained either from observations or from theory.

In this paper, we employed a purely empirical approach, instead of relying on theoretical stellar isochrones, to search for anomalously sub-luminous stars in the solar neighborhood. We utilized an observed MS of the Hyades at its well-known metallicity and distance \citep{pinsonneault:04}, and applied metallicity effects on stellar colors and magnitudes. This was done by rewriting an absolute magnitude of a star as
\begin{eqnarray}
M^{\rm MS}_{V,i} = M^{\rm Hyades}_V (X_i) + \overline{\delta M_V} ({\rm [Fe/H]}_i),\label{eq:mv3}
\end{eqnarray}
where $M^{\rm Hyades}_V (X_i)$ is a $M_V$ of the Hyades' MS at a given color (or temperature) of a star in $X$ passband such as in $\bv$. The $\overline{\delta M_V} ({\rm [Fe/H]}_i)$ represents a metallicity term, which may also depend on colors. However, the color range of our KPNO sample is sufficiently narrow that color-magnitude relations at different metallicities are almost parallel to each other (see theoretical lines in the top left-hand panel of Figure~\ref{fig:cmd})\footnote{The approximate behavior of stars with differing metallicities can be understood with a simply homology relation \citep[e.g.,][see their equation~4]{portinari:10}. Note, however, that the influence of helium enrichment should be disentangled in order to measure the metallicity effect on stellar luminosity.}, and $\overline{\delta M_V}$ can essentially be treated as a function of metallicity alone. Theoretical predictions can be used to derive a metallicity term in broadband colors, but they are still uncertain because of large remaining uncertainties in input physics and stellar model parameters, and ultimately need to be constrained against a well-defined set of observational data \citep[e.g.,][]{an:07a,an:07b,an:15}. Comparisons of our sample stars with theoretical models will be discussed in the next paper of this series.

To derive a metallicity correction term in Equation~\ref{eq:mv3} on an empirical basis, we employed the {\it Hipparcos} data themselves. Here, our core assumption in this paper is that parallaxes for the majority of stars in the {\it Hipparcos} catalog are correct, but only a small fraction of these stars, such as those in the Pleiades, have larger parallax errors than those specified in the catalog. For this, we computed a difference between $M_V$ from {\it Hipparcos} and $M_V$ from the Hyades' MS at a given color of a star:
\begin{eqnarray}
\delta M_{V,i}
&\equiv& M^{\rm HIP}_{V,i} - M^{\rm Hyades}_V (X_i) \label{eq:mv4} \\
&=& V_i + 5 \log{\pi_i} + 5 - M^{\rm Hyades}_V (X_i),\label{eq:mv5}
\end{eqnarray}
where $V_i$ and $\pi_i$ are an observed $V$ magnitude and a star's parallax, respectively. The average $\delta M_{V}$ for a sufficiently large number of stars in a narrow bin of metallicity is
\begin{eqnarray}\label{eq:mv6}
\overline{\delta M_V} ({\rm [Fe/H]}) = \langle M^{\rm HIP}_{V} \rangle - M^{\rm Hyades}_V,
\end{eqnarray}
and yields an empirical correction to be added to the $M^{\rm Hyades}_V$ of a star (Equation~\ref{eq:mv3}). The $\overline{\delta M_V}$ can be computed for various bins in [Fe/H], and these corrections can be expressed as a metallicity sensitivity function in a given color index.

Using Equations~\ref{eq:mv3} and \ref{eq:mv4}, the difference between the MS-fitting and {\it Hipparcos}-based distance in Equation~\ref{eq:mv1} can be expressed as 
\begin{eqnarray}
\Delta M_{V,i}
&\equiv& M^{\rm HIP}_{V,i} - M^{\rm Hyades}_V (X_i) - \overline{\delta M_V} ({\rm [Fe/H]}_i) \label{eq:mv7} \\
&=& \delta M_{V,i} - \overline{\delta M_V} ({\rm [Fe/H]}_i).\label{eq:mv8}
\end{eqnarray}
Since the metallicity correction term $\overline{\delta M_V} ({\rm [Fe/H]})$ depends only on metallicity, searching for anomalously sub-luminous stars with large differences in $M_V$ between the MS-fitting and the {\it Hipparcos} distance is equivalent to finding outliers in $\delta M_V$ at a given metallicity.

\begin{figure*}
\center
\includegraphics[scale=0.65]{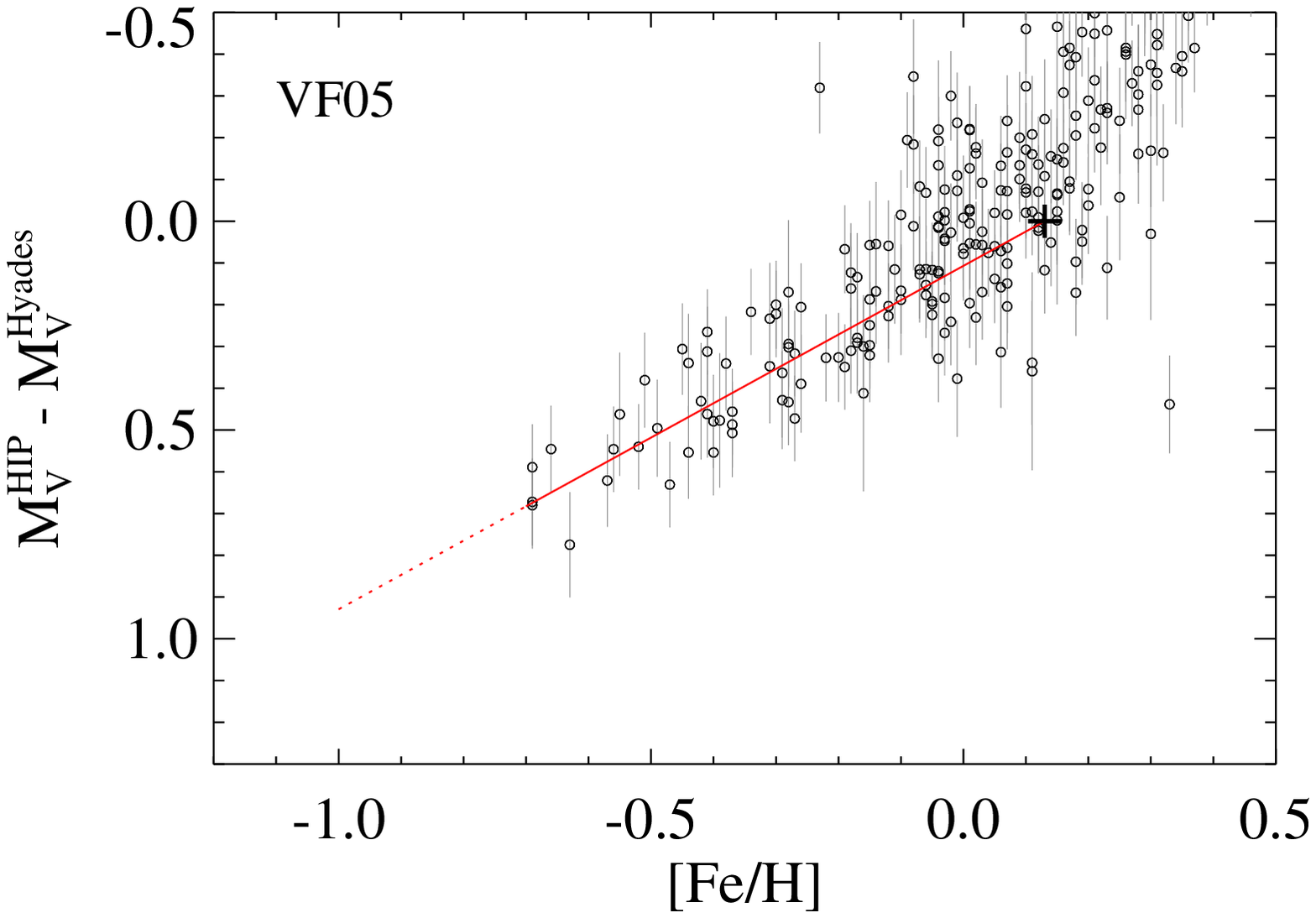}
\includegraphics[scale=0.65]{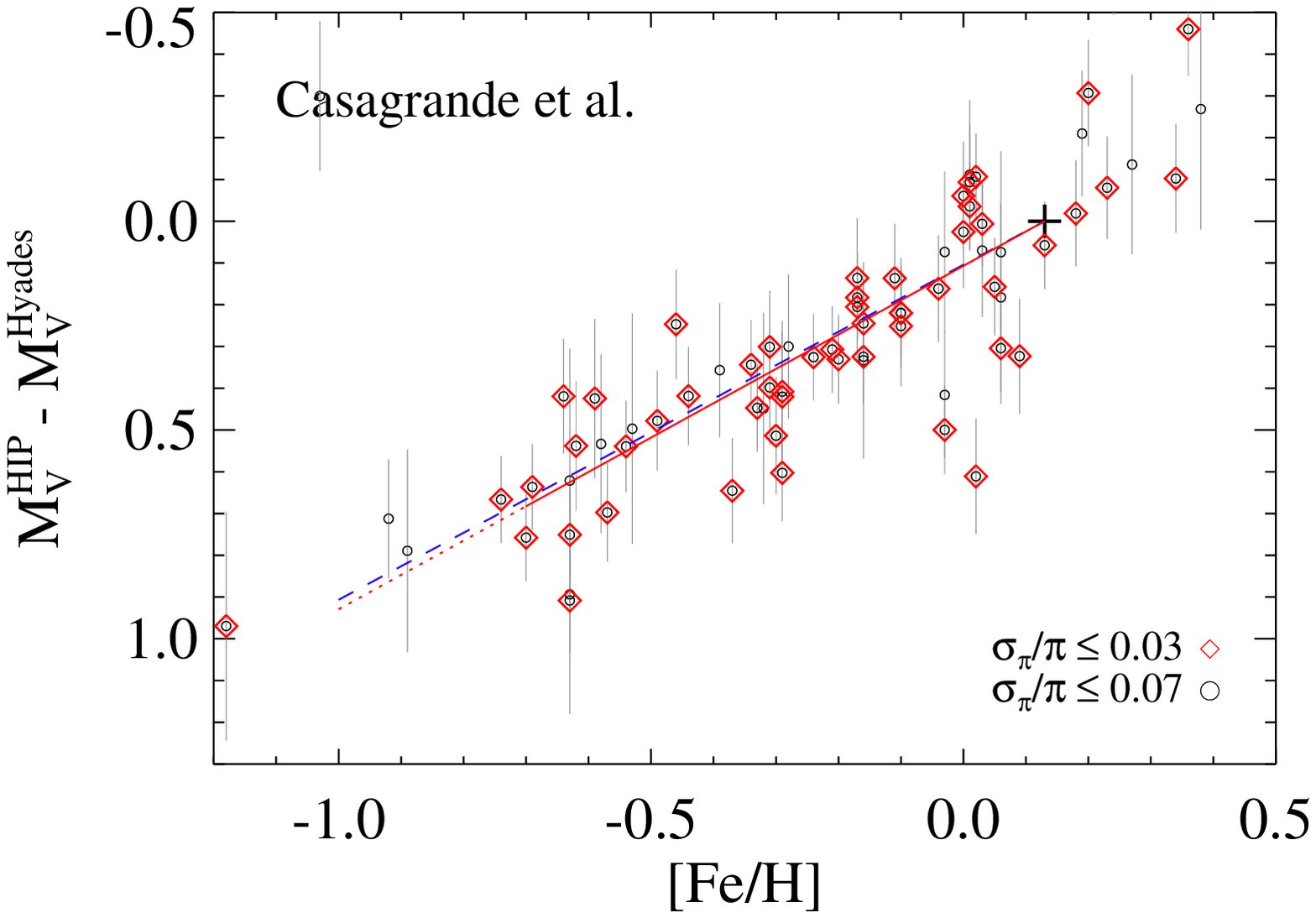} 
\caption{{\it Top:} The $\delta M_V$ of the {\it Hipparcos} field dwarfs with metallicities from VF05. Only those in $0.7 \leq \bv \leq 1.0$ with good parallaxes ($\sigma_\pi/\pi \leq 0.03$) are shown. The red solid line is a linear fit to stars in $-0.7 \leq {\rm [Fe/H]} \leq -0.1$, forced to match the position of the Hyades at ${\rm [Fe/H]} = +0.13$ with a zero magnitude difference (black cross), and is shown in both panels. {\it Bottom:} Same as in the top panel, but showing calibration stars in \citet{casagrande:10}. The blue dashed line is a linear fit to stars with fractional errors in parallax of better than $3\%$ (red diamond points), which passes through the black cross.\label{fig:sens}} \end{figure*}

We utilized stars in VF05 with good parallaxes ($\sigma_\pi / \pi \leq 0.03$) to derive an empirical metallicity sensitivity function, or the amount of offset in $M_V$ as a function of metallicity, in broadband colors. We restricted the sample to $0.7 \leq \bv \leq 1.0$ to make the sensitivity measurement suitable for our KPNO targets. The (conservative) lower limit ($\bv=0.7$) was set to eliminate potential contaminations by bright turn-off stars. The top panel in Figure~\ref{fig:sens} shows $\delta M_V$ of these stars computed using Equation~\ref{eq:mv5} in $\bv$ colors. These values decrease towards higher metallicity, because metal-rich MS stars are brighter than metal-poor stars at a given $\bv$ color. The observed trend in this panel reflects the metallicity sensitivity function (Equation~\ref{eq:mv6}).

The observed metallicity sensitivity seen in the top panel of Figure~\ref{fig:sens} tends to become steeper at higher metallicities. On the other hand, our sample is restricted to those having metallicities below solar, where the observed trend can be approximated by a straight line. Another consideration when deriving a metallicity sensitivity function is that $M_V$ estimated using Hyades' MS ($M^{\rm Hyades}_V$) must be equal to $M_V$ from {\it Hipparcos} ($M^{\rm HIP}_V$) at the metallicity of the Hyades (${\rm [Fe/H]} = +0.13$), or $\delta M_V=0$, by definition. However, there are more stars found with negative $\delta M_V$ than those with positive values at [Fe/H]$\sim+0.13$. This could be due to a scale error in metallicity or the presence of unresolved binaries in the sample, which are brighter than single MS stars, and therefore having systematically smaller $\delta M_V$. Without having a complete census on binarity, we simply proceeded with fitting a straight line to the data for stars with $-1.0 \leq {\rm [Fe/H]} \leq -0.1$, as shown by the red line, forced to pass through our adopted reference point, the Hyades (cross mark). Our metallicity sensitivity line has a slope of $-0.8$~mag dex$^{-1}$, and is used in the following analysis as a reference value for the mean metallicity sensitivity function in $\bv$.

We additionally used stars in \citet{casagrande:10} to independently check the above metallicity sensitivity function. We cross-identified stars in \citeauthor{casagrande:10} with the {\it Hipparcos} catalog using a $10''$~search radius along with stellar colors and magnitudes. In the bottom panel of Figure~\ref{fig:sens}, stars with fractional errors in parallax of better than $3\%$ are shown in diamond and those with $\leq 7\%$ in circled points, with spectroscopic metallicities as reported in \citet{casagrande:10}. As in the top panel, we selected stars in $0.7 \leq \bv \leq 1.0$. The blue dashed line is a linear fit to stars in \citeauthor{casagrande:10} with $\leq 3\%$ fractional errors in parallax (red diamond points), and shows nearly the same slope as the one obtained using stars in VF05 (red solid line).

In Figure~\ref{fig:sens}, the dispersion of the data points around the mean line is $\sigma(\delta M_V)=0.12$~mag for VF05 and $0.10$~mag for \citet{casagrande:10}, when only stars with good parallaxes ($\sigma_\pi/\pi \leq 0.03$) are used. Photometric errors in $\bv$ of $\sim0.02$~mag, which is reasonable to assume, are translated into $0.1$~mag error in $M_{V}$, because the slope of MS is about $5$ on a $\bv$ CMD. The error in the observed $V$ magnitude directly affects the error in $\delta M_V$ (see Equation~\ref{eq:mv5}), but is negligible compared to that from a color error. A parallax error likely produces an error in $\delta M_V$ of $\sim0.05$~mag, and the error in [Fe/H] of $\sim0.05$~dex is translated into $0.04$~mag error in $\delta M_V$. All together, the error in $\delta M_V$ is largely dominated by photometric color errors. The remaining errors could come from unresolved companions of binaries and/or an older age of a star than the Hyades ($550$~Myr), which would make stars look brighter than a single-star zero-age MS.

While the observed dispersion is close to our expectation, there are a few outliers in Figure~\ref{fig:sens} that are far below the metallicity sensitivity line, with the largest magnitude offset $\Delta M_V$ of about $0.5$~mag. According to Equation~\ref{eq:mv8}, this offset corresponds to a larger {\it Hipparcos}-based absolute magnitude than $M_V$ from MS fitting. A possibility is that $\bv$ color of a star is too blue by $0.1$~mag, or that $V$ mag is too large by $0.5$~mag, but the sizes of photometric errors required are uncomfortably large. The other possibility is that a star's spectroscopic metallicity was overestimated by $\Delta {\rm [Fe/H]}=0.5$, but again the expected error seems too large for a high-resolution spectroscopic analysis. In addition, this can be due to an overestimated parallax measurement in the {\it Hipparcos} catalog, which may pose a similar problem with the Pleiades stars. However, a parallax error cannot be condemned for a large $\Delta M_V$ unless other sources of errors are well understood. In the following analysis, we used our homogeneous KPNO data set with well-determined spectroscopic metallicities to better identify and characterize the properties of outliers in the $\delta M_V$ versus [Fe/H] diagram.

\subsection{Magnitude Excess of KPNO Sample}\label{sec:kpno}

\begin{figure*}
\center
\includegraphics[scale=0.65]{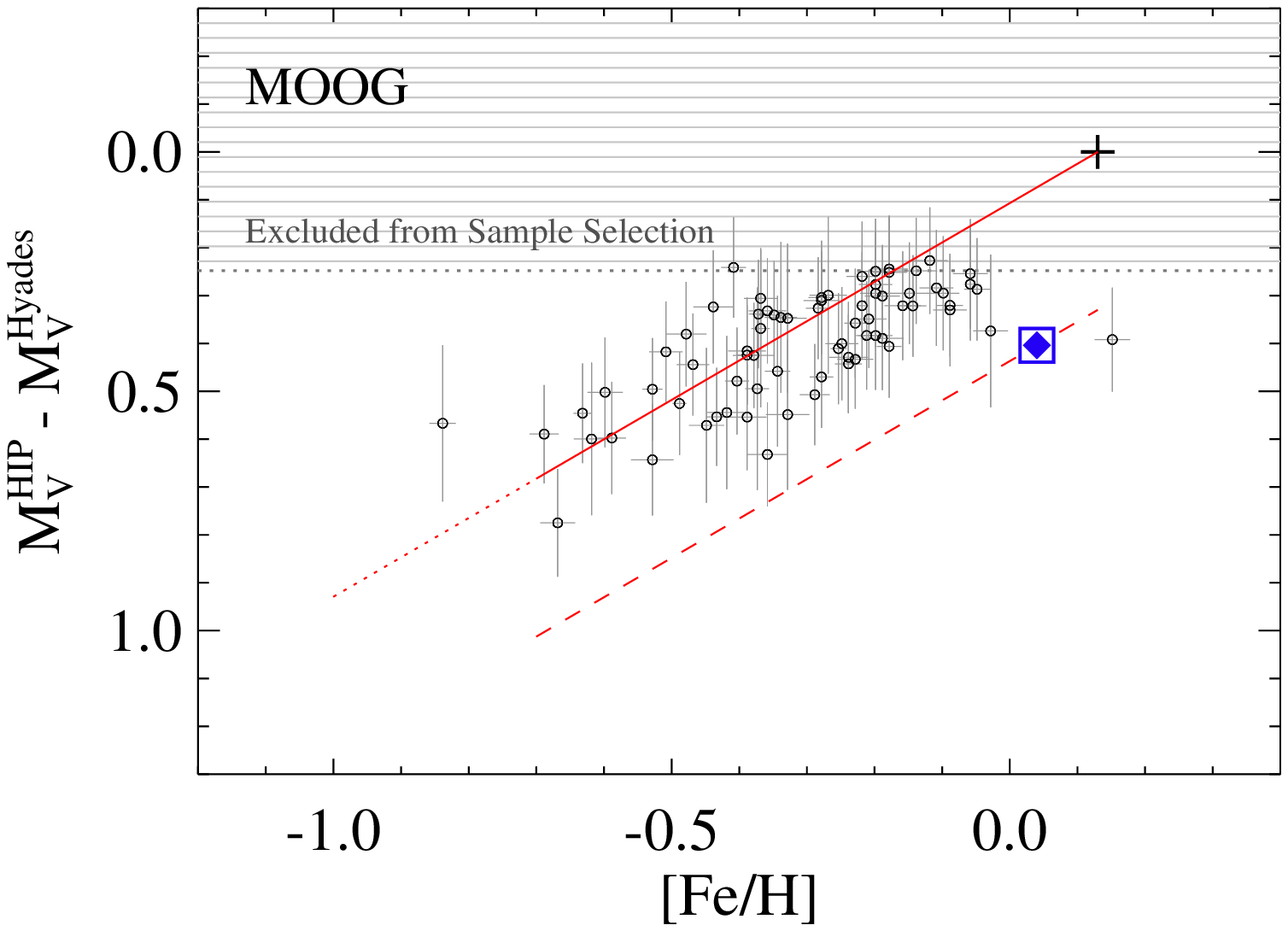}
\includegraphics[scale=0.65]{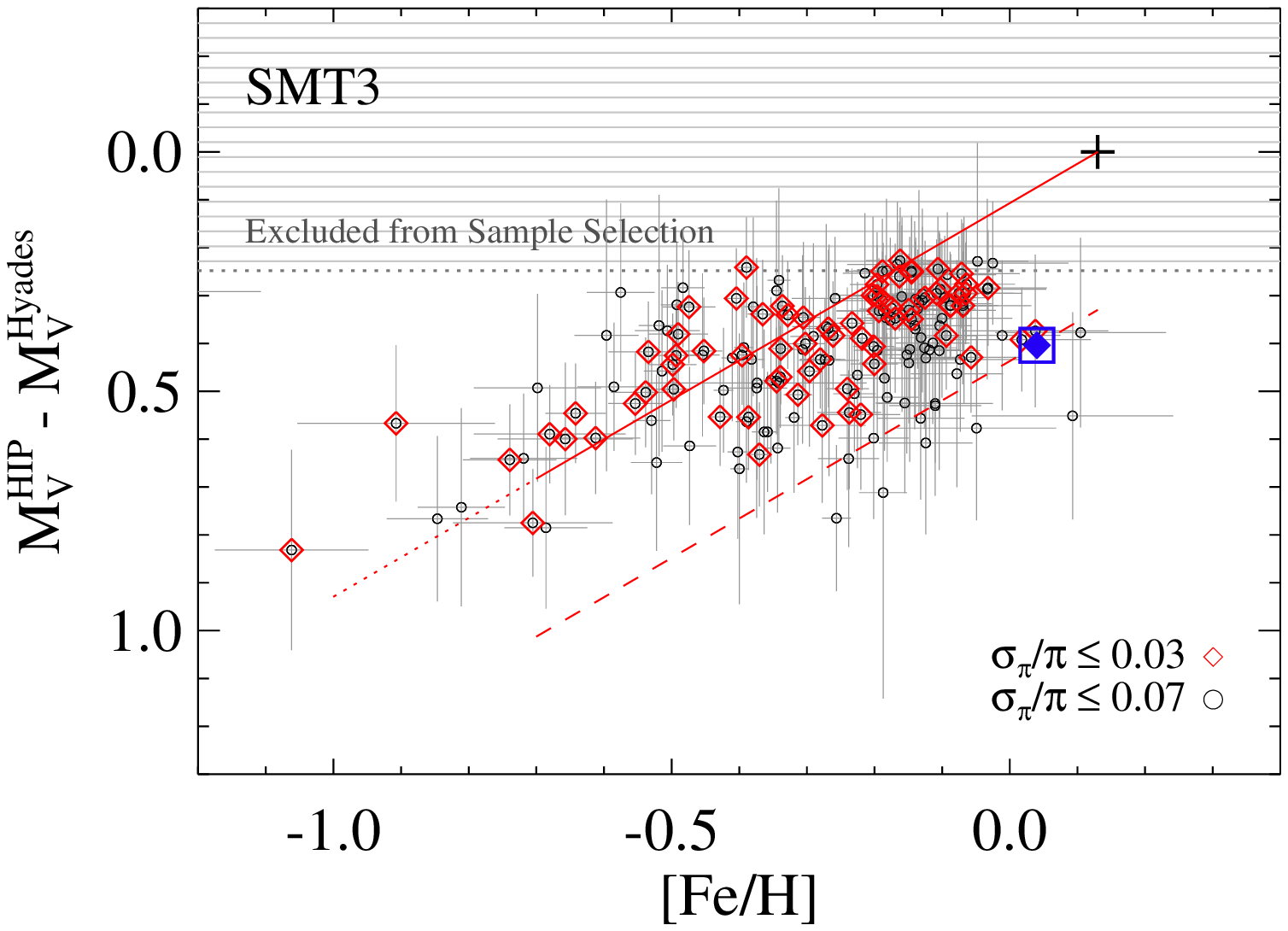}
\caption{The $\delta M_V$ of the KPNO sample with MOOG (top) and SMT3 (bottom) metallicity estimates. Stars in the top panel are those having good parallax measurements ($\sigma_\pi/\pi \leq 0.03$). The same set of stars are also shown as red diamond points in the bottom panel. The red solid line represents a mean metallicity sensitivity function as derived from the VF05 stars in the top panel of Figure~\ref{fig:sens}. The blue diamond point with a box indicates the Pleiades, and its magnitude excess ($\Delta M_V$) is shown by a red dashed line. The grey shaded area represents a $\delta M_V$ limit set by our sample selection.\label{fig:deficit}} \end{figure*}

Figure~\ref{fig:deficit} shows the same $\delta M_V$ versus [Fe/H] plot as in Figure~\ref{fig:sens}, but for our KPNO sample, where $\delta M_V$ of stars and their errors are those listed in the second and third columns in Table~\ref{tab:dmv}. Stars with MOOG and SMT3 metallicities (both scaled to match the VF05 metallicities) are used in the top and the bottom panel, respectively. In each panel, the cross mark indicates the Hyades at the cluster's distance from {\it Hipparcos} ($\Delta M_V = \delta M_V = 0$), and the same red line as in the top panel of Figure~\ref{fig:sens} is shown for the empirical metallicity sensitivity function in $\bv$ CMDs. A magnitude excess in this study is defined as having a shorter {\it Hipparcos} distance than that from MS fitting, or a positive offset of a star from the mean metallicity function ($\Delta M_V > 0$). While our sample does not include stars in the Pleiades, we mark the position of the Pleiades as a blue diamond point at its well-known metallicity ${\rm [Fe/H]}=+0.04$ \citep[see references in][]{an:07b}. We found $\delta M_V=0.40$ from $33$ single MS stars in the Pleiades using the Hyades' MS. The {\it Hipparcos} distance of the Pleiades is shorter than the MS-fitting distance by $\Delta \dmn = \Delta M_V=0.33$ \citep{pinsonneault:98,an:07b}. Below we use $\Delta M_V=0.33$ as a reference to judge whether stars in our KPNO sample are anomalously faint, or have a large magnitude excess, like those in the Pleiades at its {\it Hipparcos} distance.

In the top panel of Figure~\ref{fig:deficit}, a total of $74$ field stars with MOOG metallicities are shown, representing the entire KPNO sample with parallax errors of better than $3\%$. In the bottom panel, $167$ sample stars with SMT3 metallicities are shown with parallax errors of better than $7\%$. The red diamond points are a subset of these stars, having more accurate parallaxes ($\sigma_\pi/\pi \leq 0.03$). The vertical error bars are a quadrature sum of errors from photometry and parallax (see Equation~\ref{eq:mv5}). As in Figure~\ref{fig:sens}, we included an error in $\delta M_V$ from a photometric color, by multiplying a $\bv$ error by a slope of the Hyades' MS. In the bottom panel, horizontal error bars are metallicity errors reported by SMT; HIP~95727 is not displayed because of its large metallicity error ($0.51$~dex).

In Figure~\ref{fig:deficit} the gray shaded region represents a forbidden area that was excluded from our color-magnitude selection of the KPNO sample. As shown in Figure~\ref{fig:cmd}, the upper limit in stellar brightness (or a lower limit in $M_V$) in our sample selection was set almost parallel to the MS of the Hyades, and any stars having $\delta M_V \la 0.25$~mag were not included in our observing runs. The boundary of the shaded region is not a clear cut division, because the slope of the Hyades' MS is not exactly parallel to our color-magnitude selection. Nevertheless, our search for sub-luminous MS stars is almost unaffected by this sample bias since such stars would reveal themselves as having significantly faint $M_V$ with the {\it Hipparcos} parallax, or a large $\Delta M_V$ at a given metallicity. On the other hand, a metallicity sensitivity function cannot be constructed using our field star sample alone because of a sample bias against intrinsically bright, metal-rich stars.

Considering the above sample bias, most of our KPNO stars in Figure~\ref{fig:deficit} are distributed in the same manner as in Figure~\ref{fig:sens} and follow the empirical metallicity sensitivity function (red solid line). In the bottom panel, a distribution of the sample with $\leq 3\%$ parallax errors is tighter than that from the entire KPNO sample. A true dispersion of a magnitude excess from the mean sensitivity line [$\sigma(\Delta M_V$)] is difficult to measure for our sample due to the limit set by our sample selection (grey region). Nevertheless, metal-poor stars ($-0.7 \leq {\rm [Fe/H]} \leq -0.4$) are relatively free from the sample bias, from which we found $\sigma (\Delta M_V)=0.11$~mag for a rms dispersion of these stars with highly accurate parallaxes ($\sigma_\pi/\pi \leq 0.03$). On the other hand, the expected size of error in $\Delta M_V$ for individual stars is $0.16$~mag from $2\%$ errors in the photometry, $\sim2\%$ error in parallax, and $0.067$~dex error in metallicity (see above), which approximately matches the measured dispersion.

\begin{figure*}
\centering
\includegraphics[scale=0.65]{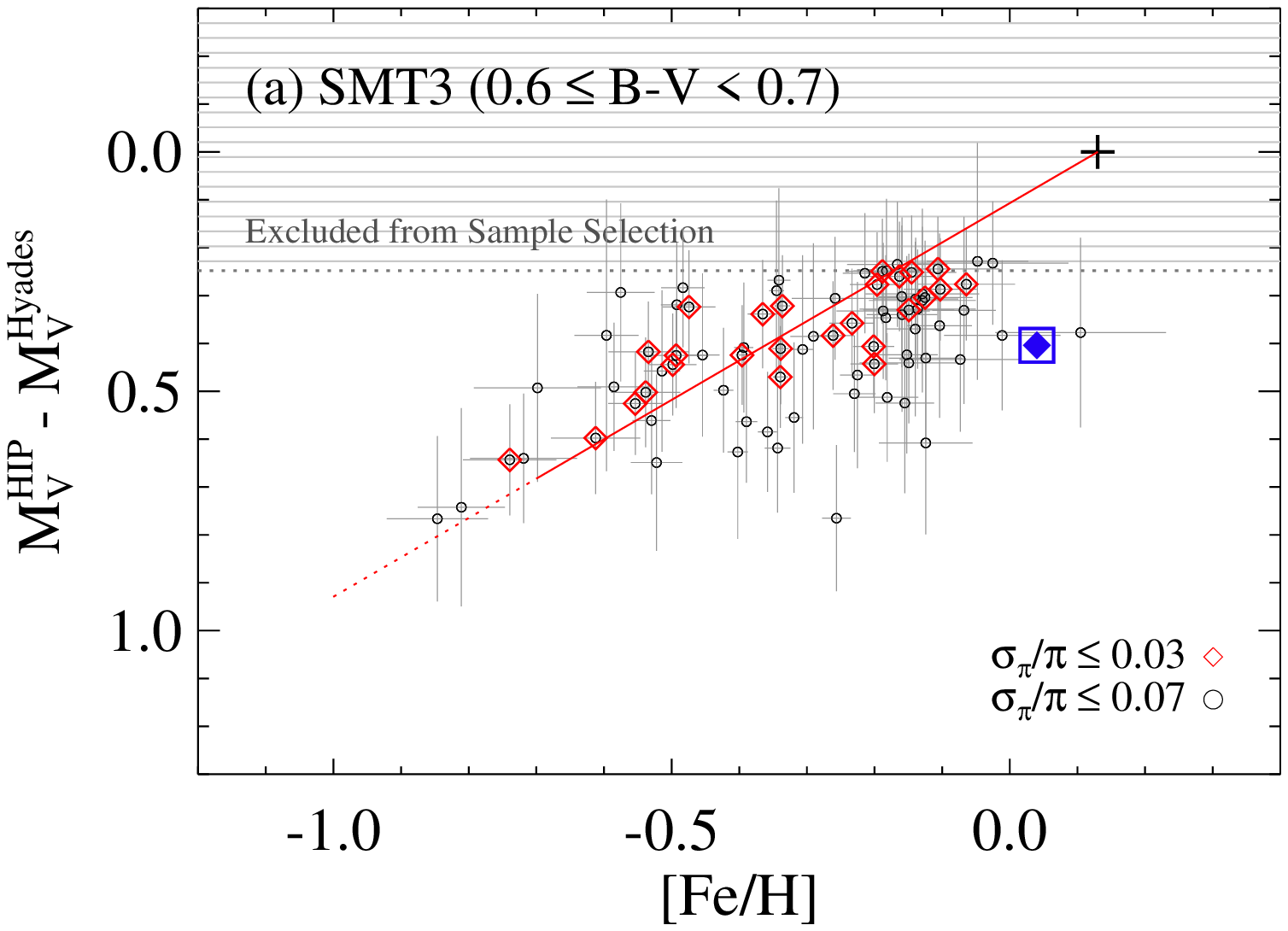}
\includegraphics[scale=0.65]{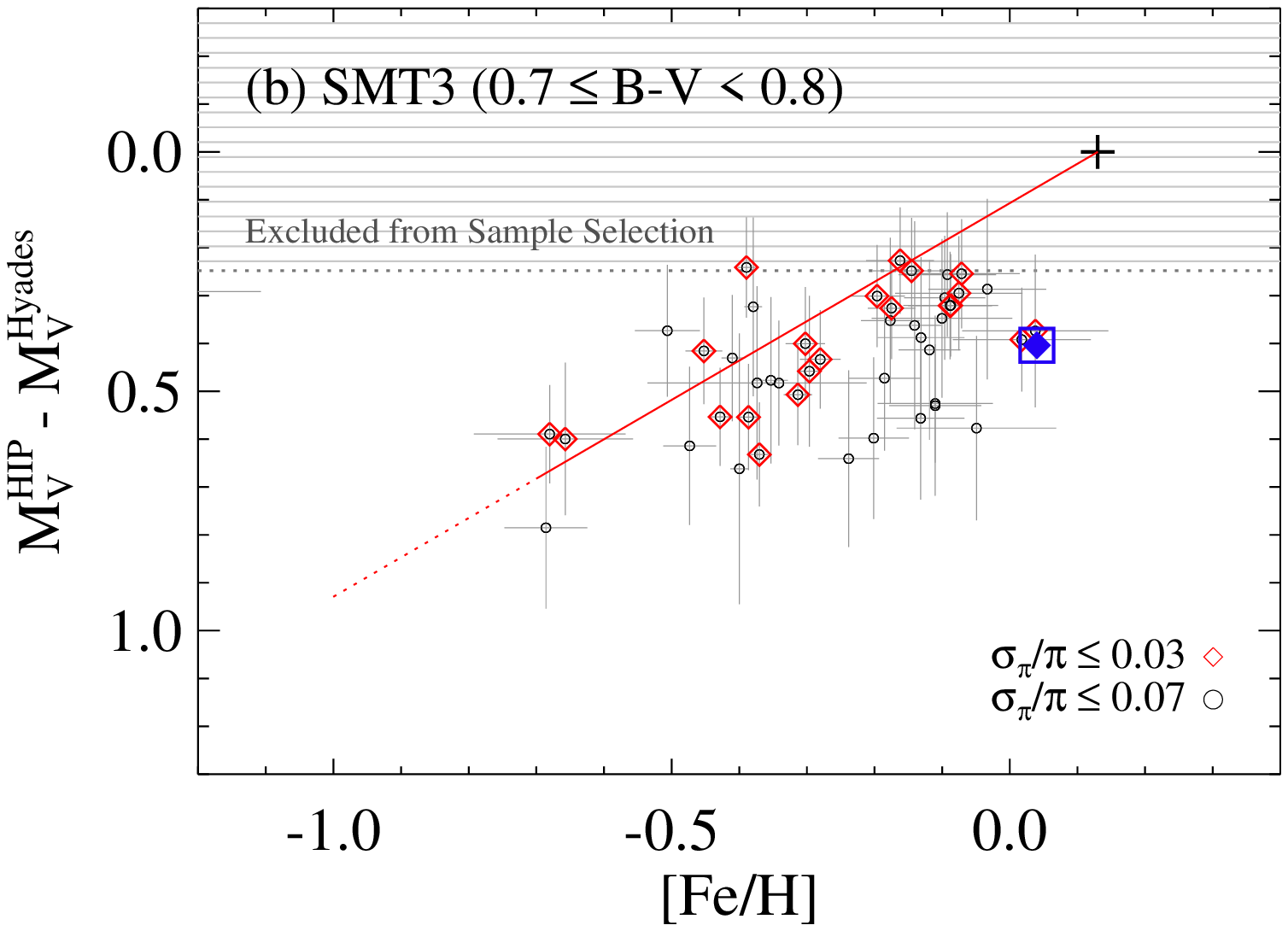}
\includegraphics[scale=0.65]{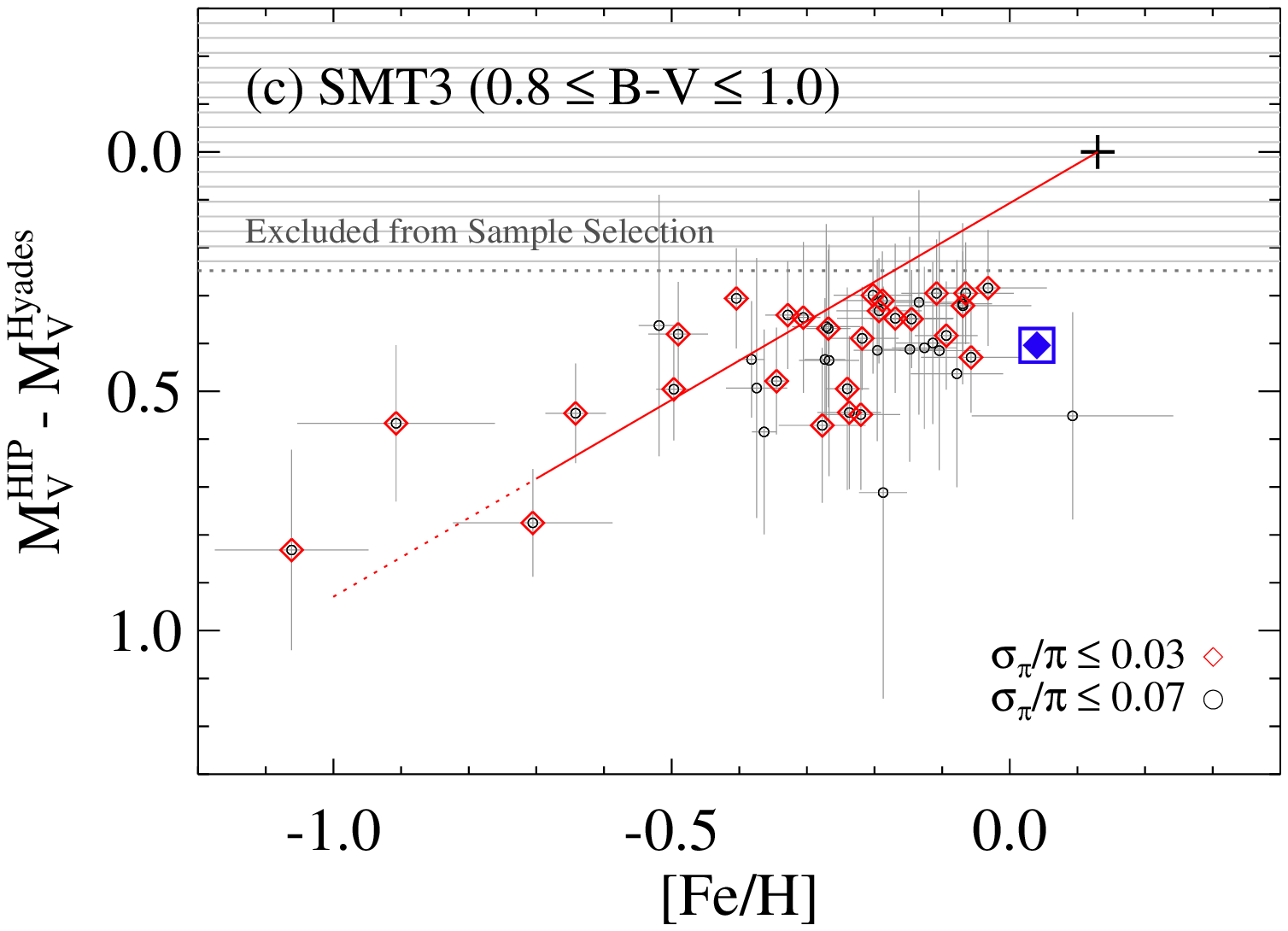}
\caption{Same as in the bottom panel of Figure~\ref{fig:deficit}, but displaying KPNO samples in different color ranges: (a) $0.6 \leq \bv < 0.7$, (b) $0.7 \leq \bv < 0.8$, and (c) $0.8 \leq \bv \leq 1.0$.\label{fig:deficit_color}} \end{figure*}

\begin{figure*}
\epsscale{0.65}
\plotone{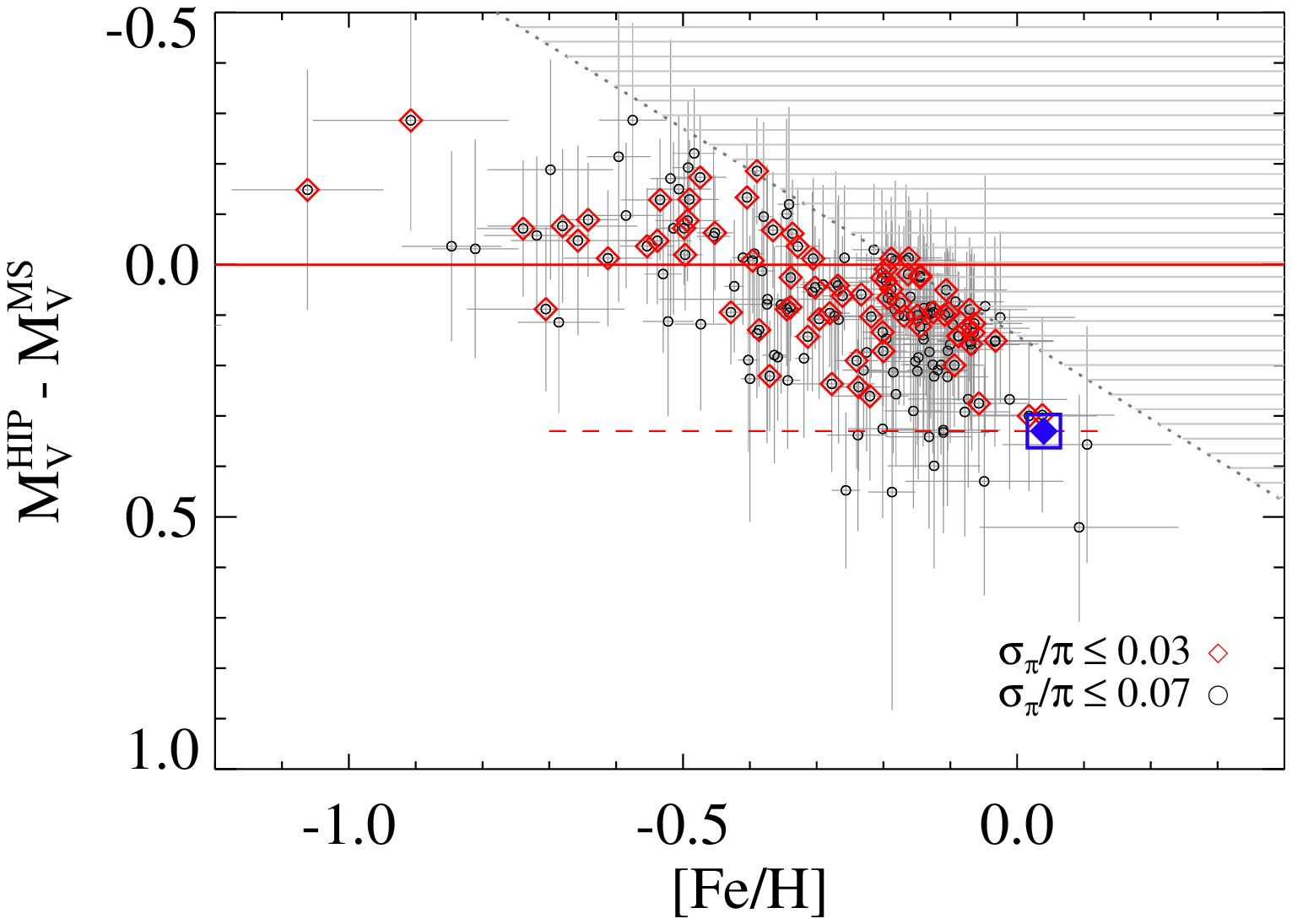}
\caption{Same as in the bottom panel of Figure~\ref{fig:deficit}, but displaying $\Delta M_V$ or a difference between a MS-fitting distance modulus and that from the {\it Hipparcos} parallax.\label{fig:distmod}} \end{figure*}

Figure \ref{fig:deficit_color} shows the same $\delta M_V$ versus [Fe/H] diagrams as in the bottom panel of Figure~\ref{fig:deficit}, but in three different color bins: $0.6 \leq \bv < 0.7$ (top), $0.7 \leq \bv < 0.8$ (middle), and $0.8 \leq \bv \leq 1.0$ (bottom). The minimum $\delta M_V$ of stars in $0.8 \leq \bv \leq 1.0$ is slightly larger than those for bluer stars. As described in \S~\ref{sec:sample}, this is because our color-magnitude selection, which was made using the Pleiades' MS, is not exactly parallel to the Hyades' MS (see the top left-hand panel in Figure~\ref{fig:cmd}). The Pleiades members progressively become fainter than older stars with the same metallicity or those predicted from standard stellar models, most likely due to stellar activities. The effect is mild at $0.9 \la \bv \la 1.0$, but is not convincingly observed for stars with $\bv < 0.9$ \citep[see Figure~20 in][]{an:07b}. We limited our sample to $\bv \leq 1$ since such activity-related change of stellar colors and magnitudes becomes severe for redder stars \citep{stauffer:03,an:07b}. Other than a slight difference in the minimum $\delta M_V$, our sample stars in Figure~\ref{fig:deficit_color} behave almost independently of color ranges.

While most of the KPNO sample stars have {\it Hipparcos} parallaxes that are consistent with MS-fitting distances, there are a few stars that have shorter {\it Hipparcos} distances. In the top panel of Figure~\ref{fig:deficit}, there is one such star found from the MOOG analysis (HIP~99965). In the bottom panel, which shows the same $\delta M_V$ with SMT3 metallicities, none of the stars with accurate parallaxes ($\sigma_\pi/\pi \leq 0.03$; red diamond) has a larger magnitude excess than the Pleiades ($\Delta M_V = 0.33$; red dashed line). For the extended sample with $\sigma_\pi/\pi \leq 0.07$, a total of $9$ stars exhibit larger $\Delta M_V$ than the Pleiades, which account for $\sim 5$\% of the entire KPNO sample. These are HIP~5313, HIP 10276, HIP~73138, HIP~73845, HIP~74126, HIP~75446, HIP~78336, HIP~81831, and HIP~97668. The $\Delta M_V$ estimates of our sample stars, including those with the largest magnitude excess, are listed in the $4^{\rm th}$ and $5^{\rm th}$ columns in Table~\ref{tab:dmv}. The errors in $\Delta M_V$ are larger than those of $\delta M_V$ due to an additional contribution from spectroscopic metallicity errors. Figure~\ref{fig:distmod} shows $\Delta M_V$ of stars as a function of the SMT3 metallicity. The symbols are the same as in the bottom panel of Figure~\ref{fig:deficit}. The statistical significance of the above $9$ stars is less ($\sim2\sigma$) than that of the Pleiades, since individual stars have larger errors in $\delta M_V$ from photometry, metallicity, and parallax than in the case of a cluster where a large number of stars can be used together to increase an internal precision of the measurements.

Because stars with large $\Delta M_V$ are all having relatively large parallax errors, it is unclear whether the large magnitude excess was induced by random parallax measurement errors, or by a hidden systematic error in the {\it Hipparcos} parallax, which was originally suggested by \citet{pinsonneault:98} to explain the short Pleiades distance. We also analyzed spectra of the above $9$ stars with the largest magnitude excess using MOOG (Table~\ref{tab:moog}), which have not been originally included in our MOOG analysis due to their large parallax errors. With MOOG metallicities, six out of the $9$ stars were still having larger $\Delta M_V$ than the Pleiades. Their $\Delta M_V$ estimates are shown in the $6^{\rm th}$ and $7^{\rm th}$ columns in Table~\ref{tab:dmv}, along with those for all stars analyzed using MOOG.

In addition to $\bv$ colors, we repeated the above experiment with $\vk$ colors using the MS of the Hyades in the $\vk$ versus $V$ CMD to compute $\delta M_V$ of individual stars. These $\delta M_V$ estimates are listed in the $8^{\rm th}$ and $9^{\rm th}$ columns in Table~\ref{tab:dmv}, which are not equal to $\delta M_V$ from $\bv$ because of a different color-$\teff$ relation in $\vk$. The $\vk$ is also less sensitive to metallicity than $\bv$, while being more sensitive to unresolved binaries, giving an independent look at the distribution of stars in the $\delta M_V$ versus [Fe/H] diagram. In addition, photometry from a uniform all sky survey like 2MASS can provide more reliable estimates in $\delta M_V$.

\begin{figure*}
\centering
\includegraphics[scale=0.65]{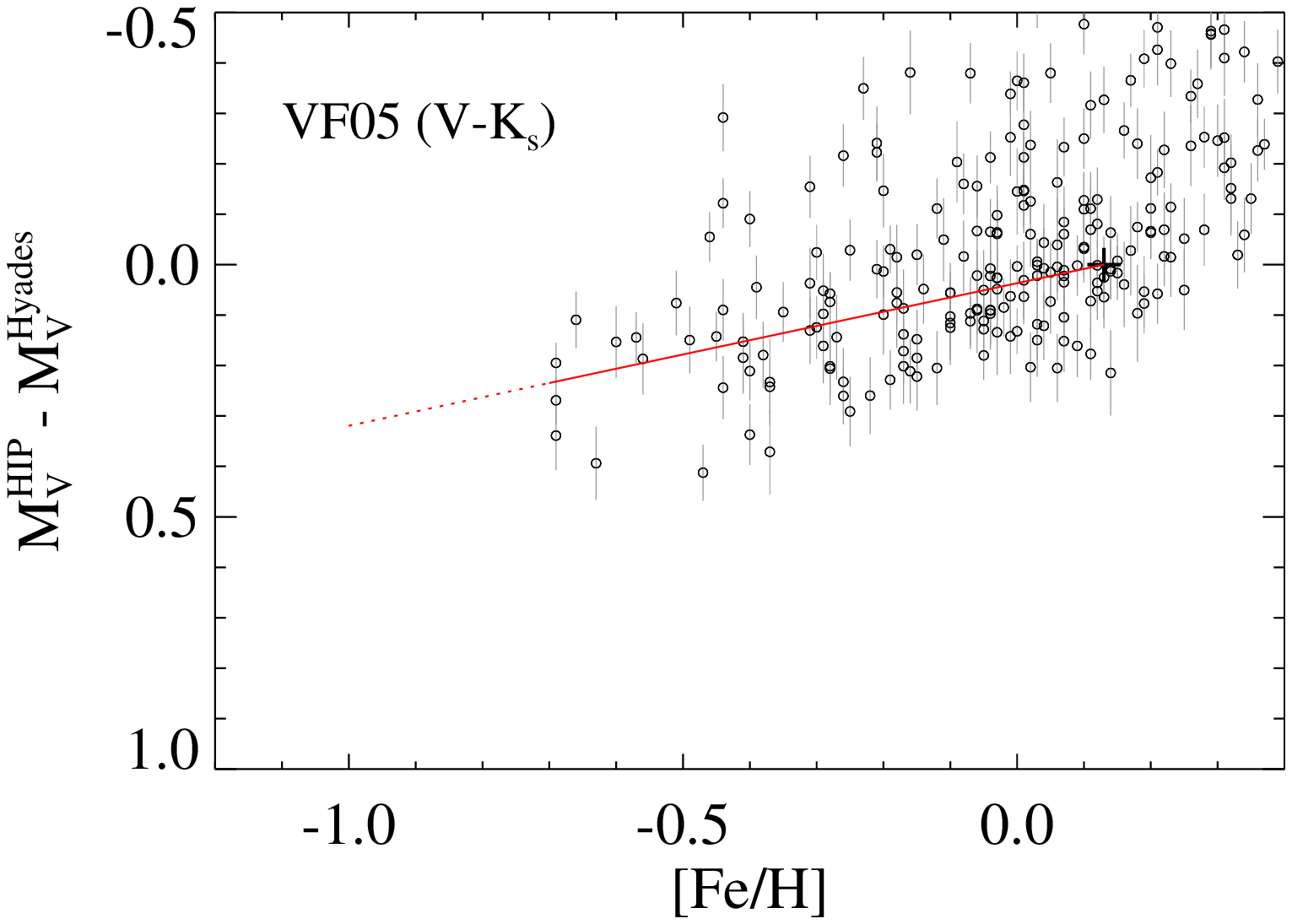}
\includegraphics[scale=0.65]{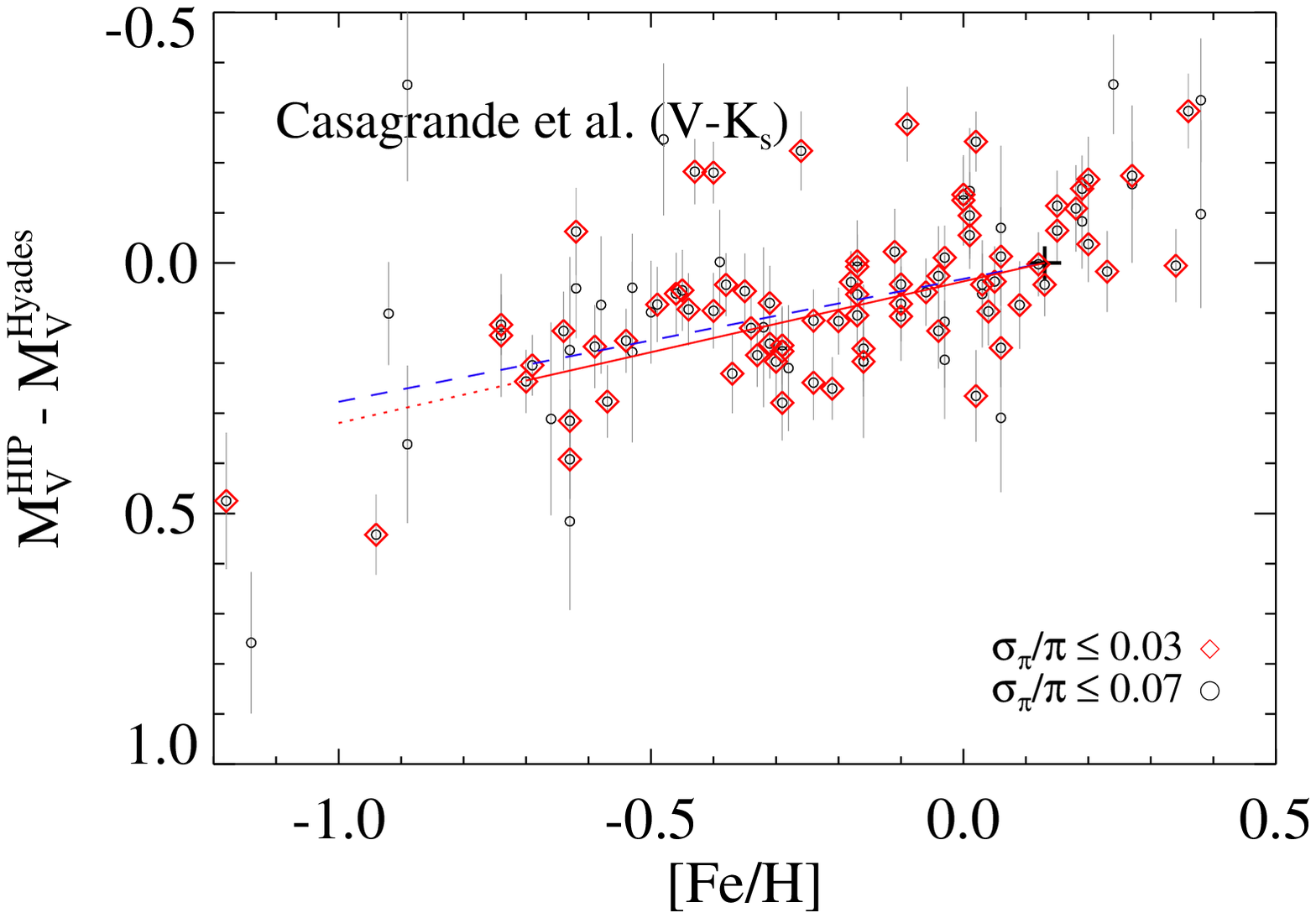}
\caption{Same as in Figure~\ref{fig:sens}, but in $\vk$ colors.\label{fig:sensvk}} \end{figure*}

As in $\bv$, we derived a mean empirical metallicity sensitivity in $\vk$ using the same sample stars in VF05 with good parallaxes ($\sigma_\pi/\pi \le 0.03$). The result is shown in the top panel of Figure~\ref{fig:sensvk}, where we used and displayed only those in $1.6 \le \vk \le 3.0$ to avoid a contamination by turn-off stars. Stars with $-1.0 \le {\rm [Fe/H]} \le -0.1$ and $\delta M_V \ge -0.5$ were used in the linear regression for the metallicity sensitivity function in $\vk$ (red solid line). The slope of this line is $-0.28$~mag~dex$^{-1}$, which is significantly shallower than that from $\bv$ (Figure~\ref{fig:sens}). The lower panel in Figure~\ref{fig:sensvk} additionally shows a metallicity sensitivity function in $\vk$ (blue line) as derived from an independent set of stars ($1.6 \le \vk \le 3.0$) in \citet{casagrande:10} with good parallaxes ($\sigma_\pi/\pi \le 0.03$), of which slope ($-0.25$~mag~dex$^{-1}$) is almost identical to the one from the top panel (red line).

\begin{figure*}
\centering
\epsscale{0.65}
\includegraphics[scale=0.65]{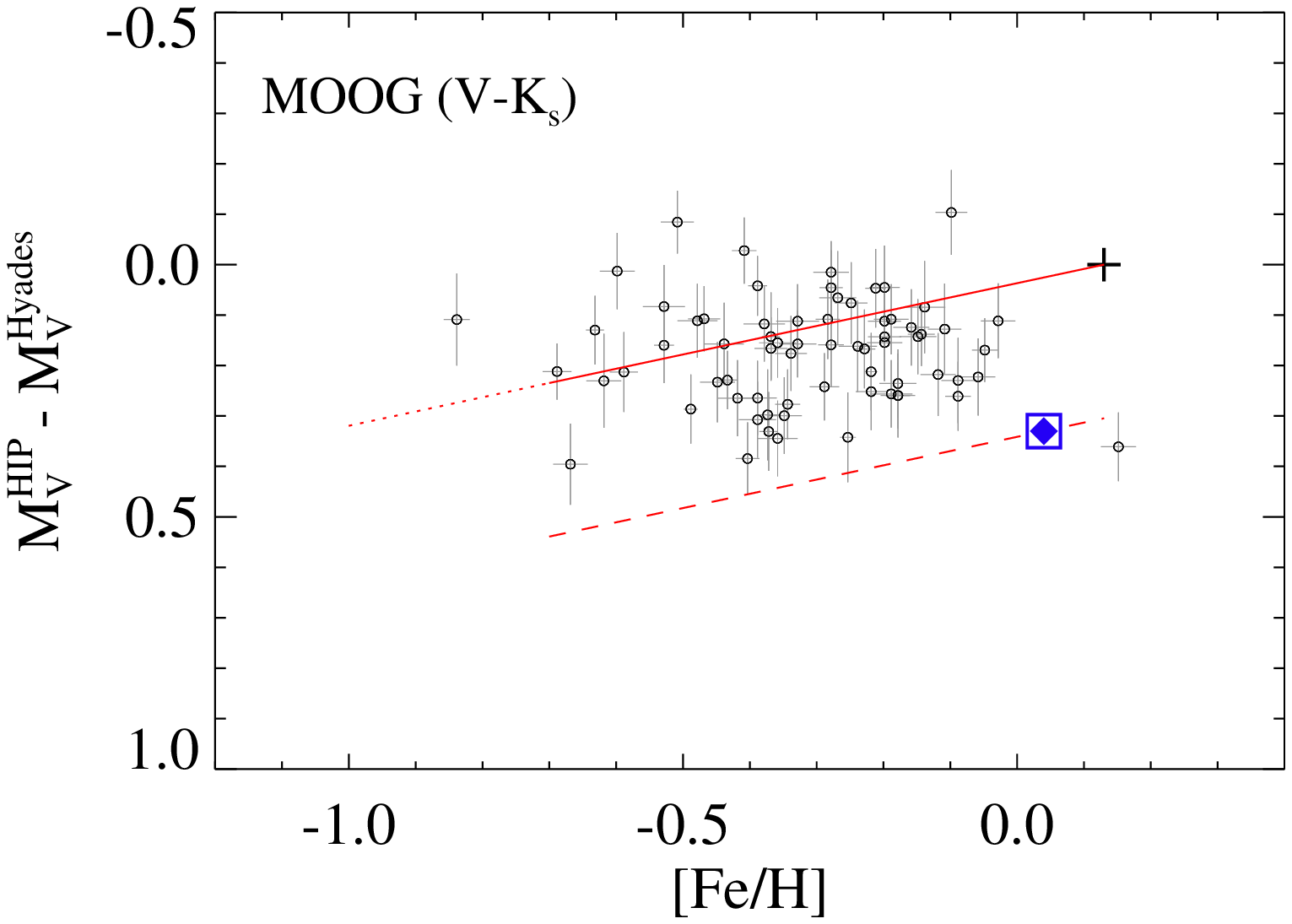}
\includegraphics[scale=0.65]{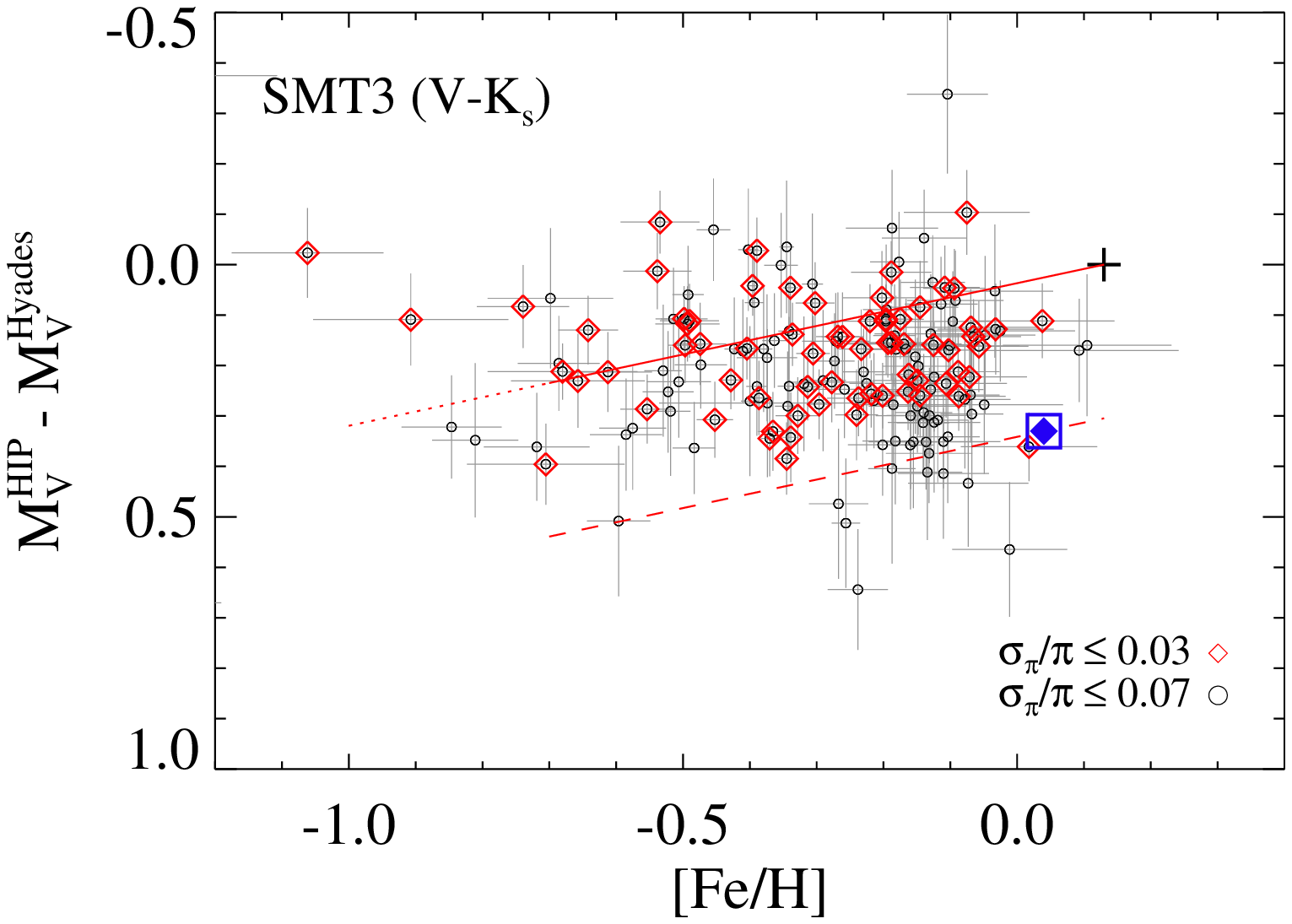}
\caption{Same as in Figure~\ref{fig:deficit}, but in $\vk$ colors.\label{fig:vkdeficit}} \end{figure*}

Our KPNO sample stars are shown in Figure~\ref{fig:vkdeficit} where $\delta M_V$ was derived from $\vk$ CMDs. As in Figure~\ref{fig:deficit}, MOOG metallicities are used in the top panel, and [Fe/H] from SMT3 are used in the bottom panel. Stars with highly accurate parallaxes ($\sigma_\pi / \pi \leq 0.03$) are shown in the top panel and indicated by red diamond points in the bottom panel. The red solid line is the mean metallicity sensitivity function from the top panel in Figure~\ref{fig:sensvk}. Most of our sample stars follow this trend, if one takes into account the fact that our sample selection was biased against stars with small (negative) $\delta M_V$. The $\delta M_V$ of the Pleiades, as determined from the Hyades' MS, is $\delta M_V=0.33$ (or $\Delta M_V=0.30$), and is shown as a blue diamond symbol.

In total, $9$ stars were identified as having larger $\Delta M_V$ than the Pleiades (red dashed line): HIP~5313, HIP~56092, HIP~71720, HIP~75446, HIP~77810, HIP~81831, HIP~97668, HIP~99965, and HIP~114385. However, only about half of them (HIP~5313, HIP~75446, HIP~81831, and HIP~97668) show as large $\Delta M_V$ as the Pleiades in $\bv$ (Table~\ref{tab:dmv}), suggesting that photometric errors would have made a significant contribution to an error in $\Delta M_V$. We also analyzed these stars using MOOG (Table~\ref{tab:moog}), since they were not originally included in our MOOG analysis (except HIP~99965) due to large parallax errors ($\sigma_\pi / \pi > 0.03$). Nevertheless, we found that the difference in metallicity between the two approaches is small. Its impact on $\Delta M_V$ is further reduced by the relatively weak metallicity dependence in $\vk$ ($-0.28$~mag~dex$^{-1}$), resulting in a negligible difference in $\Delta M_V$.

\begin{figure*}
\epsscale{0.7}
\plotone{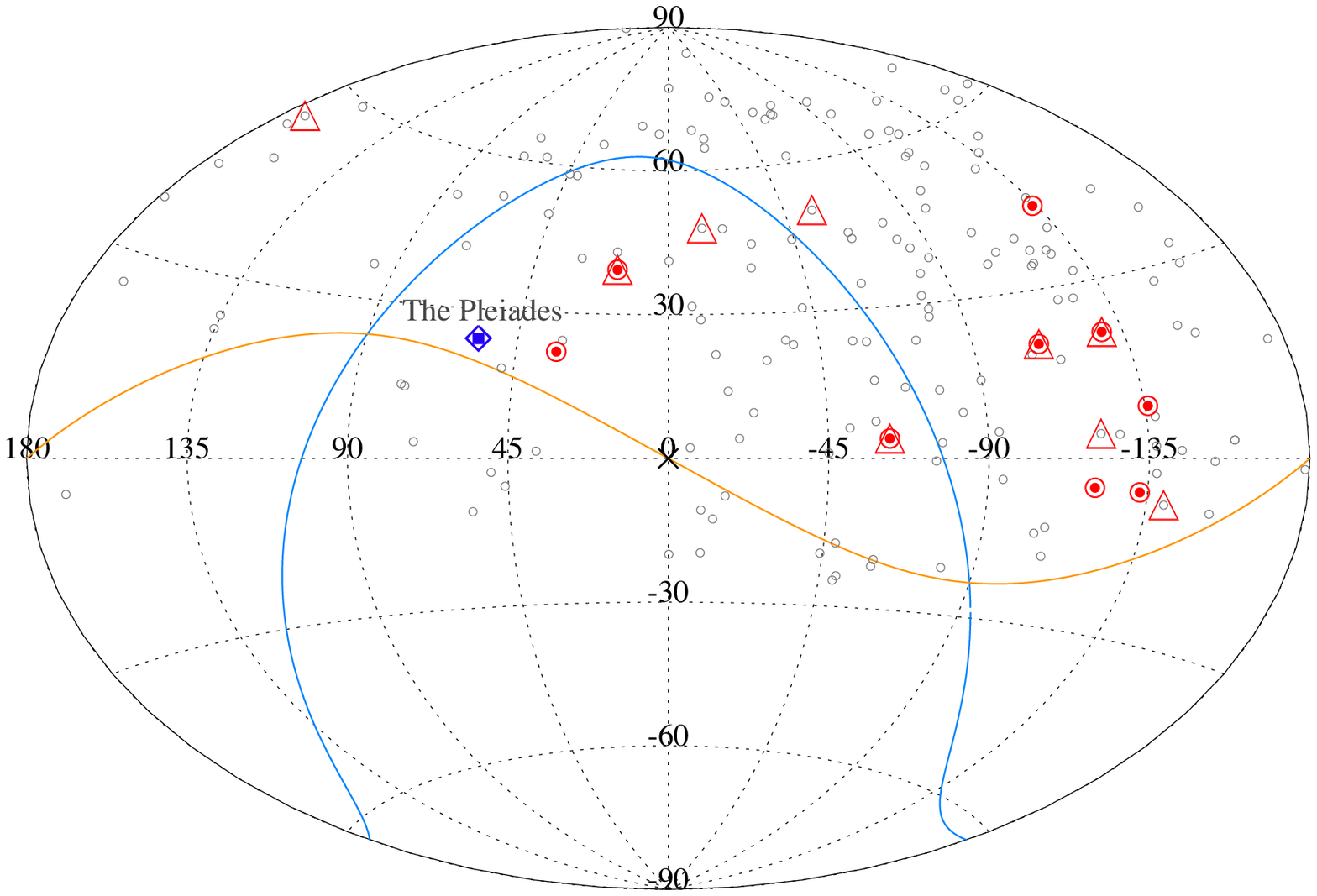}
\caption{Distribution of the KPNO sample stars in equatorial coordinates. Stars with a larger magnitude excess than those of the Pleiades (blue diamond) in $\bv$ and $\vk$ CMDs are shown in red bull's-eyes and triangles, respectively. The ecliptic and the Galactic plane are shown in an orange and a blue line, respectively.\label{fig:map}} \end{figure*}

Figure~\ref{fig:map} shows positions of the KPNO samples in equatorial coordinates, which is by design randomly distributed at $\delta > -30\arcdeg$ in our observing programs. The red bull's-eyes are the positions of the $9$ stars with the largest magnitude excess in $\bv$ CMDs, and red triangles are those from $\vk$. In either of the cases, these stars are not spatially correlated with each other, nor do they show a spatial correlation with the location of the Pleiades (blue diamond point). Also, they are not associated with any of the fundamental great circles, such as the ecliptic (orange line) or the Galactic plane (blue line).

Our result shows that there are a small fraction of stars with sufficiently large $\Delta M_V$, but their statistical significance is only marginal (a $2\sigma$ level). On the other hand, we failed to unambiguously identify stars with a large magnitude excess like the Pleiades, independently of a spectroscopic analysis technique and a color index employed, among those having good parallaxes ($\sigma_\pi/\pi < 0.03$). This may suggest that stars with the Pleiades-like phenomenon are rare, at least among nearby field stars with good parallax measurements. To confidently identify stars with (still unknown) systematic errors in parallax, it would be necessary to further shrink the size of errors in photometry and metallicity.

\subsection{Young Age and Stellar Activity}\label{sec:activity}

Young stars often exhibit chromospheric activities with large stellar spots, which are thought to be related with their large angular momenta. All together, one can naively expect that either of these effects could somehow modify stellar energy distributions, making their observed broadband colors or magnitudes deviate from those of older stars. Such color anomalies have already been observed from late K-type dwarfs in the Pleiades \citep{stauffer:03}, but our KPNO sample covers spectral types earlier than $\sim$K2, which helps to avoid issues on the potential modifications of colors and magnitudes by stellar activity and/or young age. Below we focus on young and/or active stars on the $\delta M_V$ versus [Fe/H] diagram and see if these stars have systematically shorter {\it Hipparcos} distances than older stars. This will directly test a hypothesis that the longer Pleiades distance from MS fitting is due to yet unknown physics of young/active stars in the cluster \citep{vanleeuwen:99}.

\subsubsection{Lithium Absorptions}\label{sec:lithium}

The color-magnitude selection of our sample is biased against the most metal-rich stars in the disk, and therefore probably does not favor a selection of young ($\la 200$~Myr) stars in the solar neighborhood. On the other hand, very young stars ($\la 30$~Myr) are also difficult to detect in our survey because low-mass pre-MS stars are brighter than MS at a given color.  We measured EWs of lithium absorption at $6707.70$~\AA\ for our KPNO stars, and found that about $25\%$ of the entire KPNO sample ($N=42$) show an EW of Li larger than $5$~m\AA. However, a majority of them show weak absorptions, suggesting that most of our sample stars including those with non-detections are relatively old. Among these, however, a small number of our KPNO spectra revealed a relatively strong Li absorption, implying young ages of these objects. The EWs of stars having $W {\rm (Li)}>5$~m\AA\ are listed in the last column of Table~\ref{tab:dmv}.

\begin{figure*}
\centering
\includegraphics[scale=0.7]{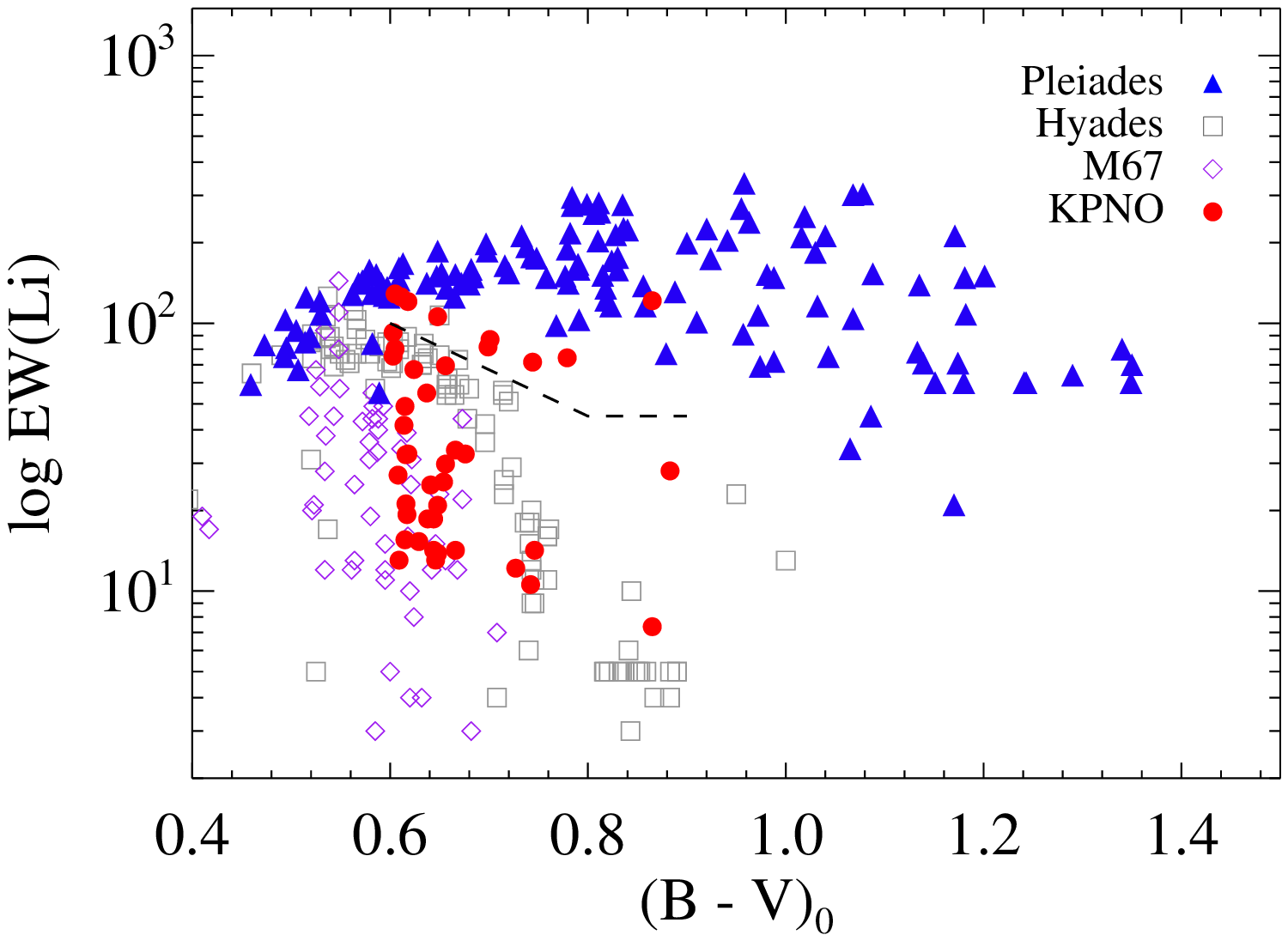}
\includegraphics[scale=0.7]{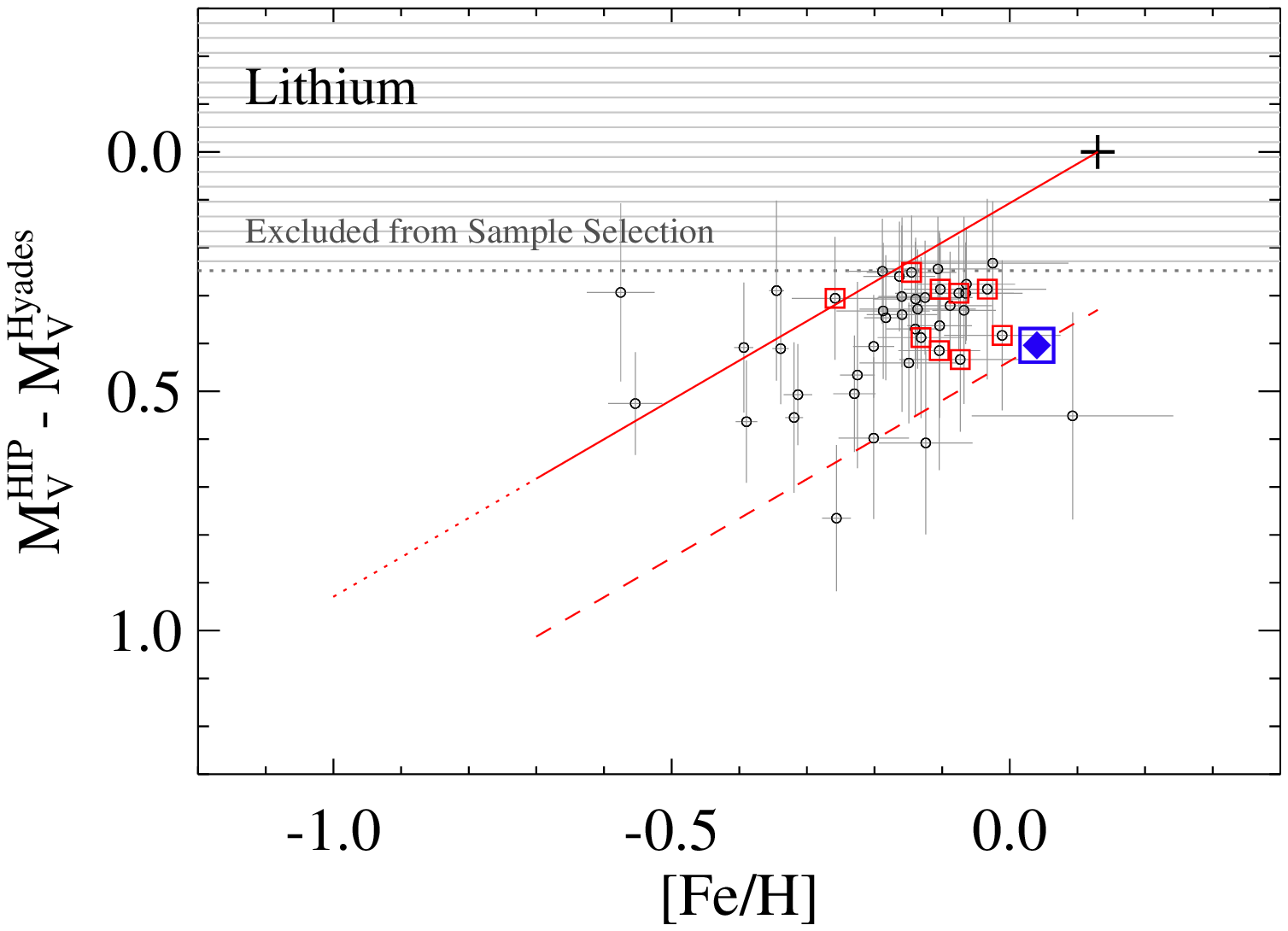}
\caption{{\it Top:} EWs of lithium $6707$ \AA\ for the KPNO sample (red filled circles). EWs from open clusters are shown for the Pleiades ($100$~Myr; blue filled triangle), the Hyades ($550$~Myr; grey open box), and M67 ($4$~Gyr; open diamond). Our selection of young stars is indicated by a dashed line. {\it Bottom:} Same as in the bottom panel of Figure~\ref{fig:deficit}, but displaying stars with lithium $6707$ \AA\ absorptions ($\ge 5$~m\AA). Stars with EWs larger than the dashed line in the top panel are shown in red squares.\label{fig:licomp}} \end{figure*}

While the Li line strength becomes weaker as a star gets older, an EW of lithium is also strongly dependent on stellar colors (or mass) by different depths of outer convective cells and by different amounts of angular momentum. This is shown in the top panel of Figure~\ref{fig:licomp}, which displays lithium EWs of MS stars in three open clusters as a function $\bv$ \citep[see][]{soderblom:10}: the Pleiades \citep[blue open circle,][]{butler:87,soderblom:93a,garcialopez:94,jones:96,jeffries:99}, the Hyades \citep[black open box,][]{soderblom:90,soderblom:95,thorburn:93}, and M67 \citep[open diamond,][]{hobbs:86,spite:87,garcialopez:88,pasquini:97,jones:99,randich:02} at the age of $100$~Myr, $550$~Myr, and $4$~Gyr, respectively. On top of these, our KPNO sample stars are marked by red closed circles. Only couples of stars in our sample fall into the range of Li EWs covered by the Pleiades members, which is not surprising given that our sample selection on a color-magnitude diagram was not designed to find youngest stars in the solar neighborhood. Due to a relatively small number of such stars, we selected young stars as those having larger Li EWs than those found in the Hyades as shown by a dashed line. This selection includes $9$ stars (HIP~21276, HIP~60074, HIP~63322, HIP~72703, HIP~76674, HIP~77810, HIP~82388, HIP~108774, and HIP~114385). The one having an exceptionally large Li EW [$W {\rm (Li)}\approx120$~m\AA] at $\bv=0.87$ is HIP~63322.

The bottom panel in Figure~\ref{fig:licomp} shows the same $\delta M_V$ versus [Fe/H] diagram of stars as in the bottom panel of Figure~\ref{fig:deficit} with SMT3 metallicities, but displaying stars with a lithium EW larger than $5$~m\AA. The red boxed points are young stars selected above as having largest Li EWs. As seen in this panel, a distribution of stars with lithium absorptions is not dissimilar to those seen from older stars (Figure~\ref{fig:deficit}). Among these, three stars (HIP~5313, HIP~74126, and HIP~78336) exhibit a larger magnitude excess than the Pleiades ($\Delta M \ge 0.33$), but they do not show a sufficiently strong Li absorption. More importantly, none of the above selected young stars exhibit larger $\Delta M_V$ than the Pleiades. Even if the sample is further restricted to five stars with $W {\rm (Li)}>100$~m\AA, having almost identical properties to those in the Pleiades, their $\Delta M_V$ values show no systematic offset with respect to the empirical metallicity sensitivity function (red solid line). On the other hand, if the Pleiades' distance is short because of the young age of its stars, we would expect that stars having similar Li strength with the Pleiades members should have systematically large $\Delta M_V$. Therefore, our result in Figure~\ref{fig:licomp} suggests that the short Pleiades' distance cannot be condemned to somehow modified color-magnitude relations for young stars in the cluster, and that MS-fitting distances for young stars, with spectral types earlier than $\sim$K2, are sufficiently close to those derived using color-magnitude relations for older stars.

\begin{figure*}
\centering
\epsscale{0.82}
\plotone{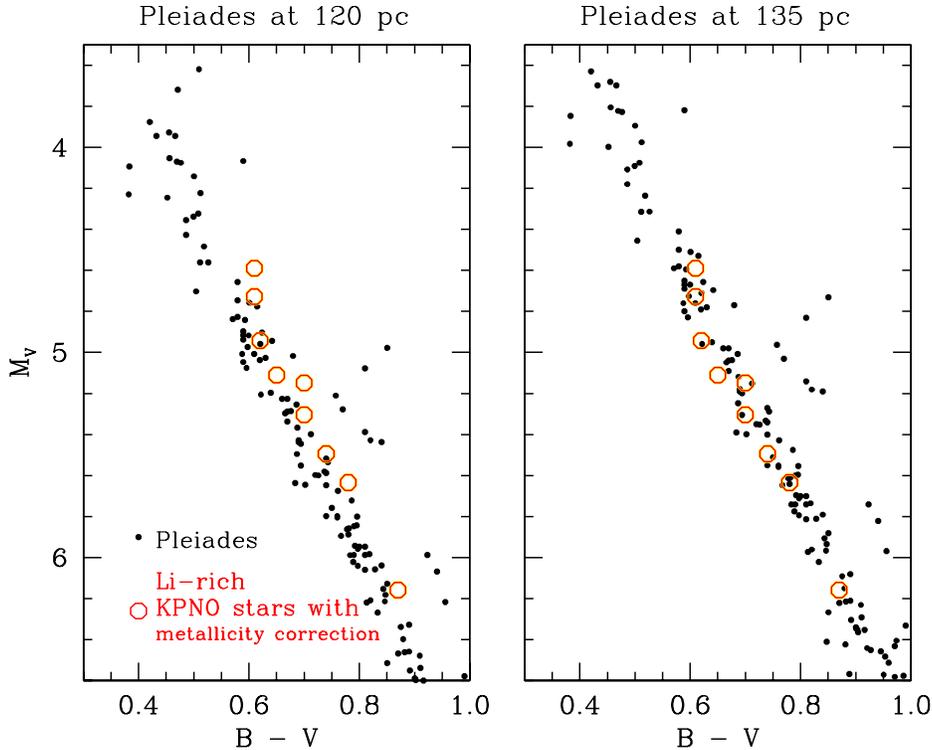}
\caption{CMDs of the Pleiades (black points) assuming the {\it Hipparcos} parallax measurement in \citet{vanleeuwen:09} (left panel) and the mean distance of the cluster from a number of geometric distance measurements except that from {\it Hipparcos} (right panel). The $\ebv=0.032$ is assumed for hypothetical zero-color stars in the cluster with color-dependent reddening laws. The red open circles are Li-rich KPNO stars (those above the dashed line in Figure~\ref{fig:licomp}) with $V$-band magnitudes corrected for a metallicity difference from the Pleiades (see text).\label{fig:compcmd}} \end{figure*}

Figure~\ref{fig:compcmd} shows CMDs of the Pleiades \citep{stauffer:07,kamai:14} assuming the short (left panel) and the long (right panel) distance scale of the cluster, respectively. The most recent {\it Hipparcos} distance in \citet{vanleeuwen:09} is used in the left panel. For the long distance scale, we used a weighted average distance $\dmn=5.647\pm0.013$ ($134.7$~pc) from a number of geometric measurements (trigonometric parallaxes and binary solutions) listed in \citet{an:07b} and the VLBI measurement \citep{melis:14}. We assumed $\ebv=0.032$ for the Pleiades \citep{an:07b} with color-dependent reddening and extinction laws \citep{an:07a}. The red open circles are our KPNO stars selected as having strong Li absorptions. For an apple-to-apple comparison with the Pleiades stars, we corrected their $V$-band magnitudes for a metallicity difference from the Pleiades ([Fe/H]$=+0.04$) using the empirical metallicity sensitivity function (i.e., red line in Figure~\ref{fig:sens}). As these stars have lower metallicities than the Pleiades, they become brighter with these corrections.

The comparison with the Pleiades' CMDs in Figure~\ref{fig:compcmd} shows that Li-rich young stars are consistently brighter than the MS of the Pleiades by $\Delta M_V\approx0.25$ when the {\it Hipparcos} distance is assumed (left-hand panel).\footnote{This does not contradict with our sample selection (\S~\ref{sec:sample}), in which we selected stars that are near or below the MS of the Pleiades assuming the \citet{vanleeuwen:09} distance, because of the metallicity corrections as described above.} On the other hand, the long distance scale of the cluster leads to an excellent match of young field stars with the observed MS of the Pleiades. An absolute metallicity scale for our sample could be in error. However, $\Delta M_V=0.08$~mag is expected if we have consistently overestimated our metallicities (or equivalently underestimated the Pleiades' metallicity) by $\Delta {\rm [Fe/H]}=0.1$, which is far smaller than what is required to explain the $\Delta M_V\approx0.25$ difference with the Pleiades' MS in the left panel.

In fact, most of these Li-rich stars are fainter than the mean MS relation at a given metallicity (with respect to the mean empirical metallicity sensitivity function) because of the sample selection bias as delineated by a shaded area in the bottom panel of Figure~\ref{fig:licomp}. Therefore, the red circles in Figure~\ref{fig:compcmd} represent approximately half of a Li-rich population in a given metallicity range, and constitute only a lower half of the brightness distribution. Nonetheless, missing Li-rich stars would be found above the red circles in the $\bv$ CMD (Figure~\ref{fig:compcmd}), and would make the agreement with the Pleiades' MS even worse if the {\it Hipparcos} parallax is assumed for the cluster's distance. Figure~\ref{fig:compcmd} simply restates our conclusion that the Pleiades distance in \citet{vanleeuwen:09} is too short and cannot be explained with the young age of cluster members.

\subsubsection{Stellar Activity Indices}\label{sec:rhk}

\begin{figure*}
\centering
\includegraphics[scale=0.72]{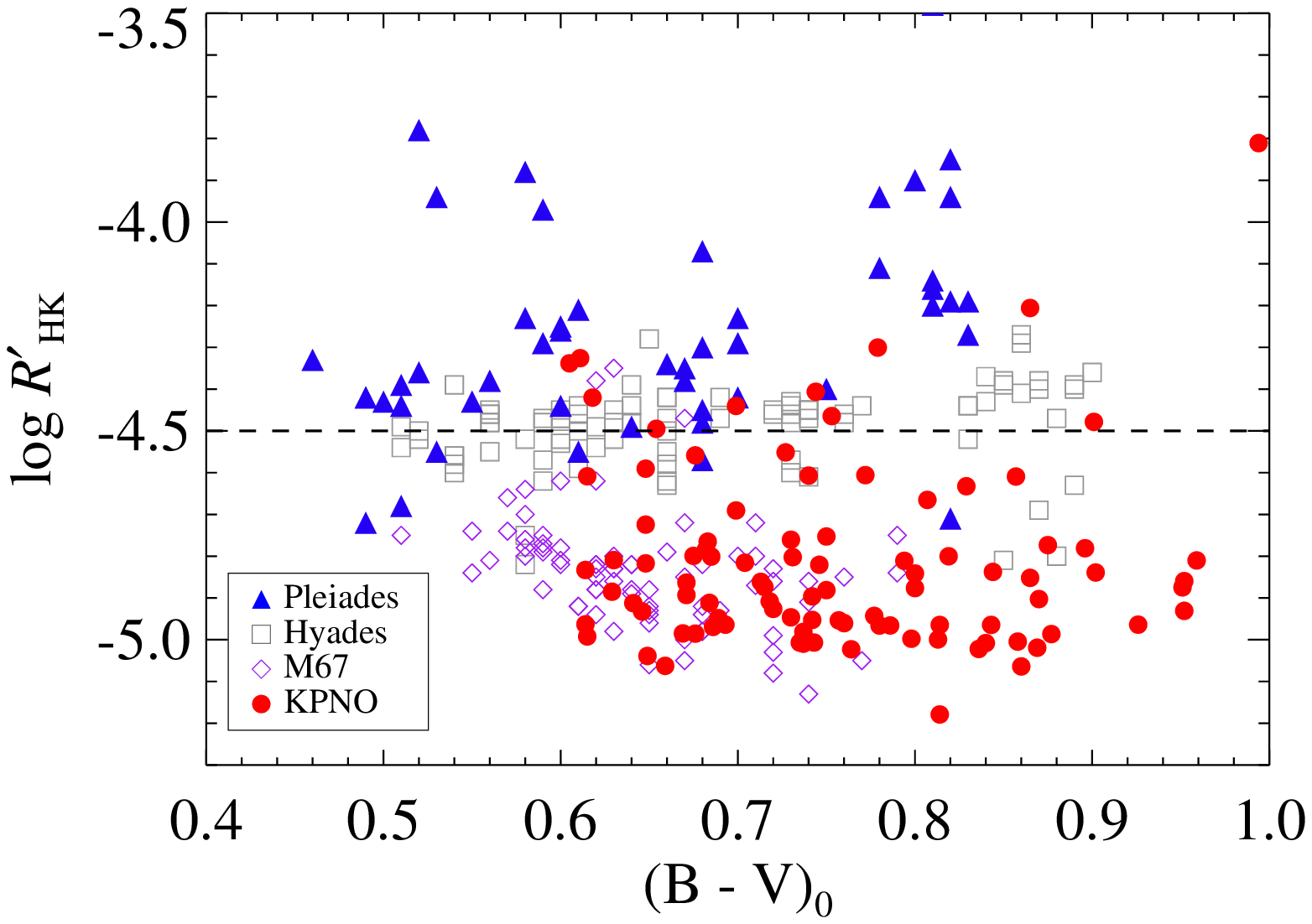}
\includegraphics[scale=0.72]{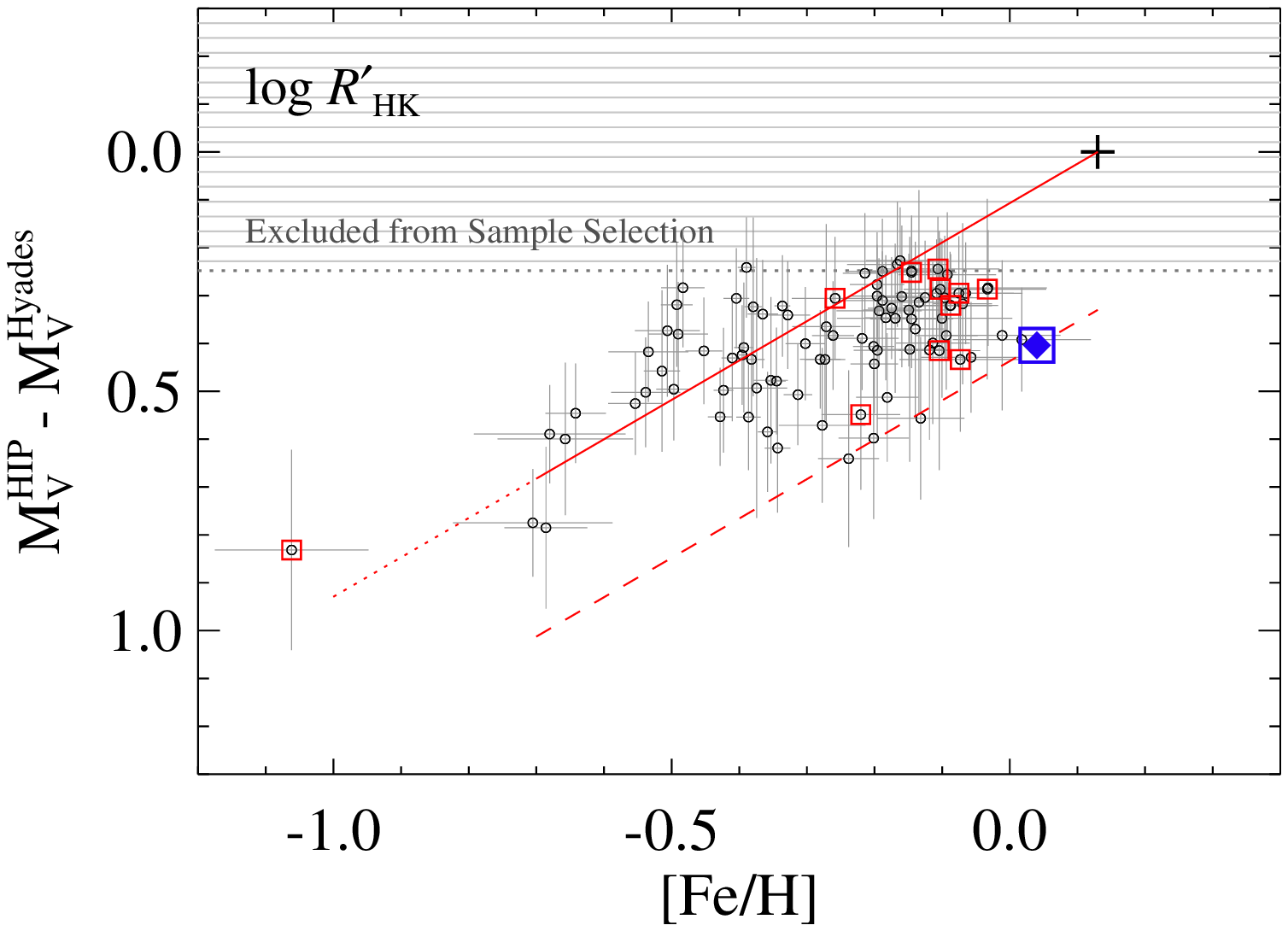}
\caption{{\it Top:} $R'_{\rm HK}$ values for the KPNO sample in comparison with cluster measurements: the Pleiades ($100$~Myr; blue filled triangle), the Hyades ($550$~Myr; grey open box), and M67 ($4$~Gyr; open diamond). The black dashed line represents our division of active/inactive stars based on $R'_{\rm HK}$. {\it Bottom:} Same as in the bottom panel of Figure~\ref{fig:deficit}, but displaying stars with $R'_{\rm HK}$ values available in the literature. The red boxed points are stars with $\log{R'_{\rm HK}}>-4.5$.\label{fig:activity}} \end{figure*}

One of the most frequently used activity indicators is $R'_{\rm HK}$ index, which measures the chromospheric emission line strength of \ion{Ca}{2} H and K at $3933.7$~\AA\ and $3968.5$~\AA\ in the central part of its broad absorption profile, normalized by photospheric continuum emissions \citep[see][]{noyes:84}. The $R'_{\rm HK}$ is a function of $\teff$ or colors, and is known to decrease with stellar ages \citep[see][]{soderblom:10}. This is shown in the top panel of Figure~\ref{fig:activity}, where $R'_{\rm HK}$ measurements from individual stars in three fiducial open clusters \citep[Pleiades, the Hyades, and M67;][]{mamajek:08} are displayed \citep[see also][]{soderblom:10}. The $R'_{\rm HK}$ measurements from the Pleiades show a large scatter, indicating a star-to-star variation in chromospheric emissions, and a separation from those in the Hyades is not clearcut. Nevertheless, the cluster observations clearly suggest a systematic change of $R'_{\rm HK}$ with age.

In the top panel of Figure~\ref{fig:activity}, our KPNO sample stars are indicated by red closed circles on top of the cluster observations. We took a large compilation of S-index measurements in the literature \citep[][see references therein]{pace:13}. In total, $93$ stars in our sample have valid S-index measurements. Following \citet{pace:13}, we took the average of a minimum and a maximum S-index values for each star, whenever there are repeat measurements, as a proxy for a time-averaged chromospheric activity level. We adopted a procedure in \citet{noyes:84} to convert S index into $R'_{\rm HK}$.

Evidently, most stars in Figure~\ref{fig:activity} have similar $R'_{\rm HK}$ values with those in M67, suggesting old ages of these stars. On the other hand, there are approximately a dozen stars with $R'_{\rm HK}$ values that are comparable to those in the Pleiades. We took $\log{R'_{\rm HK}}=-4.5$ (dashed line) to select 11 likely young stars in our sample (HIP~4907, HIP~21276, HIP~60074, HIP~63322, HIP~63636, HIP~76674, HIP~77810, HIP~82388, HIP~106231, HIP~108774, and HIP~111888). Most of the Li-rich stars were also selected as having a strong chromospheric activity. Remaining Li-rich stars (HIP~72703 and HIP~114385) either do not have a $R'_{\rm HK}$ measurement in the literature or have a value near $\log{R'_{\rm HK}}=-4.5$. Nevertheless, our selection of active stars from $R'_{\rm HK}$ includes many Hyades dwarfs, and the separation of young stars from older populations is not as clear as in Figure~\ref{fig:licomp} based on a lithium absorption.

The bottom panel of Figure~\ref{fig:activity} displays stars with $R'_{\rm HK}$ measurements on the $\delta M_V$ versus [Fe/H] diagram with SMT3 metallicities. As expected, most of these stars are old, and are found along the mean metallicity sensitivity function (red solid line). Meanwhile, the boxed points are stars with $\log{R'_{\rm HK}}>-4.5$, and none of these chromospherically active stars show a larger magnitude excess than the Pleiades ($\Delta M_V > 0.33$). This result reiterates our conclusion above based on Li-rich stars that a young age of stars does not significantly modify color-magnitude relations for MS dwarfs.

In addition to $R'_{\rm HK}$ index, we collected X-ray luminosities of our sample stars from the NASA Exoplanet archive \citep{ramirez:13}, and looked into the properties of stars selected based on X-ray luminosity. The X-ray luminosities show tight correlations with chromospheric activities such as $R'_{\rm HK}$ \citep{mamajek:08}, and can be used to trace young populations in the disk. However, measurements of X-ray luminosity are available only for $10$ stars in our sample \citep{huensch:98,huensch:99,schmitt:04}. Among these, only two star (HIP~21276, HIP~106231) exhibit a higher X-ray luminosity than those in the Pleiades \citep[$L_{\rm X} = 29.00$~ergs~s$^{-1}$;][]{daniel:02}. However, these stars have a significantly smaller magnitude excess than the Pleiades members. Although the number of stars with X-ray luminosities in the literature is small, it is clear that active stars in our sample have the same photometric properties as those for older stars. We conclude that stellar activity or young age of stars have a little impact (if any) on a large magnitude excess.

\subsection{Empirical MS-fitting Distance to the Pleiades}\label{sec:plfit}

Given the lack of evidence on anomalous color-magnitude relations for young/active stars with $0.6 \leq \bv \leq 1.0$, a purely empirical MS-fitting distance to the Pleiades can be obtained using the observed MS of the Hyades and the empirical metallicity sensitivity function as derived from the {\it Hipparcos} dwarfs. We assumed $\ebv=0.032$ for the Pleiades (see above) and obtained $\dmn=5.657\pm0.017$ and $5.669\pm0.024$ (statistical) in the $\bv$ and $\vk$ CMDs, respectively, using photometry of single MS stars of the cluster \citep{an:07b} in $0.6 \leq \bv \leq 1.0$ and the corresponding range in $\vk$. We corrected distance moduli for a metallicity difference between the Pleiades and the Hyades, based on the empirical metallicity sensitivity functions in each of the color indices (red lines in Figure~\ref{fig:sens} and \ref{fig:sensvk}), which results in $\dmn = 5.585\pm0.023$ and $5.644\pm0.025$ in the $\bv$ and $\vk$ CMDs, respectively, where the errors are a quadrature sum of errors in fitting and metallicity ($\sigma_{\rm [Fe/H]}=0.02$). The average distance modulus of the Pleiades from the two CMDs becomes $\dmn=5.615\pm0.030$; the error represents half of the difference in distance modulus. This purely empirical distance modulus is in perfect agreement with the mean geometric distance modulus of the cluster [$\dmn=5.647\pm0.013$], but is significantly longer than the {\it Hipparcos}-based distance $\dmn=5.40\pm0.03$ \citep{vanleeuwen:09}.

\section{Summary}\label{sec:summary}

The debate on the Pleiades distance has continued even after the new reduction of the {\it Hipparcos} parallaxes \citep{vanleeuwen:09}, which predicts $\sim0.3$~mag fainter magnitudes of the cluster stars than those expected from MS fitting or other independent distance determinations. In this study, we tested a hypothesis that the long photometric distance of the Pleiades is due to anomalous colors or magnitudes of its cluster members, by searching for hypothesized sub-luminous field stars in the {\it Hipparcos} catalog. For comparison with the {\it Hipparcos} parallax, we derived accurate metallicities of $170$ nearby G- and K-type field dwarfs based on high S/N ratios, high-resolution spectra, employing two independent spectral analysis techniques (MOOG and SMT). Our photometric distances based on these metallicities are purely empirical, being independent of any theoretical stellar isochrones, and relies on the observed MS of the Hyades and the metallicity dependence of colors, which was derived from the {\it Hipparcos} parallaxes and our metallicity measurements for a large number of field stars.

Among stars with highly accurate parallaxes ($\sigma_\pi/\pi\leq0.03$), we could identify only one star from a $\bv$ CMD with a larger magnitude excess ($\Delta M_V$) than the Pleiades, or a shorter {\it Hipparcos}-based distance by $\Delta \dmn = \Delta M_V \ge 0.33$, from the MOOG analysis. However, none of these stars in our sample have a larger magnitude excess than the Pleiades when the SMT metallicities are employed. For an extended sample with $\sigma_\pi/\pi\leq0.07$, we identified $9$ stars with $\Delta M_V \ge 0.33$, but the differences in distance modulus of these stars are only marginal. Furthermore, only six out of the $9$ stars remain to show a large magnitude excess with an independent analysis technique from MOOG. In addition to $\bv$, we repeated the above exercise with $\vk$ colors, and found that only $3$ stars identified as having $\Delta M_V \ge 0.33$ from $\bv$ have larger magnitude excesses than the Pleiades in $\vk$. Therefore, their oddity may not be surprising at all, and can be understood from errors in photometry, metallicity, and parallax.

Although we could not identify stars with large magnitude excesses at a statistically significant level, we were able to reject a hypothesis that these outlying stars are mostly young or active stars. We selected young/active stars based on the Li $6707$~\AA\ absorption, $R'_{\rm HK}$, or X-ray luminosity, and found that photometric distances of these young/active stars are not greatly different from the {\it Hipparcos} parallaxes. Although only a few stars in our sample may be as young as those in the Pleiades, none of these stars show larger differences in distance from {\it Hipparcos} than the Pleiades. While more young star samples can be used to better quantify the difference in distance, our result suggests that the short Pleiades distance is not at least directly related to the young age of the cluster.

The successful launch of Gaia \citep{perryman:01} has opened a new era in studies of stars in the Milky Way Galaxy, which will deliver precise astrometric data for about one billion stars with more than two orders of magnitude improvement in parallax measurements than available in the past. It is hoped that Gaia will eventually help to resolve the Pleiades distance problem and will cast new light on hidden systematic errors in the {\it Hipparcos} parallax measurements. However, the Pleiades distance controversy has revealed a practical limit in the analysis of space-based astrometric data, which leads to demand for a careful check on the future parallax measurements. Our analysis technique based on accurate spectroscopic data can be utilized to assess the accuracy of parallax measurements by Gaia.

\acknowledgements

We thank an anonymous referee for various suggestions, which helped to improve the readability of the manuscript. We thank Courtney Epstein for her assistance in the observations. B.K.\ and D.A.\ acknowledge support provided by the National Research Foundation of Korea to the Center for Galaxy Evolution Research (No.\ 2010-0027910) and by Basic Science Research Program through the National Research Foundation of Korea (NRF) funded by the Ministry of Education (2010-0025122, 2015R1D1A1A09058700). This work has developed from a master's thesis conducted by B.K.\ under the supervision of D.A.\ at Ewha Womans University. J.R.S.\ gratefully acknowledges funding support from NASA Kepler grant NNX1AV62G. Y.S.L.\ acknowledges support provided by the National Research Foundation of Korea to the Center for Galaxy Evolution Research (No.\ 2010-0027910) and the Basic Science Research Program through the National Research Foundation of Korea (NRF) funded by the Ministry of Science, ICT \& Future Planning (NRF-2015R1C1A1A02036658). D.M.T.\ acknowledges support from award AST-1411685 from the National Science Foundation to The Ohio State University. This research has made use of the NASA Star and Exoplanet Database (NStED), which was operated until 2011 by the California Institute of Technology, under contract with the National Aeronautics and Space Administration. This research has made use of the SIMBAD database, operated at CDS, Strasbourg, France.

{}

\clearpage
\LongTables
\input{tab1.tex}
\input{tab2.tex}
\clearpage
\input{tab3.tex}
\input{tab4.tex}
\clearpage
\input{tab5.tex}
\input{tab6.tex}
\clearpage
\input{tab7.tex}

\end{document}

%% file: tab1.tex
\begin{deluxetable*}{llcccccccccccc}
\tablewidth{0pt}
\tabletypesize{\scriptsize}
\tablecaption{Photometric Properties of the KPNO Sample\label{tab:phot}}
\tablehead{
  \colhead{HIP} &
  \colhead{HD} &
  \colhead{$V$} &
  \colhead{$\sigma(V)$} &
  \colhead{$\bv$} &
  \colhead{$\sigma_{\bv}$} &
  \colhead{$\vk$} &
  \colhead{$\sigma_{\vk}$} &
  \colhead{$\pi$\tablenotemark{a}} &
  \colhead{$\sigma_{\pi}$\tablenotemark{a}} &
  \colhead{$M_{V}$\tablenotemark{b}} &
  \colhead{$\sigma(M_{V})$} &
  \multicolumn{2}{c}{Number of Obs.} \\
  \cline{13-14}
  \colhead{} &
  \colhead{} &
  \colhead{(mag)} &
  \colhead{(mag)} &
  \colhead{(mag)} &
  \colhead{(mag)} &
  \colhead{(mag)} &
  \colhead{(mag)} &
  \colhead{(arcsec)} &
  \colhead{(arcsec)} &
  \colhead{(mag)} &
  \colhead{(mag)} &
  \colhead{Spring} &
  \colhead{Autumn} 
}
\startdata
{\rm 1674} & {\rm 1624}& $8.48$ & $0.02$ & $0.80$ & $0.02$ & $2.62$ & $0.03$ & $31.54$ & $0.96$ & $5.97$ & $0.11$ & \nodata & $2$ \\
{\rm 4907} & {\rm 5996}& $7.66$ & $0.02$ & $0.75$ & $0.02$ & $1.76$ & $0.03$ & $39.20$ & $0.56$ & $5.63$ & $0.10$ & \nodata & $1$ \\
{\rm 5313} & {\rm 6664}& $7.78$ & $0.02$ & $0.61$ & $0.02$ & $1.53$ & $0.03$ & $31.79$ & $1.66$ & $5.29$ & $0.13$ & \nodata & $2$ \\
{\rm 5521} & {\rm 6963}& $7.66$ & $0.02$ & $0.73$ & $0.02$ & $1.72$ & $0.03$ & $36.95$ & $0.79$ & $5.50$ & $0.10$ & \nodata & $1$ \\
{\rm 9172} & {\rm 11926} & $7.57$ & $0.02$ & $0.66$ & $0.02$ & \nodata & \nodata & $31.99$ & $0.86$ & $5.09$ & $0.06$ & \nodata & $1$ \\
{\rm 9829} & {\rm 12846} & $6.89$ & $0.02$ & $0.65$ & $0.02$ & $1.59$ & $0.03$ & $43.89$ & $0.57$ & $5.10$ & $0.04$ & \nodata & $2$ \\
{\rm 10276} & {\rm 13483} & $8.46$ & $0.02$ & $0.77$ & $0.03$ & $1.87$ & $0.03$ & $31.58$ & $1.11$ & $5.96$ & $0.11$ & \nodata & $2$ \\
{\rm 10629} & {\rm 13783} & $8.29$ & $0.02$ & $0.69$ & $0.02$ & $1.71$ & $0.03$ & $25.37$ & $0.93$ & $5.31$ & $0.08$ & \nodata & $2$ \\
\enddata 
\tablenotetext{a}{Parallaxes and parallax errors from the revised $\it Hipparcos$ catalog \citep{vanleeuwen:07a,vanleeuwen:07b}.}
\tablenotetext{b}{Absolute magnitude computed from the $\it Hipparcos$ parallax.}
\tablenotetext{c}{A spectroscopic binary listed in the 9$^{th}$ catalogue of spectroscopic binary orbits \citep[][{\tt http://sb9.astro.ulb.ac.be}]{pourbaix:04}.}
\end{deluxetable*}

%% file: tab2.tex
\begin{deluxetable}{lccccccc}
\tablewidth{0pt}
\tabletypesize{\scriptsize}
\tablecaption{Stellar Parameters derived using MOOG\label{tab:moog}}
\tablehead{
  \colhead{HIP} &
  \colhead{$\teff$} &
  \colhead{$\log {g}$} &
  \colhead{${\rm [Fe/H]}$} &
  \colhead{$\xi$} &
  \colhead{${\rm [Fe/H]_{corr}}$\tablenotemark{a}} &
  \colhead{$\sigma({\rm [Fe/H]})$} \\
  \colhead{} &
  \colhead{(K)} &
  \colhead{} &
  \colhead{} &
  \colhead{(${\rm km\ s^{-1}}$)} &
  \colhead{} &
  \colhead{}
}
\startdata
{\rm 4907}   & $5405$ & $4.34$ & $-0.16$ & $1.2$ & $-0.09$ & $0.019$ \\
{\rm 5313}\tablenotemark{b}   & $5881$ & $4.05$ & $-0.29$ & $1.3$ & $-0.22$ & $0.017$ \\
{\rm 5521}   & $5398$ & $4.00$ & $-0.29$ & $1.1$ & $-0.22$ & $0.009$ \\
{\rm 9172}   & $5638$ & $4.04$ & $-0.13$ & $1.4$ & $-0.06$ & $0.009$ \\
{\rm 9829}   & $5795$ & $4.29$ & $-0.22$ & $1.1$ & $-0.14$ & $0.020$ \\
{\rm 10276}\tablenotemark{b}  & $5368$ & $4.42$ & $-0.25$ & $1.2$ & $-0.18$ & $0.016$ \\
{\rm 15442}  & $5760$ & $4.18$ & $-0.27$ & $1.3$ & $-0.20$ & $0.026$ \\
{\rm 17265}  & $5339$ & $4.23$ & $-0.69$ & $1.3$ & $-0.62$ & $0.026$ \\
\enddata
\tablenotetext{a}{Corrected to match the metallicity scale of VF05 (see text).}
\tablenotetext{b}{Stars with less accurate parallax measurements ($\sigma_\pi/\pi>3\%$), but included as having larger magnitude excesses ($\Delta M_V$) than the Pleiades.}
\end{deluxetable}

%% file: tab3.tex
\begin{deluxetable}{ccccc}
\tablewidth{0pt}
\tabletypesize{\scriptsize}
\tablecaption{Atomic Line List\label{tab:line}}
\tablehead{
  \colhead{Element} &
  \colhead{Ion} &
  \colhead{Wavelength} &
  \colhead{E.P.} &
  \colhead{$\log{gf}$} \\
  \colhead{} &
  \colhead{} &
  \colhead{(\AA)} &
  \colhead{(eV)} &
  \colhead{}
}
\startdata
{\rm Fe}  &  $1$  &  $5775.08$  &  $4.22$  &  $-1.30$  \\
{\rm Fe}  &  $1$  &  $5809.22$  &  $3.88$  &  $-1.74$  \\
{\rm Fe}  &  $1$  &  $5849.69$  &  $3.69$  &  $-2.89$  \\
{\rm Fe}  &  $1$  &  $5852.22$  &  $4.55$  &  $-1.23$  \\
{\rm Fe}  &  $1$  &  $5855.08$  &  $4.61$  &  $-1.48$  \\
{\rm Fe}  &  $1$  &  $5856.09$  &  $4.29$  &  $-1.33$  \\
{\rm Fe}  &  $1$  &  $5858.78$  &  $4.22$  &  $-2.16$  \\
{\rm Fe}  &  $1$  &  $5859.59$  &  $4.55$  &  $-0.30$  \\
\enddata
\end{deluxetable}

%% file: tab4.tex
\begin{deluxetable}{lcccc}
\tablewidth{0pt}
\tablecaption{Systematic Errors in Abundance Measurements from MOOG Analysis\label{tab:sys}}
\tablehead{
  \colhead{Ion} &
  \colhead{$\Delta \teff$} &
  \colhead{$\Delta \log{g}$} &
  \colhead{$\Delta$ [Fe/H]} &
  \colhead{$\Delta \xi$} \\
  \colhead{} &
  \colhead{$\pm100$~K} &
  \colhead{$\pm0.3$~dex} &
  \colhead{$\pm0.1$~dex} &
  \colhead{$\pm0.2$~km~s$^{-1}$}
}
\startdata
\ion{Fe}{1} & $\pm0.06$ & $\mp0.01$ & $\pm0.01$ & $\mp0.04$ \\
\ion{Fe}{2} & $\mp0.06$ & $\pm0.14$ & $\pm0.02$ & $\mp0.04$ \\
\enddata
\end{deluxetable}

%% file: tab5.tex
\begin{deluxetable}{llllr}
\tablewidth{0pt}
\tablecaption{Comparisons between Various Spectroscopic Parameter Estimates\label{tab:stat}}
\tablehead{
  \colhead{Comparison} &
  \colhead{$\langle \Delta \teff \rangle$} &
  \colhead{$\langle \Delta \log{g} \rangle$} &
  \colhead{$\langle \Delta {\rm [Fe/H]} \rangle$} &
  \colhead{$N_{\rm star}$} \\
  \colhead{} &
  \colhead{(K)} &
  \colhead{} &
  \colhead{} &
  \colhead{} 
}
\startdata
MOOG $-$ VF05 & $+53\pm10$ & $-0.45\pm0.08$ & $-0.07\pm0.01$ & $30$\\
MOOG $-$ C10  & $+32\pm11$ & $-0.35\pm0.12$ & $-0.02\pm0.01$ & $9$ \\
SMT3 $-$ MOOG & $+37\pm4$  & $+0.35\pm0.04$ & $+0.16\pm0.02$ & $74$ \\
SMT3 $-$ SMT2 & $+51\pm4$  & $+0.09\pm0.01$ & $+0.03\pm0.00$ & $170$ \\
SMT3 $-$ VF05 & $+85\pm14$ & $-0.10\pm0.02$ & $+0.07\pm0.01$ & $36$\\
SMT3 $-$ C10  & $+84\pm25$ & $-0.08\pm0.02$ & $+0.14\pm0.04$ & $11$
\enddata
\tablecomments{All comparisons made using original metallicity estimates.}
\end{deluxetable}

%% file: tab6.tex
\begin{deluxetable*}{lcccccccccccccc}
\tablewidth{0pt}
\tabletypesize{\scriptsize}
\tablecaption{Stellar Parameters derived using SMT\label{tab:smt}}
\tablehead{
  \colhead{} &
  \multicolumn{7}{c}{SMT3} &
  \colhead{} &
  \multicolumn{6}{c}{SMT2} \\
  \cline{2-8}
  \cline{10-15}
  \colhead{HIP} &
  \colhead{$T_{\rm eff}$} &
  \colhead{$\sigma (T_{\rm eff})$} &
  \colhead{$\log {g}$} &
  \colhead{$\sigma (\log {g})$} &
  \colhead{${\rm [Fe/H]}$} &
  \colhead{$\sigma ({\rm [Fe/H]})$} &
  \colhead{${\rm [Fe/H]_{corr}}$\tablenotemark{a}} &
  \colhead{} &
  \colhead{$T_{\rm eff}$} &
  \colhead{$\sigma (T_{\rm eff})$} &
  \colhead{$\log {g}$} &
  \colhead{$\sigma (\log {g})$} &
  \colhead{${\rm [Fe/H]}$} &
  \colhead{$\sigma ({\rm [Fe/H]})$} \\
  \colhead{} &
  \colhead{(K)} &
  \colhead{(K)} &
  \colhead{} &
  \colhead{} &
  \colhead{} &
  \colhead{} &
  \colhead{} &
  \colhead{} &
  \colhead{(K)} &
  \colhead{(K)} &
  \colhead{} &
  \colhead{} &
  \colhead{} &
  \colhead{}
}
\startdata
 {\rm 1674} &   $5127$ &  $ 71$ & $4.45$ & $0.054$ & $-0.31$ & $0.019$ & $-0.38$ && $5248$ &  $37$ & $4.67$ & $0.011$ & $-0.24$ & $0.009$ \\
 {\rm 4907} &   $5528$ &  $ 24$ & $4.67$ & $0.128$ & $-0.01$ & $0.070$ & $-0.09$ && $5429$ &  $57$ & $4.54$ & $0.016$ & $-0.07$ & $0.013$ \\
 {\rm 5313} &   $5865$ &  $ 72$ & $4.49$ & $0.053$ & $-0.18$ & $0.021$ & $-0.26$ && $5839$ &  $47$ & $4.44$ & $0.011$ & $-0.19$ & $0.010$ \\
 {\rm 5521} &   $5708$ &  $ 48$ & $4.81$ & $0.133$ & $-0.02$ & $0.062$ & $-0.09$ && $5477$ &  $58$ & $4.42$ & $0.029$ & $-0.17$ & $0.015$ \\
 {\rm 9172} &   $5670$ &  $ 54$ & $4.58$ & $0.089$ & $+0.01$ & $0.072$ & $-0.06$ && $5743$ &  $69$ & $4.61$ & $0.015$ & $+0.05$ & $0.014$ \\
 {\rm 9829} &   $5634$ &  $ 31$ & $4.11$ & $0.030$ & $-0.26$ & $0.012$ & $-0.34$ && $5687$ &  $44$ & $4.19$ & $0.009$ & $-0.23$ & $0.009$ \\
 {\rm 10276} &  $5465$ &  $ 18$ & $4.67$ & $0.125$ & $-0.06$ & $0.065$ & $-0.13$ && $5360$ &  $40$ & $4.48$ & $0.013$ & $-0.13$ & $0.011$ \\
 {\rm 10629} &  $5635$ &  $ 26$ & $4.34$ & $0.033$ & $-0.42$ & $0.023$ & $-0.49$ && $5493$ &  $43$ & $4.12$ & $0.011$ & $-0.54$ & $0.011$ \\
\enddata
\tablenotetext{a}{Corrected to match the metallicity scale of VF05 (see text).}
\end{deluxetable*}

%% file: tab7.tex
\begin{deluxetable*}{lccccccccc}
\tablewidth{0pt}
\tabletypesize{\scriptsize}
\tablecaption{$\delta M_{V}$, $\Delta M_{V}$, and Lithium EWs of the KPNO Sample \label{tab:dmv}}
\tablehead{
  \colhead{} &
  \multicolumn{6}{c}{$\bv$} &
  \multicolumn{2}{c}{$\vk$} &
  \colhead{} \\
  \cline{2-7}
  \cline{8-9}
  \colhead{} &
  \multicolumn{2}{c}{} &
  \multicolumn{2}{c}{SMT3} &
  \multicolumn{2}{c}{MOOG} &
  \multicolumn{2}{c}{} &
  \colhead{} \\
  \cline{4-5}
  \cline{6-7}
  \colhead{HIP} &
  \colhead{$\delta M_{V}$} &
  \colhead{$\sigma(\delta M_{V})$} &
  \colhead{$\Delta M_{V}$} &
  \colhead{$\sigma(\Delta M_{V})$} &
  \colhead{$\Delta M_{V}$} &
  \colhead{$\sigma (\Delta M_{V})$} &
  \colhead{$\delta M_{V}$} &
  \colhead{$\sigma (\delta M_{V})$} &
  \colhead{W(Li)} \\
  \colhead{} &
  \colhead{(mag)} &
  \colhead{(mag)} &
  \colhead{(mag)} &
  \colhead{(mag)} &
  \colhead{(mag)} &
  \colhead{(mag)} &
  \colhead{(mag)} &
  \colhead{(mag)} &
  \colhead{(m\AA)}
}
\startdata
{\rm 1674} & $0.43$ & $0.12$ & $0.01$ & $0.12$ & \nodata & \nodata & $-0.93$ & $0.09$ &\nodata \\
{\rm 4907} & $0.32$ & $0.11$ & $0.14$ & $0.13$ & $0.14$ & $0.11$ & $0.26$ & $0.07$ & \nodata \\
{\rm 5313} & $0.77$ & $0.15$ & $0.45$ & $0.15$ & $0.48$ & $0.15$ & $0.51$ & $0.13$ & $80.20$ \\
{\rm 5521} & $0.32$ & $0.11$ & $0.14$ & $0.13$ & $0.03$ & $0.11$ & $0.21$ & $0.08$ & $12.17$ \\
{\rm 9172} & $0.28$ & $0.12$ & $0.12$ & $0.14$ & $0.12$ & $0.12$ & \nodata & \nodata & $69.40$ \\
{\rm 9829} & $0.32$ & $0.11$ & $-0.06$ & $0.11$ & $0.10$ & $0.11$ & $0.14$ & $0.06$ & \nodata \\
{\rm 10276} & $0.56$ & $0.17$ & $0.34$ & $0.18$ & $0.30$ & $0.17$ & $0.37$ & $0.10$ & \nodata \\
{\rm 10629} & $0.32$ & $0.13$ & $-0.19$ & $0.13$ & \nodata & \nodata & $0.06$ & $0.10$ & \nodata \\
\enddata
\end{deluxetable*}